\documentclass[useAMS,usenatbib]{mn2e}
\pdfoutput=1
\usepackage{natbib}
\usepackage{amsmath}
\usepackage{graphicx} 
\usepackage[usenames,dvipsnames,svgnames,table]{xcolor}
\usepackage{ulem}
\usepackage{url}

\newcommand{\guy}[1]{{#1}}
\newcommand{\fix}[1]{{#1}}

\newcommand{\MCMC}{Markov Chain Monte Carlo (MCMC)}

\newcommand{\Keplers}{\textit{Kepler's} }
\newcommand{\Kepler}{\textit{Kepler} }
\newcommand{\param}{\vec{\theta }}
\newcommand{\data}{D}
\newcommand{\model}{M}
\newcommand{\posterior}{p(\param | \data , \model )}
\newcommand{\prior}{p(\param | \model)}
\newcommand{\bikelihood}{p( \data | \param, \model)}
\newcommand{\evidence}{p(\data | \model)}
\newcommand{\modes}{O(\nu)}
\newcommand{\back}{B(\nu)}
\newcommand{\white}{W(\nu)}
\newcommand{\nunyq}{\nu_{\rm Nyquist}}
\newcommand{\PS}{\model (\nu)}
\newcommand{\sinc}[1]{{\rm sinc} \left( #1 \right)}
\renewcommand{\split}{\delta \nu_{\rm s}}
\newcommand{\height}{H}
\newcommand{\width}{\Gamma}
\newcommand{\mro}{n}
\newcommand{\kernel}{K_{n,l,m}(r, \theta)}
\newcommand{\rotprof}{\Omega(r, \theta)}
\newcommand{\amplitude}{A}

\title[Oscillations of exoplanet host stars]{Oscillation frequencies for 35 \Kepler solar-type planet-hosting stars using Bayesian techniques and machine learning.}
\author[G. R. Davies et al.]{G.R.~Davies$^{1,2}$, V.~Silva Aguirre$^{2}$, T.R.~Bedding$^{3,2}$, R.~Handberg$^{2,1}$, \newauthor M.N.~Lund$^{2,1}$, W.J.~Chaplin$^{1,2}$, D.~Huber$^{3,4,2}$, T.R.~White$^{5}$, O.~Benomar$^{6}$, \newauthor S.~Hekker$^{7,2}$, S.~Basu$^{8,2}$, T.L.~Campante$^{1,2}$, J.~Christensen-Dalsgaard$^{2}$, \newauthor Y.~Elsworth$^{1,2}$, C.~Karoff$^{9,2}$, H.~Kjeldsen$^{2}$, M.S.~Lundkvist$^{2}$, \newauthor T.S.~Metcalfe$^{10,2}$,  D.~Stello$^{3}$. \\
$^{1}$School of Physics and Astronomy, University of Birmingham, Birmingham, B15 2TT, United Kingdom.\\
$^{2}$Stellar Astrophysics Centre (SAC), Department of Physics and Astronomy, \\ Aarhus University, Ny Munkegade 120, DK-8000 Aarhus C, Denmark \\
$^{3}$ Sydney Institute for Astronomy (SIfA), School of Physics, University of Sydney, NSW 2006, Australia. \\
$^{4}$ SETI Institute, 189 Bernardo Avenue, Mountain View, CA 94043.\\
$^{5}$ Institut f\"{u}r Astrophysik, Georg-August-Universit\"{a}t G\"{o}ttingen, Friedrich-Hund-Platz 1, 37077 G\"{o}ttingen, Germany.\\
$^{6}$ Department of Astronomy, School of Science, The University of Tokyo, Bunkyo-ku, Tokyo 113-0033, Japan.\\
$^{7}$ Max-Planck-Institut fur Sonnensystemforschung, Justus-von-Liebig-Weg 3, 37077 Goettingen, Germany\\
$^{8}$ Department of Astronomy, Yale University, 274 J.W.Gibbs Lab, 260 Whitney Ave., New Haven, CT 06511, USA. \\
$^{9}$ Department of Geoscience, Aarhus University, H{\o}egh-Guldbergs Gade 2, 8000, Aarhus C, Denmark.\\
$^{10}$ Space Science Institute, 4750 Walnut St. Suite 205, Boulder CO 80301 USA.\\
} 
\begin{document}
\date{Accepted 1988 December 15. Received 1988 December 14; in original form 1988 October 11}
\pagerange{\pageref{firstpage}--\pageref{lastpage}} \pubyear{2014}
\maketitle
\label{firstpage}
\begin{abstract}
\Kepler has revolutionised our understanding of both exoplanets and their host stars.  Asteroseismology is a valuable tool in the characterisation of stars and \Kepler is an excellent observing facility to perform asteroseismology.  Here we select a sample of 35 \Kepler solar-type stars which host transiting exoplanets (or planet candidates) with detected solar-like oscillations. 
Using available \Kepler short cadence data up to Quarter 16 we create power spectra optimised for asteroseismology of solar-type stars.  We identify modes of oscillation and estimate mode frequencies by ``peak bagging'' using a Bayesian MCMC framework.  In addition, we expand the methodology of quality assurance using a Bayesian unsupervised machine learning approach. We report the measured frequencies of the modes of oscillation for all 35 stars and frequency ratios commonly used in detailed asteroseismic modelling. Due to the high correlations associated with frequency ratios we report the covariance matrix of all frequencies measured and frequency ratios calculated. These frequencies, frequency ratios, and covariance matrices can be used to obtain tight constraint on the fundamental parameters of these planet-hosting stars.\\ 
\end{abstract}
\section{Introduction}
The NASA \Kepler space telescope has discovered numerous extra-solar planets \citep{2014ApJS..210...19B, 2014ApJS..210...20M, 2014ApJ...784...45R, 2015ApJS..217...31M}.  It is desirable to know the ages of these exoplanets for studies of planetary formation and evolution, which generally requires estimating the ages of the host stars.  Where host stars have outer convective envelopes and measurable solar-like oscillations, it is possible to obtain tight constraints on the stellar age using asteroseismology \citep[e.g.][]{2012ApJ...748L..10M,2013ARA&A..51..353C,2013ApJ...769..141S,2014A&A...569A..21L}.
\par
The high-precision and long temporal baseline of the \Kepler observations allow the frequencies of modes of solar-like oscillators to be measured with high precision.  These oscillation frequencies can be used to estimate stellar mass, radius, and age \citep[see][]{2005A&A...441..615M, 2012ApJ...748L..10M, 2014ApJS..214...27M, 2015arXiv150407992S}, the depths of the He ionisation zone and convective zone \citep{2014ApJ...782...18M, 2014ApJ...790..138V}, and even to constrain models of stellar structure and evolution \citep{2013ApJ...769..141S}.
\par 
Ages for solar-like oscillators have been determined from asteroseismic grid modelling  for more than 500 main-sequence and sub-giant stars \citep{2014ApJS..210....1C}.  The typical precision achieved was of order $25 \%$ with a precision of better than 1 Gyr for one third of the sample.  
That method used the average properties of the frequencies, such as the mean separation between overtones (the large spacing $\Delta \nu$), or the frequency of maximum power $\nu_{\rm max}$.
However, a factor of two improvement in constraints on age are available when detailed stellar modelling can be used to compare observed individual mode frequencies (or combinations thereof) to mode frequencies from stellar models \citep{2013ApJ...769..141S,2014ApJS..214...27M,2015arXiv150407992S}. 
\par
A number of studies have estimated individual mode frequencies of \Kepler solar-type stars \citep{2012ApJ...748L..10M, 2013ApJ...766..101C, 2014ApJ...790...12B, 2015ApJ...799..170C, 2015MNRAS.446.2959D}.  \cite{2012A&A...543A..54A} have estimated frequencies for a large ensemble of solar-type stars.  The precision on individual mode frequencies is now so high that it is necessary to correct these frequencies for stellar line-of-sight velocity shifts when comparing results with model frequencies \citep{2014MNRAS.445L..94D}.  In addition, near-surface effects that cannot be easily replicated in stellar models need to be accounted for \citep{2008ApJ...683L.175K, 2013MNRAS.435..242G, 2014A&A...568A.123B}.  It has been shown that this near-surface effect can be minimised by considering combinations of mode frequencies \citep{2005A&A...434..665R, 2005MNRAS.356..671O, 2011A&A...529A..63S} or by phase matching \citep{2014arXiv1406.6491R}.
\par
In this paper we analyse \Kepler light curves for 35 stars thought to host planets and produce precise estimates of their oscillation frequencies.  We use Markov Chain Monte Carlo (MCMC) techniques \citep{2011A&A...527A..56H} to estimate the mode frequencies and Bayesian machine-learning techniques to assess the quality of the output.  The output frequencies will be used in a companion paper for detailed asteroseismic modelling, to estimate fundamental properties of each system \citep{2015arXiv150407992S}.  Therefore, we take care to provide the information necessary to assess a detailed modelling ``goodness of fit'' statistic by reporting the covariance matrices of the observational parameters.
\par
The paper is laid out as follows.  Section \ref{sec::targets} introduces the targets and their basic properties.  Section \ref{sec::method} details the method used to move from \Kepler time series to the description of the mode frequencies, frequency combinations, and quality control statistics.  Section \ref{sec::ratios_def} defines the frequency combinations we will present.  Section \ref{sec::results} gives example results for a single star with the full set of results available \guy{online}.  We draw our conclusions in section \ref{sec::conclusion}.

\section{List of targets}
\label{sec::targets}
Our sample of stars is drawn from known solar-like oscillators that are planet host candidates, designated Kepler Objects of Interest (KOIs), or are confirmed planet hosts.  The signal-to-noise ratio of the oscillation signal in each target was required to be sufficient that individual modes of oscillation are clearly present and that the angular degrees ($l$) of the modes can be identified.  We have selected a set of 35 stars that meet our criteria.  Table \ref{tab::basic} lists the basic properties of the targets.
\par
\cite{2012A&A...543A..54A} defined three categories of stars from their sample; ``simple'', ``F-like'', and ``mixed mode''.  The definitions are based on visual inspection of the \'{e}chelle diagram, in which the Fourier power spectrum is divided into segments of length of the large frequency spacing that are stacked vertically \citep{1983SoPh...82...55G}.  Simple stars have clear ridges corresponding to the radial, dipole, and quadrupole modes.  F-like stars present two broad ridges, each representing odd and even degree modes, respectively, with a potential ambiguity in the degree identification of these ridges \citep[see][]{2010ApJ...713..935B, 2008A&A...488..705A, 2009A&A...507L..13B, 2012ApJ...751L..36W} due to the larger line widths displayed by hotter solar-like pulsators \citep{2014A&A...566A..20A}.  Mixed-mode stars show dipole modes that are bumped from their respective ridges due to coupling between pressure and gravity modes.
\par
\begin{figure}
\includegraphics[width=88mm,clip]{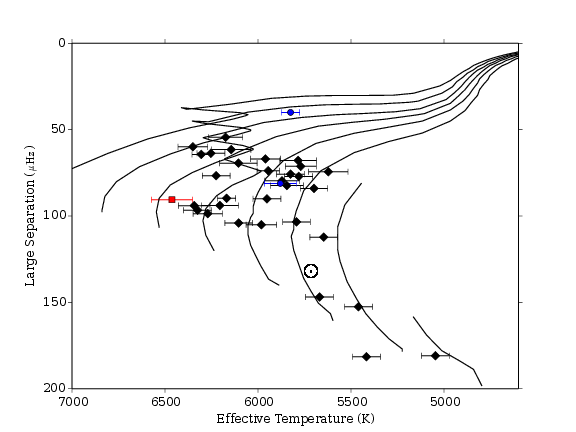}
  \caption{Large separation as a function of effective temperature for the stars in this study.  The solid lines are evolutionary tracks \citep{2008A&A...482..883M} for stars of mass from $0.8 \; \rm M_{\odot}$ (farthest right) to $1.5 \; \rm M_{\odot}$ (farthest left), in steps of $0.1 \; \rm M_{\odot}$.   The 33 {\it simple} stars are shown as black diamonds, the {\it mixed-mode} star is shown as a pale blue dot, and the {\it F-like} star is shown as a red square. }
  \label{fig::tracks}
\end{figure}

\begin{figure*}
\includegraphics[width=55mm,clip]{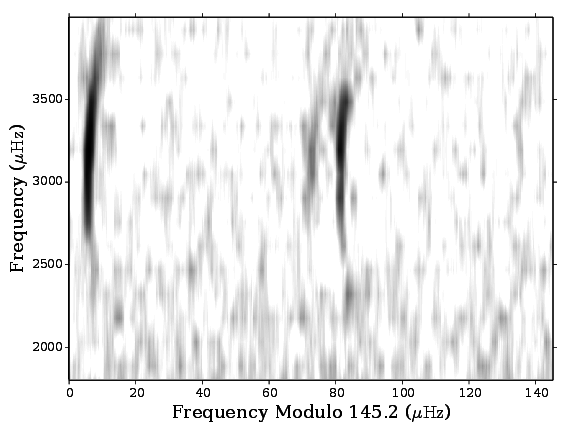}
\includegraphics[width=55mm,clip]{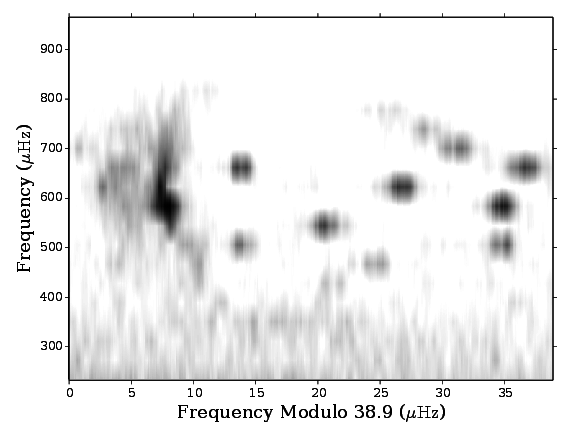}
\includegraphics[width=55mm,clip]{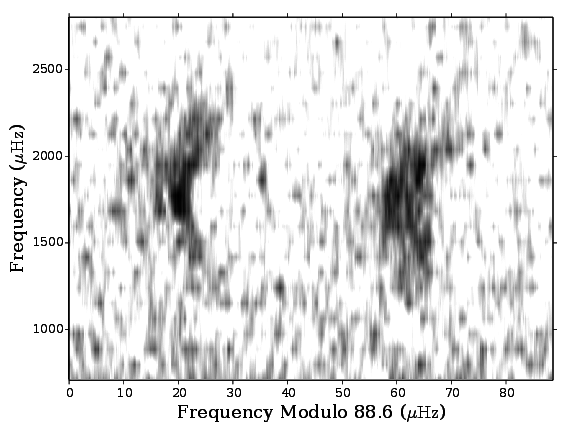}
  \caption{Echelle diagrams for three targets demonstrating the three classifications used.  Left: A ``simple'' star KIC 3544595. Centre: The ``mixed mode'' star KIC 7199397.  Right: The ``F-like'' star KIC 7670943.}
  \label{fig::types}
\end{figure*}

 
Figure \ref{fig::tracks} shows our 35 stars plotted with the global large-frequency separation as a function of effective temperature.
The sample includes one ``F-like'' star (KIC 7670943) and one ``mixed-mode'' star (KIC 7199397).  The remainder of the sample are of the ``simple'' star type.  As examples of \'{e}chelle diagrams and the differences in classification, Figure \ref{fig::types} shows \'{e}chelle diagrams for each of the three classes of star.  \guy{The greyscale of the echelle diagram represents power spectral density.  We
have rebinned the power spectrum in the horizontal direction so that each
order contains 70 bins. We did not smooth or rebin the data in the vertical
direction but we did interpolate between orders to create a smoother image.} 

\guy{\begin{table*}
   \caption{Targets and basic properties for the stars considered.  The asteroseismic parameters $\nu_{\rm max}$ and $\Delta \nu$ are taken from \citep{2013ApJ...767..127H}.  The $T_{\rm eff}$ and $\rm [Fe/H]$ parameters are taken from the companion paper \citep{2015arXiv150407992S}  .$^{a}$ \citep{2012Natur.486..375B}. $^{b}$ This work.}
   \begin{tabular}{lllllllr}
   \hline
   \hline
   KOI & KIC & ID & $\nu_{\rm max}$ & $\Delta \nu$ & $T_{\rm eff}$ & $\rm[Fe/H]$ & Notes \\
   & & & $\rm \mu Hz$ & $\rm \mu Hz$ & K & dex & \\
   \hline
   268 & 3425851 &  & $2038 \pm 60$ & $92.6 \pm 1.5$ & $6343 \pm 85$ & $-0.04 \pm 0.10$ &  \\
   69 & 3544595 & Kepler-93 & $3366 \pm 81$ & $145.77 \pm 0.45$ & $5669 \pm 75$ & $-0.18 \pm 0.10$ &  \citep{2014ApJ...790...12B} \\
   975 & 3632418 & Kepler-21 & $1153 \pm 32$ & $60.86 \pm 0.55$ & $6305 \pm 50$ & $-0.03 \pm 0.10$ &  \citep{2012ApJ...746..123H} \\
   280 & 4141376 &  & $2928 \pm 97$ & $128.8 \pm 1.3$ & $6134 \pm 91$ & $-0.24 \pm 0.10$ &  \\
   281 & 4143755 &  & $1458 \pm 57$ & $77.2 \pm 1.3$ & $5622 \pm 106$ & $-0.4 \pm 0.11$ &  \\
   244 & 4349452 & Kepler-25 & $2106 \pm 50$ & $98.27 \pm 0.57$ & $6270 \pm 79$ & $-0.04 \pm 0.10$ &  \citep{2012MNRAS.421.2342S} \\
   108 & 4914423 & Kepler-103 & $1663 \pm 56$ & $81.5 \pm 1.6$ & $5845 \pm 88$ & $0.07 \pm 0.11$ &  \citep{2014ApJS..210...20M} \\
   123 & 5094751 & Kepler-109 & $1745 \pm 117$ & $91.1 \pm 2.3$ & $5952 \pm 75$ & $-0.08 \pm 0.10$ &  \citep{2014ApJS..210...20M} \\
   85 & 5866724 & Kepler-65 & $1880 \pm 60$ & $89.56 \pm 0.48$ & $6169 \pm 50$ & $0.09 \pm 0.08$ &  \citep{2013ApJ...766..101C} \\
   285 & 6196457 & Kepler-92 & $1299 \pm 53$ & $66.6 \pm 1.1$ & $5871 \pm 94$ & $0.17 \pm 0.11$ &  \citep{2014ApJS..210...25X} \\
   3158 & 6278762 & Kepler-444 & $4538 \pm 144$ & $179.64 \pm 0.76$ & $5046 \pm 74$ & $-0.37 \pm 0.09$ &  \citep{2015ApJ...799..170C} \\
   41 & 6521045 & Kepler-100 & $1502 \pm 31$ & $77.0 \pm 1.1$ & $5825 \pm 75$ & $0.02 \pm 0.10$ &  \citep{2014ApJS..210...20M} \\
   75 & 7199397 &  & $644 \pm 8^{b}$ & $38.9 \pm 0.8^{b}$ & $5824 \pm 50^{a}$ & $-0.22 \pm 0.08^{a}$ &  \\
   269 & 7670943 &  & $1895 \pm 73$ & $88.6 \pm 1.3$ & $6463 \pm 110$ & $0.09 \pm 0.11$ &  \\
   274 & 8077137 & Kepler-128 & $1324 \pm 39$ & $68.8 \pm 0.64$ & $6105 \pm 100$ & $-0.11 \pm 0.07$ &  \citep{2014ApJS..210...25X} \\
   260 & 8292840 & Kepler-126 & $1983 \pm 37$ & $92.85 \pm 0.35$ & $6205 \pm 100$ & $-0.26 \pm 0.07$ &  \citep{2014ApJ...784...45R} \\
   122 & 8349582 & Kepler-95 & $1677 \pm 90$ & $83.6 \pm 1.4$ & $5699 \pm 74$ & $0.3 \pm 0.10$ &  \citep{2014ApJS..210...20M} \\
   245 & 8478994 & Kepler-37 & $4660 \pm 50$ & $178.7 \pm 1.4$ & $5417 \pm 75$ & $-0.32 \pm 0.07$ &  \citep{2013Natur.494..452B} \\
   370 & 8494142 & Kepler-145 & $1133 \pm 81$ & $61.8 \pm 0.76$ & $6144 \pm 106$ & $0.13 \pm 0.10$ &  \citep{2014ApJS..210...25X} \\
   5 & 8554498 &  & $1153 \pm 76$ & $61.98 \pm 0.96$ & $5945 \pm 60$ & $0.17 \pm 0.05$ &  \\
   319 & 8684730 &  & $938 \pm 72^{b}$ & $52.2 \pm 0.7^{b}$ & $5924 \pm 50^{a}$ & $0.17 \pm 0.08^{a}$ &  \\
   42 & 8866102 & Kepler-410 & $2014 \pm 32$ & $94.5 \pm 0.60$ & $6325 \pm 75$ & $0.01 \pm 0.10$ &  \citep{2014ApJ...782...14V} \\
   974 & 9414417 &  & $1115 \pm 32$ & $60.05 \pm 0.27$ & $6253 \pm 75$ & $-0.13 \pm 0.10$ &  \\
   288 & 9592705 &  & $1008 \pm 21$ & $53.54 \pm 0.32$ & $6174 \pm 92$ & $0.22 \pm 0.10$ &  \\
   1925 & 9955598 & Kepler-409 & $3546 \pm 119$ & $153.18 \pm 0.14$ & $5460 \pm 75$ & $0.08 \pm 0.10$ &  \citep{2014ApJS..210...20M} \\
   263 & 10514430 & False-positive & $1303 \pm 30$ & $70.0 \pm 1.0$ & $5784 \pm 98$ & $-0.11 \pm 0.11$ &  \\
   275 & 10586004 & Kepler-129 & $1395 \pm 40$ & $69.2 \pm 1.4$ & $5770 \pm 83$ & $0.29 \pm 0.10$ &  \citep{2014ApJ...784...45R} \\
   2 & 10666592 & HAT-P7 & $1115 \pm 110$ & $59.22 \pm 0.59$ & $6350 \pm 80$ & $0.26 \pm 0.08$ &  \citep{2008ApJ...680.1450P} \\
   1612 & 10963065 & Kepler-408 & $2193 \pm 48$ & $103.2 \pm 0.63$ & $6104 \pm 74$ & $-0.2 \pm 0.10$ &  \citep{2014ApJS..210...20M} \\
   276 & 11133306 &  & $2381 \pm 95$ & $107.9 \pm 1.9$ & $5982 \pm 82$ & $-0.02 \pm 0.10$ &  \\
   246 & 11295426 & Kepler-68 & $2154 \pm 13$ & $101.57 \pm 0.10$ & $5793 \pm 74$ & $0.12 \pm 0.07$ &  \citep{2013ApJ...766...40G} \\
   277 & 11401755 & Kepler-36 & $1250 \pm 44$ & $67.9 \pm 1.2$ & $5960 \pm 80$ & $-0.24 \pm 0.06$ &  \citep{2012Sci...337..556C} \\
   262 & 11807274 & Kepler-50 & $1496 \pm 56$ & $75.71 \pm 0.31$ & $6225 \pm 75$ & $-0.0 \pm 0.08$ &  \citep{2013MNRAS.428.1077S} \\
   7 & 11853905 & Kepler-4 & $1436 \pm 42$ & $74.4 \pm 1.1$ & $5781 \pm 76$ & $0.09 \pm 0.10$ &  \citep{2010ApJ...713L.126B} \\
   72 & 11904151 & Kepler-10 & $2730 \pm 280$ & $118.2 \pm 0.20$ & $5647 \pm 74$ & $-0.15 \pm 0.10$ &  \citep{2011ApJ...729...27B} \\
   \hline
   \end{tabular}
   \label{tab::basic}
\end{table*}
}

\section{Method}
\label{sec::method}

\subsection{Data preparation}

\Kepler short cadence pixel data spanning observing quarters Q0--Q16 were prepared following the procedures described by \citet{2014MNRAS.445.2698H}, to yield photometric time series ready for asteroseismic analyses. Planetary transits were removed using a smoothing filter on the timeseries without removing in-transit data points, thereby maintaining \Keplers exceptional duty cycle and not introducing sidelobes in the spectral window function.

\subsection{Mode identification}
\label{sec::modeid}
 ``Peak bagging'' is the process of fitting a model to the observed power spectrum in order to estimate the frequency of each mode and this is the focus here.
 \par
Mode identification is an important part of the peak bagging work we have undertaken.  The correct identification of the mode degree, $l$, is important for the model of the power spectrum defined in later sections.  For all but the hottest main-sequence solar-like pulsators, mode identification is usually straightforward.  For F-type and subgiant stars, it requires some care.  Let us first define some terminology.  For a basic introduction to solar-like oscillations see \cite{2014aste.book...60B}.
\par
\fix{
Solar-like oscillators can show p modes, g modes and mixed modes. P-mode oscillations are standing acoustic waves where pressure supplies the restroring force. Solar-like oscillators may also show g-mode oscillations, of the character of standing gravity waves where buoyancy is the restoring force. Typically we do not observe the g modes directly, but for the more evolved stars it is possible to observe mixed modes with a combined gravity-wave and acoustic character.
\par
The observable p modes are most sensitive to the conditions in the outer regions of stars.  Since gravity waves can propagate only in regions that are not convective, i.e., the radiative interior, g modes are most sensitive to the deeper regions of a star.  As a result of their p and g mixed character, mixed modes take on properties that are a combination of the p and g properties. Mixed modes therefore can be sensitive to both the deep interior and the outer region.
\par   
P-mode frequencies $\nu_{n,l}$ that have approximate regularity in the power spectrum can be expressed in terms of the radial order $n$, the angular degree $l$, and a combination of frequency separations.  This familiar asymptotic relation is 
\begin{equation}
\nu_{n,l} \approx \Delta \nu \left( n + \frac{l}{2} + \epsilon \right) - \delta \nu_{0l},
\label{eq::asy1}
\end{equation} 
where $\Delta \nu$ is the large separation, the difference in frequency between modes of consecutive $n$ but the same $l$.  The small separations $\delta \nu_{0l}$ give the difference in frequency for each $l$ relative to the radial mode of the same order.  It is easy to see using the asymptotic relation that p modes of the same $l$ produce ridges in the \'{e}chelle diagram.  The separation between these ridges of constant $l$ is described by the small separations.  Finally, the position of all the ridges in the \'{e}chelle diagram can be shifted by the $\epsilon$ term.  Each of the terms in this asymptotic expression has physical significance but we refer the reader to other works for those details \citep{1980ApJS...43..469T, 1986hmps.conf..117G, 2003Ap&SS.284..165G}.
\par
Main-sequence solar-like oscillators typically display high-order p-mode oscillations.  For main-sequence stars of K, G, or cool-F spectral type, mode identification is trivial and performed by visual inspection of the \'{e}chelle diagram of the power spectrum \cite[see][]{2012A&A...543A..54A, 2013ARA&A..51..353C}.
For hotter F-type stars mode identification is made by using the relationship between the phase term in the asymptotic relation ($\epsilon$ in equation \ref{eq::asy1}) and effective temperature \cite[see][]{2012ApJ...751L..36W}.
\par
Subgiants have mixed modes that display a more complex pattern where some modes can be strongly bumped away from the regular pattern seen in main-sequence stars.  Here, we used visual inspection to determine the radial and octupole modes ($l=0,3$) which are largely unaffected by mixing and the model of \cite{2011A&A...535A..91D, 2014ApJ...781L..29B} for dipole and quadrupole modes ($l=1,2$).
\par
At this stage of the process (mode identification) we are not considering rotational splitting, and so from a theoretical point of view each mode (p or mixed) can be described by the combination of radial order $n$ and spherical degree $l$.  For p modes, the asymptotic description gives one mode of each $l$ for each acoustic radial order, hence each p mode can be uniquely described by $n$ and $l$.  For stars with mixed modes, more than one mode of each degree can be observed in the frequency range described by each acoustic order.    The modes of a model star have a unique and well defined order \citep{2010Ap&SS.328..259D,2012ApJ...745L..33B} but for each bumped mode, the appropriate $n$ is not obvious from observations alone.
\par
For convenience and to uniquely describe all observed modes we introduce some new notation.  As we may observe more than a single mixed mode of the same degree per order we require extra notation in addition to the acoustic radial order $n_{p}$ and the degree $l$.  We introduced an identifier (or counter), $j$, that is included to differentiate between modes with the same $n_p$ and $l$.  Modes that displayed significant bumping caused by their mixed nature are labelled using $n',l$ where $n' = n_p(j), l=l$.  The identifier $j$ bears no relation to $n$, $n_{p}$, or $l$ but simply labels the mixed modes that have been fitted.  For example, $n'=20(02),l=1$ is the second observable dipole bumped mode in the 20th acoustic radial order.  Similarly, $n'=17(01), l=2$ is the first observable quadrupole bumped mode in the 17th acoustic order.  With this notation all p and mixed modes are uniquely identified by $l$ and $n'$.  For the remainder of this work we drop the subscript $p$ and use $n$ as either the acoustic order or $n'$.
\par 
We used an observational definition of mixed character in that we required bumped modes to display significant deviation from the non-bumped modes.  We insisted that bumped modes are shifted in frequency away from the nominal p-mode ridges in the \'{e}chelle diagram and showed different mode properties (i.e., mode line width and height) from nearby radial modes.  Modes that displayed only very small differences in properties were treated as non-bumped modes.
}
\par
The process of mode identification and the determination of initial frequencies provide a starting point for the peak bagging process.  We also required initial values for the mode amplitudes, line widths, rotational splitting, and angle of inclination, all of which are provided by inspection.  We used MCMC techniques to estimate mode parameters, which gives solutions that are insensitive to the initial values and hence provide excellent levels of repeatability.  
\subsection{Peak bagging}
\label{sec::pb}
We used a Bayesian approach to estimate the frequencies of oscillation modes.  Bayes' theorem for the {\it posterior} probability density function for a set of parameters, $ \param $, given observed data, $ \data $, and a model, $ \model $ states:
\begin{equation}
\posterior = \frac{\prior \, \bikelihood }{ \evidence },
\end{equation} 
where $\prior$ are the prior probabilities of the model parameters, $\bikelihood$ is the likelihood of obtaining the data given a particular choice of parameters, and $\evidence$ is the normalisation factor, known as the evidence.
\par
In the following section, we will describe our choice of model, priors, and the assessment of the likelihood, $\bikelihood$.  We finish the section by briefly describing our implementation of the \MCMC $\;$ sampler used to estimate $\posterior$.
\subsubsection{Likelihood function}
The power spectrum of Gaussian noise in the time domain is a $\chi^{2}$ distribution with 2 degrees of freedom.  To model this noise distribution we used the log likelihood function from \cite{1990ApJ...364..699A}:
\begin{equation}
\ln {\bikelihood} = \sum_{i} \left[ \ln \left[ \model_{i}(\param) \right] + \frac{\data_{i}}{\model_{i}(\param)} \right],
\end{equation}
where $i$ indexes a bin in the frequency-power spectrum.  This treatment assumes that all frequency bins are statistically independent of each other.  This is true for observations that are regularly spaced in time. However, \Kepler data are transformed from the regular cadence of the spacecraft to the frame of the solar-system barycentre.  This produces irregularly spaced time stamps and hence statistical independence can only ever be an approximation. In addition, Kepler observations contain temporal gaps in the data due to the regular data-transfers to Earth and dumping of momentum from the reaction wheels, and random events like spacecraft safe-modes, cosmic-ray strikes, and CMEs from the Sun.  \cite{2008SoPh..251...31S} have shown how to treat the correlations between bins for gapped time series but this full treatment is computationally expensive.  In reality, the full treatment provides little actual improvement in frequency estimation for high duty-cycle for \Kepler data, where correlations are weak and data behave like independent observations.  Hence, we adopted a procedure that ignores irregular sampling and gaps in the data.  We have performed tests on simulated data, with a duty-cycle taken from real \Kepler data. We found no bias in the measured mode frequencies, but some small increases in the estimated uncertainties, which is consistent with the results of \cite{2008SoPh..251...31S}.
\subsubsection{Model}
We modelled the observed power spectrum as the sum of three components, the oscillation modes $\modes$, the stellar granulation background $\back$, and a white noise component $\white$.  The high fractional duty cycle of \Kepler observations, coupled with a cadence of $\sim 58.85 \rm \, s$ causes the apodization of signals in the frequency-power spectrum \citep{2011ApJ...732...54C}.  The apodization in amplitude is given by
\begin{equation}
\eta(\nu) = \sinc{\frac{\pi}{2} \frac{\nu}{\nunyq}},
\end{equation}  
where $\nunyq$ is the Nyquist frequency, which for \Kepler short cadence data is $\approx 8496 \rm \, \mu Hz$.\\
This apodization is a result of averaging the signal over the integration time, and hence is only applicable to signals where the instantaneous measurement depends on (or is correlated with) the last state of the system.  This means that oscillation modes and granulation, which both have characteristic time scales much longer than the integration time are apodized.  However, white noise has no characteristic time scale or memory and is independent of the last state of the system, so no apodization occurs.  Hence, our model of the observed power spectrum is the sum of the three components, with modes and granulation modulated by the apodization function:
\renewcommand{\white}{W}
\begin{equation}
\PS = \white + \eta^{2}(\nu) \left[ \modes + \back \right].
\end{equation}
Note we have dropped the frequency dependency of the white noise because the function is constant across all frequencies.
\par
Oscillation modes are modelled as Lorentzian profiles.  For each mode identified in Section \ref{sec::modeid} specified by $n$ and $l$, we now assume the mode to be rotationally split into $(2l +1)$ components labelled with $m$ \citep[e.g.,][]{2004SoPh..220..269G, 2014MNRAS.439.2025D, 2014ApJ...790..121L}.  The nominal model for the set of oscillation modes in power is then \citep{2008A&A...488..705A,2013ApJ...766..101C}:
\begin{equation}
\modes = \sum_{n} \sum_{l} \sum_{m=-l}^{l} \frac{\height_{n,l,m}}{1 + 4/\width_{n,l,m}^{2} \left(\nu - \nu_{n,l,m} \right)^{2}}, 
\end{equation} 
where $\height_{n,l,m}$ and $\width_{n,l,m}$ are the height and width of each rotationally split component, respectively, and $\nu_{n,l,m}$ is the frequency.  To this nominal model we make a number of generally accepted simplifications to limit the complexity of the model fitted and to reduce the correlation between parameters.  We discuss these in turn in the following.

\paragraph*{Rotation}
Given the modest rates of rotation observed in solar-like oscillators we have neglected the contribution of centrifugal distortion to the rotational splitting in frequency \citep{2006A&A...455..621R}.  We also did not account for the possibility of a differential rotation splitting in frequency within an azimuthal order, $m$, for example due to the presence of magnetic fields.  In this prescription the frequency of each rotationally split component satisfies
\begin{equation}
\nu_{n,l,m} = \nu_{n,l} + m \, \delta \nu_{n,l}
\end{equation} 
with
\begin{equation}
\delta \nu_{n,l} \simeq \frac{1}{2 \pi} \int_{0}^{R} \int_{0}^{\pi} \kernel \, \rotprof \, r \, {\rm d}r \, {\rm d}\theta.
\end{equation}
Here, $\rotprof$ is the internal angular velocity profile in radius $r$ and co-latitude $\theta$, and $\kernel$ is a weighting kernel that represents the sensitivity of the mode to the internal rotation profile.  For modes of mixed character, the weighting kernel is non-zero in both the core and outer regions, and hence radial differential rotation can be seen in the frequency splitting \citep{2012Natur.481...55B, 2012ApJ...756...19D, 2014A&A...564A..27D}.  To allow for this in the fitting model we set   $\delta \nu_{n',l}^{\rm mixed} = \delta \nu_{n',l}$, so that the model ignores latitudinal differential rotation but allows for radial differential rotation \citep[see][]{2014ApJ...790..121L, 2015MNRAS.446.2959D}.  For all pure p modes we have assumed the rotational frequency splitting can be well approximated by a single value, which ignores differential rotation (both latitudinal and radial), by setting $\delta \nu_{n,l}^{\rm pmode} = \split$. 
\paragraph*{Mode linewidths}
The p-mode linewidth, to a good approximation, is a function only of mode frequency and is independent of degree and azimuthal order.  
Mode linewidth for the p modes changes slowly as a function of frequency.  To reduce the number of fitted parameters, we have defined a single linewidth for each order in the \'{e}chelle diagram, $\width_{n,l,m	}^{\rm pmode} = \width_{\mro}$, which will still reproduce the underlying trend in line width.
For mixed modes the linewidth depends on the strength of the p or g character of the mode \citep{2009A&A...506...57D}.  As such, it can vary with frequency strongly across a single acoustic order.  For mixed character modes, we therefore set the linewidth to be one parameter for each value of $l$ and $n$, giving $\width_{n',l,m}^{\rm mixed} = \width_{n',l}$ (i.e., one value for all rotationally split multiplets). \\
\paragraph*{Angle of inclination and mode height}
The height in power of each mode is determined by the underlying amplitude of the oscillation, but modified by a geometric weighting across the stellar disk due to partial (or complete) cancellation.
\par
The angle of inclination, $i$, of the rotation axis of the star affects the relative heights of each azimuthal component \citep[see][]{2003ApJ...589.1009G, 2013ApJ...766..101C}.  We assume that the intensity across the stellar disk depends only on angular distance from the centre, which is true for photometric observations. We also assume equipartition of energy between different azimuthal components, which is reasonable given the small predicted asymmetries (approximately $1\%$) \citep{2009A&A...508..345B} and that we are typically observing over many mode lifetimes.  The relative heights of the azimuthal components are given by \citep{2003ApJ...589.1009G}:
\begin{equation}
\varepsilon_{l,m}(i) = \frac{(l - |m|)!}{(l + |m|)!} \left[ P_{l}^{|m|} (\cos i)\right]^{2}, 
\label{eq::angle}
\end{equation}  
where $P_{l}^{|m|}$ is the Legendre function, and the sum over $\varepsilon_{l,m}(i)$ is normalized to unity.
\par
The visibility for p modes is then the height of the radial mode for the given acoustic order $H_{n}$, multiplied by the visibility $V_{l}$ and $\varepsilon_{l,m}(i)$:
\begin{equation}
\height_{n,l,m}^{\rm pmode} = \height_{\mro} \, V_{l}^{2} \, \varepsilon_{l,m}(i).
\end{equation}    
For mixed character modes we used a different parameter for each rotationally split multiplet:
\begin{equation}
\height_{n',l,m}^{\rm mixed} = \height_{n',l} \, \varepsilon_{l,m}(i).
\end{equation}    

\paragraph*{Reparametrisation}
To improve the exploration of the parameter space we made two re-parametrisations.  Mode linewidth and height are highly anti-correlated, but mode linewidth and amplitude are less strongly correlated.  Hence we re-parameterised mode height and instead explored mode amplitude $\amplitude$, where $\amplitude^{2} = (\pi / 2) \height \width$. 
\par
In terms of rotation, $\split$ and $i$ are highly anti-correlated with a curved dependence.  For angles of inclination greater than around $20^{\circ}$ the projected rotational splitting $\split \sin i$ and $i$ are virtually uncorrelated.  Below about $20^{\circ}$ there is little signature of rotation in the power spectrum, with only $m=0$ azimuthal components detectable.  Consequently, the power spectrum evidence does not normally allow us to discriminate between a true low angle of inclination and an absence of stellar rotation \citep{2006MNRAS.369.1281B, 2013ApJ...766..101C, 2014A&A...570A..54L, 2014PASJ...66...94B}.  Therefore we re-parametrised to explore projected rotation and the angle of inclination i.e., we varied $\split \sin i$ and $i$.
\paragraph*{Background}
To model the stellar background we used a single Harvey component \citep{1985ESASP.235..199H}:
\begin{equation}
B(\nu) = \frac{4 \sigma^{2} \tau}{1 + \left( 2 \pi \tau \nu \right)^{c}},
\end{equation}  
where $\sigma$ is the amplitude and $\tau$ is the characteristic time-scale of the granulation.  We followed \cite{2012A&A...543A..54A} in that we performed a first fit to the power spectrum using a simple maximum likelihood estimator in order to determine the background parameters.  The fit was made using the following model:
\begin{equation}
M_{\rm bg}(\nu) = W + \eta^{2}(\nu) B(\nu).
\end{equation} 
The region of the power spectrum below $100 \, \rm  \mu Hz$ and the region that contains the oscillation modes were excluded from the likelihood calculation.  This provided the parameters for the background which were fixed in the subsequent peak-bagging analysis.  This provided a satisfactory fit for the background in the region of the oscillation modes \citep{2012A&A...543A..54A}.  Furthermore, since the mode frequency parameters are virtually uncorrelated with the background parameters (for all sensible solutions for the background), this approach will not influence our results for estimated mode frequencies.
\subsubsection{Prior functions}
It is necessary to specify prior probability distributions to produce marginalized posterior probability distributions for the mode frequencies.  To some extent, and considering the computational resources that are available to us, it is inevitable that specifying priors on mode frequencies will have some element of subjectivity.  Very broad uninformative priors lead to instability in the peak-bagging fit.  A pragmatic approach is to define prior probability distributions based on inspection of the power spectrum and a certain degree of judgement that balances maximising the stability of the fit and minimising the impact of the prior on the posterior distribution.  It is however possible to specify a general prior on mode frequencies that can be applied once mode identification has been made. 
\paragraph*{General prior probabilities for mode frequencies}
For the ``simple'' and ``F-type'' stars in our sample, and for the radial ($l=0$) modes in the ``mixed mode'' stars, mode frequencies can be approximated by the asymptotic relation \cite{1980ApJS...43..469T}
\begin{equation}
\nu_{n,l} \approx \Delta \nu \left(n + \frac{l}{2} + \epsilon \right) - \delta \nu_{0,l},
\label{eq::asy}
\end{equation}
where $\epsilon$ is a phase offset that is observed to be correlated with the stellar effective temperature \citep{2012ApJ...751L..36W}.
\par
This asymptotic relation describes sequences, or ridges in the \'{e}chelle diagram, of frequencies for a given $l$ separated by $\Delta \nu$.  In addition, there are differences between the sequences or ridges for each $l$ that are described by $\delta \nu_{0,l}$.  We used this relation to define prior probability functions of mode frequencies that can be applied whenever a series of modes contains no mixed character modes.  The prior formulation is simply that the ridges are smoothly varying by order $n$ for a given $l$, or in other words, that the separation of the ridges is smoothly varying by order $n$.
\par
We used an approximation from our prior knowledge that the first derivative of the observed large separation $\Delta \nu_{l}(n)$ is approximately equal to zero  \citep{2012ApJ...745L..33B},
\begin{equation}
\frac{\partial \Delta \nu_{l}(n)}{\partial n} \approx 0.
\end{equation}
We applied this prior knowledge as a Gaussian prior on the large separation which has the form
\begin{equation}
- \ln p(\param _{l}) = \ln (\sqrt{2 \pi} \sigma) + \frac{1}{2} \left(\frac{\Delta \nu_{l}(n) - \Delta \nu_{l}(n-1)}{\sigma}\right)^{2},
\end{equation}
where $\sigma$ represents the uncertainty in our prior knowledge.  We rewrite this prior function in terms of a sum of the second differences in mode frequencies and rewrite the uncertainty in terms of a tolerance $\lambda_{\Delta}$ across all $\nu_{n,l}$.  This gives the prior, which constrains the large separation to be near constant, as
\begin{equation}
- \ln p(\param _{l}) = \lambda_{\Delta} \sum_{n_{\rm min}}^{n_{\rm max}} \left( \frac{\partial^{2} \nu_{n,l}}{\partial n^{2}}\right)^{2}.
\end{equation}
We applied the same concept as before to the small separation in frequency.  Here we assume that the small separation does not change from order to order,
\begin{equation}
\frac{\partial d_{l,l}(n)}{\partial n} \approx 0.
\end{equation}
which is applied as before but in the first difference of the small separation
\begin{equation}
- \ln p(\param _{l}) = \lambda_{\delta} \sum_{n_{\rm min}}^{n_{\rm max}} \left( \frac{\partial d_{l,l}(n)}{\partial n}\right)^{2}.
\end{equation}
We applied the smoothness conditions as follows.
\par
The prior on the large separation was applied to all sets of modes with $l \leq 3$ from $n_{\rm min}$ to $n_{\rm max}$.  We computed $\partial^{2} \nu_{n,l} / \partial n^{2}$ as second differences between modes of the same $l$ and varying $n$.  This calculation requires at least 5 modes of the same $l$ and consecutive $n$ to be stable.  Where this condition was not met we did not apply this prior
\par
The prior on the small separation is applied to $\delta \nu_{0,2}(n)$.  Equally, a prior could be applied to $\delta \nu_{0,1}(n)$ but this extra constraint is typically not required to achieve a stable fit.  There are not enough modes of $l=3$ to apply the $\delta \nu_{1,3}(n)$ prior.  We used a tolerance of $\lambda_{\Delta} = 0.125 \; \rm \mu Hz$ and $\lambda_{\delta} = 0.25 \rm \; \mu Hz$ which in testing on simulated data was found to work well.  For tolerances much smaller than these, it was possible for the prior to dominate and so remove information on small changes in frequency differences.  
The selected tolerances are sufficiently wide such that the data dominated the posterior information for all good signal-to-noise ratio modes.  Modes with either higher line widths or low signal-to-noise and hence large uncertainties, could be influenced by the smoothness condition.  We are satisfied that the smoothing condition did not noticeably impact the posterior probability distributions, other than to eliminate bad solutions, for modes with good signal-to-noise-ratios.
\par 
Smoothness conditions on low signal-to-noise modes can introduce undesirable correlations between mode frequencies.  When mode frequencies are dominated by the data rather than the prior, no noticeable correlation is introduced.  However, at very low signal-to-noise, where the prior dominates the data, a correlation is necessary to satisfy the prior.  We choose to apply a smoothness conditions in order to maximize the number of modes that are measured in the power spectrum.  By including modes that are somewhat dominated by the prior probability we are able to report more frequency combinations than if we were to remove the mode frequency from our results.  Because we provide the covariance matrices for all frequencies and frequency combinations, there is no concern that inclusion will influence the results of detailed stellar modelling.

\paragraph*{General priors on non-frequency parameters}
For all parameters apart from mode frequencies we applied prior probability distributions that were uniformly distributed for each parameter.  For mode line width the prior was unity in probability space between 0 and $12 \; \rm \mu Hz$ and zero elsewhere.  For mode amplitude the prior was unity between $0$ and $100 \; \rm ppm$ and zero elsewhere.  For the rotational parameters,the prior on angle of inclination was unity between 0 and $90 ^{\circ}$ and the prior on projected rotation was unity from 0 to $3 \rm \; \mu Hz$ with both set as zero elsewhere.  Priors on the mode visibility ratios were defined as unity from 0 to $2$ and zero elsewhere.  These prior probability distributions represent very wide priors designed to only eliminate bad solutions.

\subsubsection{MCMC details}

For an introduction to the techniques and terminology of Markov Chain Monte Carlo techniques see e.g., \cite{gelman2013bayesian}.
We used a standard Metropolis-Hastings MCMC algorithm.  Each fit was initiated using a maximum likelihood estimator in order to provide rapid convergence to a good solution.  After this, the MCMC algorithm was run to optimise proposal distributions and ``burn in''.  When the optimisation of the proposal distributions achieves good levels of step independence in the chain, assessed using the autocorrelation function, we switched to a mode where the proposal distributions are no longer modified.  We then ran for 200,000 steps, after which the chains are inspected for good convergence to the posterior distributions.  If the convergence is deemed acceptable then the chains were stored as the final results.  If convergence is not satisfactory we re-ran the optimisation of the proposal distributions before attempting another 200,000 step final run.  This process was repeated until the inspected results were satisfactory.

\subsection{Quality assurance}
In this section we describe a method to assess whether the mode frequencies we have obtained are of high quality and constrained by the data rather than our prior knowledge only.  We are not testing whether or not a star shows a particular mode of oscillation - our understanding of the regular structure seen in the power spectra of solar-like oscillators is sufficient that we can extrapolate where modes of low signal-to-noise ratio are likely to be located in frequency.  We {\it are} however, testing the detectability of a given mode.  Mode frequencies that have a low probability of having been detected are not necessarily wrong (the uncertainty in the mode frequency will be large to reflect this) but are probably dominated by our prior information.  It is important to regard these mode frequencies as plausible because we will use them to construct frequency combinations for which the low signal-to-noise component contributes only a small part because of the larger uncertainty and hence a small weighting in the frequency combination.
\par
\renewcommand{\det}{{\rm Det_{n,l}}}
We present a detailed method of assessing the probability of detection of a low signal-to-noise ratio mode.  The method is relatively computationally expensive, especially if all fitted modes are assessed, and explicitly estimates the probability we have detected a mode of a given $n$ and $l$ in the data, $p(\det | \data)$.  In many cases the detection is unambiguous and with $ p(\det | \data) > 0.99$.  To save computational time, we screened for these higher signal-to-noise modes with a simple and fast null hypothesis ($H_{0}$) test.
\par
In what follows we describe both the $H_{0}$ test and the procedure for estimating $p(\det | \data)$. 
\subsubsection{Null hypothesis test}
\newcommand{\chitwo}{\chi_{2}^{2}}
The null hypothesis, $H_{0}$, states that,  in a particular narrow range of frequency, the data are consistent with the presence of only noise and no signal.  A power spectrum with only broadband noise and no oscillation modes, is made up of the background noise level multiplied by a $\chi^{2}$ distribution with 2 degrees of freedom, $\chitwo$.  Dividing the spectrum by the background noise level produces an SNR spectrum that is a $\chitwo$ with a mean of unity.  It is trivial to test the probability of obtaining a given SNR for this case.  \cite{2010MNRAS.406..767B} and \cite{2012A&A...543A..54A} give the probability of observing a certain value of SNR, $s$, as
\begin{equation}
p(s | H_{0}) = \exp{(-s)}.
\end{equation}
\par
Because modes in solar-like oscillators typically have life times that are shorter than the data sets used here.  Hence, the mode line widths are greater than the power spectrum frequency resolution.  This spreads signal from each mode over a number of frequency bins.  To account for this we rebinned the SNR spectrum $O$ over $t$ bins where $t$ is an odd number, so that the resulting spectrum was
\begin{equation}
O_{i} = \frac{1}{t}\sum_{j=t(i - 1/2)}^{t(i + 1/2)} O^{j}.
\end{equation} 
\newcommand{\chitwon}{\chi^{2}_{2t}}
\newcommand{\nsmoo}{99}
\newcommand{\pdet}{0.001}
\par
This rebinned spectrum has the same noise distribution as the spectrum of $\chitwon$.  The probability of observing a value $s$ is now
\begin{equation}
p(s | H_{0}, t) = \frac{s^{t-1} \exp (-s)}{\gamma(t)},
\end{equation}
where $\gamma(t)$ is the gamma function.  To screen for modes of high SNR we tested $H_{0}$ for that mode frequency and across a number of rebinning values.  If the probability of observing the data given $H_{0}$ was below a certain threshold we accepted the mode as detected.  It is easier to determine a detection level in terms of SNR, $s_{\rm det}$.  The probability of observing a value at $s_{\rm det}$ or greater at a given rebinning is
\begin{equation}
p_{\rm det} = \frac{1}{\gamma(t)} \int_{s_{\rm det}}^{\infty} u^{t-1} \exp (-u) {\rm d}u,
\end{equation} 
where $u$ is a dummy variable representing what may be observed.  We solved the above equation to find $s_{\rm det}$ for each $t$ in $t = 1, 3, 5, ..., \nsmoo$ given a probability of rejection of $H_{0}$ of $p_{\rm det} = \pdet$.
\subsubsection{Bayesain quality control}
We implemented an unsupervised machine-learning Bayesian scheme that provided a set of probabilities to assess whether or not a mode had been detected in the data.  This quality control is similar to the work of \cite{2012A&A...543A..54A}, which we describe here in order to demonstrate the advancements.  For each mode frequency, a test was performed that asked for the probability that the mode is present in the data $p(H_{1} | \data)$. This quantity was formed from the probability of measuring the data given the null hypothesis, $p(\data | H_{0})$, and the probability of measuring the data given the detection hypothesis, $p(\data | H_{1})$.  $H_{0}$ has already been discussed.  $H_{1}$ was assessed by integrating (or marginalising) the probability of measuring the data given a model over a range of mode parameters.  \cite{2012A&A...543A..54A} marginalized over the mode height and linewidth using uniform priors from zero to some predetermined limit.
\par
Here we have expanded on this approach.  Firstly, we marginalized over the distribution of mode frequencies summarized from the posterior distributions of the peak bagging work.  In addition, we marginalized over a range of stellar rotation scenarios (rotation rate and angle of inclination) and the likely values of the local background around the mode.
\par
This method is more complete than that of \cite{2012A&A...543A..54A} in that more parameters of the peak bagging model are accounted for, but this also  makes it more complicated.  The increased number of parameters means an increased number of dimensions over which we integrate.  This lends itself to a more complicated approach rather than simply using brute force numerical integration.  We used an mixture model unsupervised machine learning algorithm comprised of either models for a background only, a single mode, or a pair of modes and assessed this using the {\it emcee hammer} \citep{2013PASP..125..306F}, an affine-invariant MCMC method.  Each mode pair was selected as the pair closest in frequency (even or odd $l$ pairs).  If any of the fitted modes in a pair of $l=2,0$ or $l=3,1$ failed the above null hypothesis test then the pair were subjected to this Bayesian test to determine the probability of detection for each of the modes of oscillation.
\par
Before we complicate the above scheme by applying it to the real world, let us consider a simple example where we wish to estimate the probability that we have detected just one mode of oscillation in a narrow range in frequency of a power spectrum.  For our purposes there are just two possibilities: there is only a background signal, $H_{0}$; and there is a background signal plus a mode of oscillation, $H_{1}$.  We can define the probability of observing the data given our model and some set of parameters $\param$ as
\begin{equation}
p(\data | \param, p_{a}) = (1 - p_{a}) \; p(\data | H_{0}, \param) + p_{a} \; p(\data | H_{1}, \param),
\label{eq::mix}
\end{equation}
where $p_{a}$ is a value between zero and one and is the probability we have detected the mode of oscillation. 
\par
Now $p_{a}$ is an unknown value that can be treated as a parameter that can be estimated.  By using Bayesian methods to estimate the posterior probability distribution and marginalized parameter estimates, we can estimate the marginalized posterior probability distribution of $p_{a}$ \citep{2010arXiv1008.4686H, 2015PhRvD..91b3005F} and hence assess the likelihood we have detected the mode of oscillation.  We must supply prior probability distributions on all parameters within $\param$.  For $p(p_{a})$ we use a uniform prior distribution between zero and unity which assigns equal probability to each possibility.
\par
We can see that in the extreme cases of an obviously detected mode, or an obviously absent mode, Equation \ref{eq::mix} reduces to the simple likelihood functions for each case.  In the case of an obviously detected mode of oscillation
\begin{equation}
\lim_{p_{a}\to1} L(\data | \param) \approx \ln p(\data | H_{1}, \param).
\end{equation}
And in the case of a clear absence of a mode of oscillation
\begin{equation}
\lim_{p_{a}\to0} L(\data | \param) \approx \ln p(\data | H_{0}, \param).
\end{equation}
For the in between cases where neither case is decisively favoured, the parameters ($\param$) for each model are (in this application) nuisance parameters and are marginalized over.  We then have access to the estimated posterior distribution of the probability that the mode of oscillation has been detected.
\par
We have applied our method to pairs of modes that are closest in frequency and over a range of frequency that is not expected to contain other modes.  This provides an easily tractable set of possibilities over which we consider.  We defined the even $l$ modes, $\nu_{n=n-1,l=2}$ and $\nu_{n=n, l=0}$, and the odd $l$ modes, $\nu_{n=n-1,l=3}$ and $\nu_{n=n, l=1}$.  From our pairs, the radial and dipole modes are much more easily detected than the quadrupole and octupole modes.  Hence, we assume that in any given pair, it is not possible to detect these weaker modes ($l=2,3$) without having detected the stronger modes ($l=0,1$).  Given this, there are three possible combinations of models:  firstly, there may be a background but no modes of oscillation;  secondly, there is a background plus a stronger ($l=0,1$) mode; finally, there is a background plus a pair of modes.  In these cases, the probability of observing the data given the set of models for the even pair is
\begin{align*}
p(\data | \param, p_{l0}, p_{l0+l2}) = &(1 - p_{l0} - p_{l0+l2}) \; p(\data | H_{0}, \param) \\ 
&+ p_{l0} \; p(\data | H_{l0}, \param) \\ &+ p_{l0+l2} \; p(\data | H_{l0+l2}, \param),
\end{align*}
where $p_{l0}$ is the probability of having detected just the $l=0$ mode and $p_{l0+l2}$ is the probability of having detected both the $l=0$ and the $l=2$ mode.  The same form applies to the odd pair
\begin{align*}
p(\data | \param, p_{l1}, p_{l1+l3}) = &(1 - p_{l1} - p_{l1+l3}) \; p(\data | H_{0}, \param) \\ 
&+ p_{l1} \; p(\data | H_{l1}, \param) \\ &+ p_{l1+l3} \; p(\data | H_{l1+l3}, \param),
\end{align*}
where $p_{l1}$ is the probability of having detected just the $l=1$ mode and $p_{l1+l3}$ is the probability of having detected both the $l=1$ and the $l=3$ mode.  The probability of having detected the stronger of the two modes is then the probability of having only detected the strong mode plus the probability of having detected both modes, that is
\begin{align}
p({\rm Det}_{l=0} | \data) &= p_{l0} + p_{l0+l2}, \\
p({\rm Det}_{l=1} | \data) &= p_{l1} + p_{l1+l3}, 
\end{align}
while the probability of having detected the weaker modes is simply
\begin{align}
p({\rm Det}_{l=2} | \data) &= p_{l0+l2}, \\
p({\rm Det}_{l=3} | \data) &= p_{l1+l3}. 
\end{align}
\par
For each hypothesis we defined the models to be tested as follows.  For the background only model ($H_{0}$), as we are testing across a narrow range in frequency, we treat the background as flat and so trivially, $M_{H_{0}}(\nu, \param) = W(\param)$.  For the model of a radial mode plus background only, we have a simple Lorentzian plus a background:
\begin{multline}
M_{H_{l0}}(\nu, \param) = \frac{H}{1 + 4/\Gamma^{2}\left( \nu - \nu_{l0}\right)^{2}} + M_{H_{0}}(\nu, \param).
\end{multline}
And the model for the even pair of modes plus a background is
\begin{multline}
M_{H_{l0+l2}}(\nu, \param) = \sum_{m=-2}^{2} \frac{0.5 \; \varepsilon_{2,m}(i) \; H}{1 + 4/\Gamma^{2}\left( \nu - \nu_{l2} - m \; \nu_{s}\right)^{2}}\\ + M_{H_{l0}}(\nu, \param).
\end{multline}
This defines the mode linewidth as a single parameter for the pair, and the mode height (cf. amplitude) as a single parameter but divided by 2 for the $l=2$ mode of oscillation \citep[see][]{2011A&A...531A.124B}.  Note that the height of each rotationally split component within the mode is regulated by the angle of inclination of the rotational axis ($i$) following equation \ref{eq::angle}.
\par
For the odd $l$ pair, the model for observing the background and just the strong mode is
\begin{multline}
M_{H_{l1}}(\nu, \param) = \sum_{m=-1}^{1} \frac{1.5 \; \varepsilon_{1,m}(i) \; H}{1 + 4/\Gamma^{2}\left( \nu - \nu_{l1} - m \; \nu_{s}\right)^{2}} \\ + M_{H_{0}}(\nu, \param),
\end{multline}
and the model for detecting the pair of modes is
\begin{multline}
M_{H_{l1+l3}}(\nu, \param) = \sum_{m=-3}^{3} \frac{0.1 \; \varepsilon_{3,m}(i) \; H}{1 + 4/\Gamma^{2}\left( \nu - \nu_{l3} - m \; \nu_{s}\right)^{2}}\\ + M_{H_{l1}}(\nu, \param).
\end{multline}
Again the mode linewidth is common across the pair, the rotational splitting is common across the pair, and again the mode heights (cf. amplitudes) have been modified according to the expected mode visibility.
\par
We defined the prior probability distributions to test only the mode frequency obtained from the peak-bagging exercises.  The prior on mode frequency was defined by the summary statistics of the posterior probability distribution obtained for each frequency.  The prior probability was then a normal distribution with mean $\nu_{n,l}$ and standard deviation $\sigma_{\nu_{n,l}}$.  If only one of the modes in a pair (the stronger) has been fitted then we guessed the prior probability for the unfitted mode from the frequency of the fitted mode and estimate of the small separation.  We used a value of ten times the uncertainty in the strong mode for the distribution of the prior on a weak mode that has not been fitted.  For all other types of parameter, we adopted generic prior probability distributions that were uniform.
\par
Mode line widths are smaller at lower frequency and larger at higher frequency, as well as being a function of the stellar effective temperature \citep{2014A&A...566A..20A}.  We placed a lower limit on mode line width of $0.1 \rm \; \mu Hz$.  The upper limit was determined by the mode frequency relative to $\nu_{\rm max}$.  For pairs of modes where $\nu_{nl} < \nu_{\rm max}$ (using the strong mode to determine $\nu_{n,l}$) the upper limit on line width was $5.0 \rm \; \mu Hz$.  For pairs of modes with $\nu_{n,l} \geq \nu_{\rm max}$ the upper limit was $12 \rm \; \mu Hz$.  Both of these values are higher than the line widths expected for even the hottest solar-like oscillators with the largest linewidths.  Consistent with \cite{2012A&A...543A..54A}, we set the prior on amplitude to be uniform over the range zero to $15 \rm \; ppm^{2} \; \mu$.  Priors on rotational splitting were uniform from zero to twice the value estimated in the peak bagging stage, and the prior on angle of inclination was uniform over zero to $90^{\circ}$.
\par
As we have stated already, we used the parallel-tempered affine-invariant MCMC ensemble sampler \citep{2013PASP..125..306F} to explore the parameter space.  This Bayesian quality control machinery outputs the probabilities that we have detected the mode of oscillation.  We report the median of the posterior probability distributions of the natural log of the ratio of the probability of a detection over the probability of no detection, i.e., the Bayes factor $K$,
\begin{equation}
\ln K = \ln{p(\data | {\rm Det}_{l})} - \ln{p(\data | H_{0})}.
\end{equation}
The reported values then can be qualitatively assessed on the \cite{kass1995bayes} scale, such that:
\begin{equation*}
\ln K = \begin{cases}
	< 0 &\text{favours $H_{0}$}\\
	\text{0 to 1} &\text{not worth more than a bare mention}\\
	\text{1 to 3} &\text{positive}\\
	\text{3 to 5} &\text{strong}\\
	> 5 &\text{very strong}.
\end{cases}
\end{equation*} 
\par
We report all modes that have been fitted.  Because we report not only the value and uncertainty for the frequency, but also the probability that we have made a detection, decisions about the use of each mode frequency can be assessed at the stage of detailed modelling.

\section{Defining frequency combinations}
\label{sec::ratios_def}
It is well known that acoustic oscillations in the outer layers of solar-type stars are poorly modelled by stellar oscillation codes.  Corrections for the ``surface effect'' are possible \citep{2008ApJ...683L.175K, 2013MNRAS.435..242G, 2014A&A...568A.123B} but combinations of frequencies have been shown \citep{2013A&A...560A...2R} to be useful because they are approximately independent of the structure in the outer layers.  In addition to being insensitive to problems related to modelling the near-surface layers, frequency ratios are also insensitive to the Doppler frequency shifts caused by the intrinsic line-of-sight motion of a star \citep{2014MNRAS.445L..94D}.
\par  
These frequency combinations are measurable quantities that can then be compared with a stellar model.  Because of the way the frequency combinations (or ratios) are constructed each measurement is not independent of all others.  It is therefore important to have access to the covariance matrix for these combinations.  For this paper we estimated mode frequencies using a Markov Chain Monte Carlo (MCMC) approach.  A consequence of having access to the Markov chains for all parameters is that we can construct covariance matrices for measured parameters and their combinations.  Hence, we present the covariance matrix as an intrinsic result that must be considered and incorporated when performing likelihood calculations using frequency combinations.
\par
We used the definition of the large separations, $\Delta \nu_{l}(n)$, as
\begin{align}
\Delta \nu_{0}(n) &= \nu_{n,0} - \nu_{n-1,0}, \\
\Delta \nu_{1}(n) &= \nu_{n,1} - \nu_{n-1,1}.
\end{align}
The small separations $d_{l,l}(n)$ are defined as
\begin{align}
\delta \nu_{0,2}(n) &= \nu_{n,0} - \nu_{n-1,2}, \\
\delta \nu_{0,1}(n) &= \frac{1}{8} \left( \nu_{n-1,0} - 4\nu_{n-1,1} + 6\nu_{n,0} - 4\nu_{n,1} + \nu_{n+1,0} \right), \\
\delta \nu_{1,0}(n) &= \frac{1}{8} \left( \nu_{n-1,1} - 4\nu_{n,0} + 6\nu_{n,1} - 4\nu_{n+1,0} + \nu_{n+1,1} \right).
\end{align}
The frequency combinations we used are those of \cite{2003A&A...411..215R}, i.e.,
\begin{align}
r_{02}(n) &= \frac{\delta \nu_{0,2}(n)}{\Delta \nu_{1}(n)}, \\
r_{01}(n) &= \frac{\delta \nu_{0,1}(n)}{\Delta \nu_{1}(n)}, \\
r_{10}(n) &= \frac{\delta \nu_{1,0}(n)}{\Delta \nu_{0}(n+1)}.
\end{align}  
We report not only the raw fitted frequencies but also the frequency ratios and the complete covariance matrix of the measured frequencies, and frequency combinations.

\section{Results}
\label{sec::results}
In this section we describe the outputs using KIC 11295426 as an example.  Results for all other stars can be found in the \guy{online} appendices.
\subsection{Frequencies}
The MCMC peak bagging method calculates the marginalized posterior probability distribution for mode frequencies.  We have to summarise the posterior probability distribution.  We use only the sections of the Markov chain that have already been ``burnt in'', that is the MCMC sampler has reached the stage where the stationary posterior distribution is being explored.  For the central value of the distribution we quote the median of the values of the fully ``burned in'' Markov chain that represents the posterior distribution.  The $68\%$ confidence intervals are estimated and quoted as the standard deviation of the same Markov chain.
\par
For modes of oscillation with good signal-to-noise ratios in the power spectrum it is common for the marginalized posterior probability distribution to have the form of a Gaussian distribution with negligible higher-order moments.  In this scenario, the use of the median and standard deviation to describe the posterior make excellent choices for summary statistics.
\par
In the case of modes with very low signal-to-noise ratios in the power spectrum, the posterior distributions may not appear as clean Gaussian distributions.  A posterior distribution may contain multiple modes of solutions (often more than just bimodal) or a single solution that is approximately Gaussian with an elevated background.  In these cases it is difficult to define simple summary statistics that properly describe the posterior distributions.  Here the use of the median value and the standard deviation of the Markov Chains provides a compromise between accurately reproducing the posterior distribution and finding a summary statistic that can be widely applied.  Typically the standard deviation of the posterior distributions for low signal-to-noise ratio modes is very large compared to uncertainties in the frequencies of other modes which mitigates the poor approximation of the posterior distribution.
\par
\begin{figure}
   \includegraphics[width=80mm]{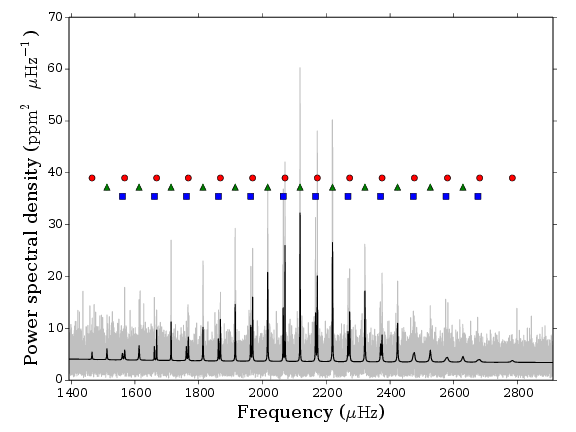}
   \includegraphics[width=80mm]{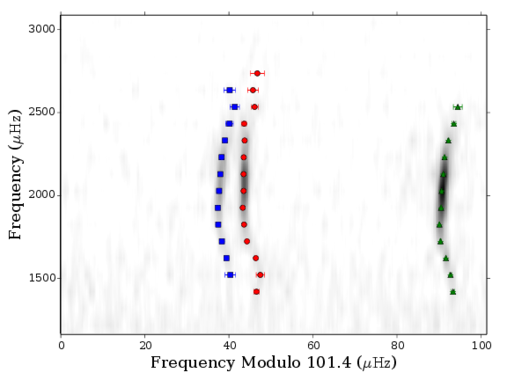}
   \caption{Power spectrum and echelle diagram for KIC 11295426.  Top: Power spectrum with data in grey smoothed over $3 \; \rm \mu Hz$ and best model in black.  Bottom: Echelle diagram with power in grey-scale.  Both: Mode frequencies are marked as: radial modes with red circles; dipole modes with green diamonds; quadrapole modes with blue squares; and octopole modes with yellow pentagons.}   \label{fig::11295426}\end{figure}

Table \ref{tab::11295426} contains the degree and summary statistics for all modes found on inspection of the frequency \'{e}chelle diagram and then fitted using the model described in section \ref{sec::pb} for KIC 11295426.  For modes that fail the null hypothesis tests and are hence deemed to be significant  enough to not require further $H_{\rm Det}$ tests we assigned a value of $ \log{p(\data|{\rm Det})/p(\data|H_{0})} = 6.91$ that would be consistent with $p(\data|{\rm Det}) = 1 - p(\data|H_{0}) \approx 0.999$.  \guy{For $\ln K$ greater than 6 we simply record a value of $> 6$.}  Figure \ref{fig::11295426} shows the region of the power spectrum that contains the modes of oscillation but presented as the power spectrum with the ``best fitting model'' and the \'{e}chelle diagram.  In each plot the mode frequencies shown in table \ref{tab::11295426} are over plotted, together with $68\%$ confidence intervals in the \'{e}chelle diagram.  It is a testament to the precision of these measurements that even in the \'{e}chelle diagram, the confidence intervals are often much smaller than the symbol that is plotted.
\par
Note that in table \ref{tab::11295426} we include a value of $n$.  This $n$ is not necessarily the same $n$ as would be used to describe the radial order of the eigenfunction of the oscillation.  Here we persist in displaying an ``observer's'' $n$ as it is useful when following which mode frequencies are used to construct the frequency ratios dealt with next.
\begin{table}
   \caption{Mode frequencies and statistics for KIC 11295426.}
   \begin{tabular}{ccccc}
   \hline 
   \hline 
   $n$ & $l$ & Frequency & 68\% credible & $ \ln{K}$  \\ 
 & & ($\rm \mu Hz$) & ($\rm \mu Hz$) & \\ 
   \hline 
   14 & 0 & 1465.55 & 0.58 & 1.58 \\ 
   14 & 1 & 1512.33 & 0.4 & $>6$ \\ 
   14 & 2 & 1560.71 & 1.25 & -0.14 \\ 
   15 & 0 & 1567.82 & 0.98 & 2.46 \\ 
   15 & 1 & 1613.09 & 0.29 & $>6$ \\ 
   15 & 2 & 1661.12 & 0.17 & $>6$ \\ 
   16 & 0 & 1668.11 & 0.12 & $>6$ \\ 
   16 & 1 & 1713.36 & 0.08 & $>6$ \\ 
   16 & 2 & 1761.4 & 0.19 & $>6$ \\ 
   17 & 0 & 1767.37 & 0.2 & $>6$ \\ 
   17 & 1 & 1813.43 & 0.12 & $>6$ \\ 
   17 & 2 & 1861.87 & 0.15 & $>6$ \\ 
   18 & 0 & 1868.02 & 0.11 & $>6$ \\ 
   18 & 1 & 1914.52 & 0.08 & $>6$ \\ 
   18 & 2 & 1963.14 & 0.14 & $>6$ \\ 
   19 & 0 & 1969.07 & 0.09 & $>6$ \\ 
   19 & 1 & 2016.27 & 0.07 & $>6$ \\ 
   19 & 2 & 2064.8 & 0.07 & $>6$ \\ 
   20 & 0 & 2070.63 & 0.05 & $>6$ \\ 
   20 & 1 & 2117.76 & 0.05 & $>6$ \\ 
   20 & 2 & 2166.43 & 0.11 & $>6$ \\ 
   21 & 0 & 2171.99 & 0.07 & $>6$ \\ 
   21 & 1 & 2219.52 & 0.07 & $>6$ \\ 
   21 & 2 & 2268.12 & 0.13 & $>6$ \\ 
   22 & 0 & 2273.38 & 0.1 & $>6$ \\ 
   22 & 1 & 2321.17 & 0.1 & $>6$ \\ 
   22 & 2 & 2370.27 & 0.22 & $>6$ \\ 
   23 & 0 & 2374.95 & 0.13 & $>6$ \\ 
   23 & 1 & 2423.44 & 0.15 & $>6$ \\ 
   23 & 2 & 2472.68 & 0.9 & $>6$ \\ 
   24 & 0 & 2476.19 & 0.53 & $>6$ \\ 
   24 & 1 & 2526.09 & 0.47 & $>6$ \\ 
   24 & 2 & 2575.31 & 1.11 & 0.51 \\ 
   25 & 0 & 2580.03 & 0.84 & 3.32 \\ 
   25 & 1 & 2628.39 & 1.05 & 2.33 \\ 
   25 & 2 & 2675.45 & 1.38 & 0.65 \\ 
   26 & 0 & 2680.99 & 1.3 & 2.42 \\ 
   27 & 0 & 2783.39 & 1.65 & 1.15 \\ 
   \hline 
   \end{tabular} 
   \label{tab::11295426}\end{table} 

\subsection{Ratios}
We computed the frequency ratios defined in section \ref{sec::ratios_def} using the full information in the ``burned in'' Markov chains.  That is, we took the chain for each mode frequency, and created a new chain that is the combination of many chains.  This method allows us to propagate all correlations in mode frequency into the frequency ratios presented here.  As for the mode frequencies, we quote the central value of the distribution as the median of the frequency ratio Markov chain and the standard deviation as the $68\%$ credible region.
\par
\begin{figure}
   \includegraphics[width=80mm]{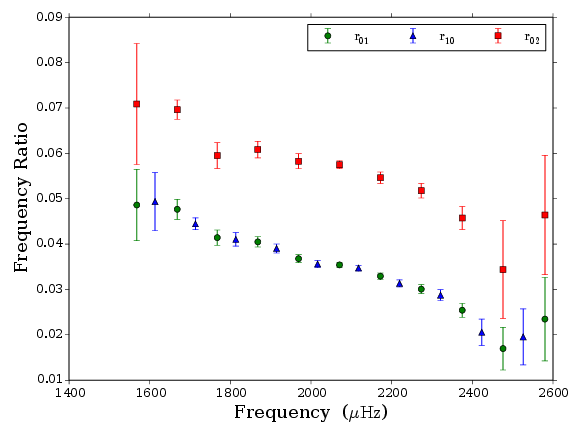}
   \caption{Ratios and $67 \%$ confidence intervals as a function of frequency for KIC 11295426.}   \label{fig::rat_11295426}\end{figure}

Table \ref{tab::rat_11295426} gives the values for the calculated mode frequency ratios that can be fully described by the set of mode frequencies identified and fitted.  The values in table \ref{tab::rat_11295426} are plotted in figure \ref{fig::rat_11295426} as a function of the frequency of the dominant mode $\nu_{n,d}$, defined as,
\begin{equation}
\nu_{n,d} = \begin{cases}
			\nu_{n,0} & \text{if } r = r_{01}(n)\\
			\nu_{n,1} & \text{if } r = r_{10}(n)\\
			\nu_{n,0} & \text{if } r = r_{02}(n).
\end{cases}
\end{equation}

\begin{table} 
   \caption{Ratios for KIC 11295426.} 
   \begin{tabular}{cccc} 
   \hline 
   \hline 
   Ratio type & $n$ & Ratio & 68\% credible interval \\ 
   \hline 
   $r_{01}$ & 15 & 0.049 & 0.008 \\ 
   $r_{10}$ & 15 & 0.049 & 0.006 \\ 
   $r_{01}$ & 16 & 0.048 & 0.002 \\ 
   $r_{10}$ & 16 & 0.044 & 0.001 \\ 
   $r_{01}$ & 17 & 0.041 & 0.002 \\ 
   $r_{10}$ & 17 & 0.041 & 0.001 \\ 
   $r_{01}$ & 18 & 0.04 & 0.001 \\ 
   $r_{10}$ & 18 & 0.039 & 0.001 \\ 
   $r_{01}$ & 19 & 0.037 & 0.001 \\ 
   $r_{10}$ & 19 & 0.036 & 0.001 \\ 
   $r_{01}$ & 20 & 0.035 & 0.001 \\ 
   $r_{10}$ & 20 & 0.035 & 0.001 \\ 
   $r_{01}$ & 21 & 0.033 & 0.001 \\ 
   $r_{10}$ & 21 & 0.031 & 0.001 \\ 
   $r_{01}$ & 22 & 0.03 & 0.001 \\ 
   $r_{10}$ & 22 & 0.029 & 0.001 \\ 
   $r_{01}$ & 23 & 0.025 & 0.002 \\ 
   $r_{10}$ & 23 & 0.021 & 0.003 \\ 
   $r_{01}$ & 24 & 0.017 & 0.005 \\ 
   $r_{10}$ & 24 & 0.02 & 0.006 \\ 
   $r_{01}$ & 25 & 0.023 & 0.009 \\ 
   $r_{02}$ & 15 & 0.071 & 0.013 \\ 
   $r_{02}$ & 16 & 0.07 & 0.002 \\ 
   $r_{02}$ & 17 & 0.06 & 0.003 \\ 
   $r_{02}$ & 18 & 0.061 & 0.002 \\ 
   $r_{02}$ & 19 & 0.058 & 0.002 \\ 
   $r_{02}$ & 20 & 0.057 & 0.001 \\ 
   $r_{02}$ & 21 & 0.055 & 0.001 \\ 
   $r_{02}$ & 22 & 0.052 & 0.002 \\ 
   $r_{02}$ & 23 & 0.046 & 0.002 \\ 
   $r_{02}$ & 24 & 0.034 & 0.011 \\ 
   $r_{02}$ & 25 & 0.046 & 0.013 \\ 
   \hline 
   \end{tabular} 
   \label{tab::rat_11295426}\end{table} 

\subsection{Correlations}
Frequency ratios, as constructed here with up to 5 mode frequencies contributing to the calculated ratio, are heavily correlated with each other as well as with mode frequencies.  In the process of detailed asteroseismic modelling we wish to compare the observed mode frequencies and ratios with the same frequencies and ratios obtained from models of stellar evolution and pulsation.  To perform a rigorous calculation of the likelihood of getting the observed data given some model it is essential to account for the correlations between observations.  These correlations can be accounted for with the covariance (or correlation) matrix of the observations.\par
\begin{figure}
   \includegraphics[width=80mm]{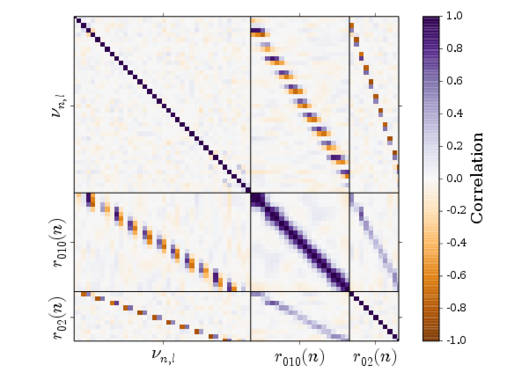}
   \caption{Correlation matrix of all frequencies and ratios for KIC 11295426.  The grid represents the matrix and hence the identity elements are all correlation 1.0.  The matrix is constructed so that frequencies and ratios are grouped separately.  If the each matrix element is labelled $[i,j]$ then the first set of $i$ contains the mode frequencies, the second set contains the $r_{010}$, and the final set the $r_{02}$.}   \label{fig::cor_11295426}\end{figure}

Figure \ref{fig::cor_11295426} shows a visualisation of the correlation matrix of all measured frequencies and calculated frequency ratios.  These correlation matrices are calculated from the ``burned in'' Markov chains and the chains of the derived quantities.  This method of using the Markov chains will account for correlations between individual mode frequencies and how these carry through into the calculation of the frequency ratios.

\section{Conclusion}
\label{sec::conclusion}
In this paper we have determined the mode frequencies for 35 solar-type stars that are thought to host planets.  These mode frequencies are to be used in the accompanying paper \citep{2015arXiv150407992S} to determine fundamental properties of these host stars.  We have performed mode identification, taken care to report all necessary properties from the peak bagging process, and estimated the probability of the mode detection.  Mode frequencies have been used to construct frequency combinations often used in detailed asteroseismic modelling and the covariance matrices of frequencies and frequency combinations are supplied.
\par
We have used Bayesian Markov Chain Monte Carlo (MCMC) methods to estimate frequency parameters.  We have defined general prior probabilities that encode the ridge-like structure of solar-like p modes into the parameter estimation.  These general priors provide an objective methodology that helps constrain the fit parameter space and penalises solutions that do not agree with the asymptotic theory of p modes.  In most cases, the introduction of these smoothness condition priors does not introduce significant correlation between frequencies.  However, in cases of low signal-to-noise ratios and or large mode line widths additional undesirable correlations are introduced.  We mitigate this by providing the magnitude of this correlation in the covariance matrix and hence, provided the covariance matrix is used in asteroseismic modelling, there is no cause for concern.
\par
\guy{We have made improvements} in the tests of mode detection originally presented in \cite{2012A&A...543A..54A} by expanding the marginalisation of mode parameters to include properties of stellar rotation and the local background noise levels.  By expanding the possible set of models included in the test of mode detections we expect more robust detections at even lower signal-to-noise ratios.  To cope with the expanded parameter dimensions we evaluate the test using mixture models assessed using a Bayesian framework.  With each mode frequency we report the natural log of the Bayes factor of detection and include details of a qualitative assessment of the measure.

\section*{Acknowledgements} 
\guy{We thank Will M. Farr and Enrico Corsaro for helpful discussions.}
GRD, WJC, TLC, amd YE acknowledge the support of the UK Science and Technology Facilities Council (STFC). 
The computations described in this paper were performed using the University of Birmingham's BlueBEAR HPC service, which provides a High Performance Computing service to the University's research community. 
Funding for the Stellar Astrophysics Centre is provided by The Danish National Research Foundation (Grant DNRF106). The research is supported by the ASTERISK project (ASTERoseismic Investigations with SONG and Kepler) funded by the European Research Council (Grant agreement no.: 267864).
DH acknowledges support by the Australian Research Council's Discover Projects funding scheme (project number DE140101364) and support by the National Aeronautics and Space Administration under Grant NNX14AB92G issued through the Kepler Participating Scientist Program.
The research leading to the presented results has received funding from the European Research Council under the European Community's Seventh Framework Programme (FP7/2007-2013) / ERC grant agreement no 338251 (StellarAges).
CK acknowledge support from the Villum Foundation.
TSM is supported by NASA grant NNX13AE91G.
\bibliographystyle{mn2e_new}
\bibliography{refs}

\appendix
\label{app::appendix}
\clearpage
\section{Plots and tables of results for all stars}
\subsubsection{3425851}
\begin{figure}
   \includegraphics[width=80mm]{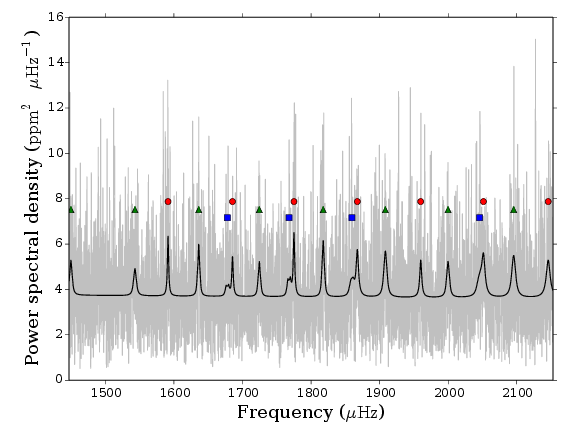}
   \includegraphics[width=80mm]{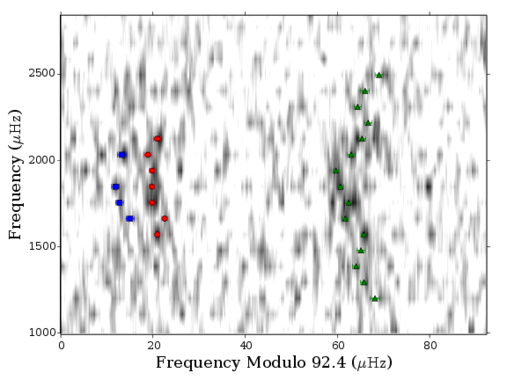}
   \caption{Power spectrum and echelle diagram for KIC 3425851.  Top: Power spectrum with data in grey smoothed over $3 \; \rm \mu Hz$ and best model in black.  Bottom: Echelle diagram with power in grey-scale.  Both: Mode frequencies are marked as: radial modes with red circles; dipole modes with green diamonds; quadrapole modes with blue squares; and octopole modes with yellow pentagons.}   \label{fig::3425851}\end{figure}
\begin{figure}
   \includegraphics[width=80mm]{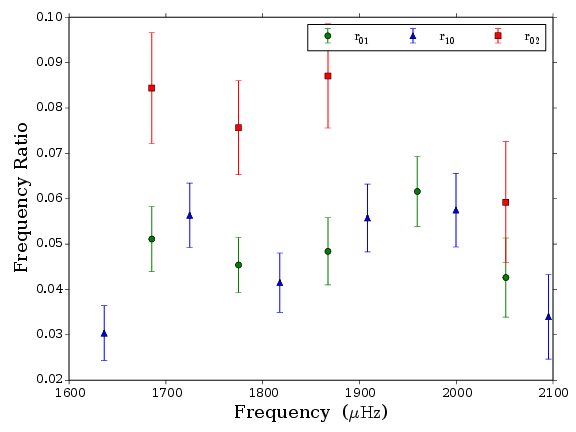}
   \caption{Ratios and $67 \%$ confidence intervals as a function of frequency for KIC 3425851.}   \label{fig::rat_3425851}\end{figure}
\begin{figure}
   \includegraphics[width=80mm]{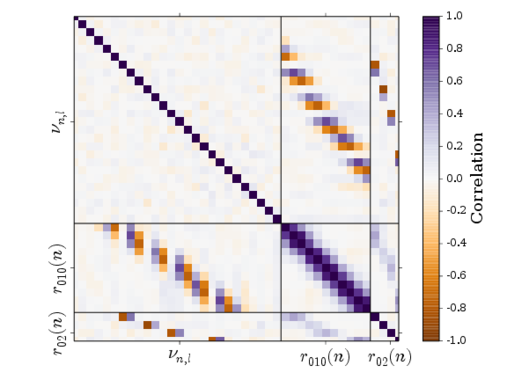}
   \caption{Correlation matrix of all frequencies and ratios for KIC 3425851.  The grid represents the matrix and hence the identity elements are all correlation 1.0.  The matrix is constructed so that frequencies and ratios are grouped separately.  If each matrix element is labelled $[i,j]$ then the first set of $i$ contains the mode frequencies, the second set contains the $r_{010}$, and the final set the $r_{02}$.}   \label{fig::cor_3425851}\end{figure}
\clearpage
\begin{table}
   \caption{Mode frequencies and statistics for KIC 3425851.}
   \begin{tabular}{ccccc}
   \hline 
   \hline 
   $n$ & $l$ & Frequency & 68\% credible & $ \ln{K}$  \\ 
 & & ($\rm \mu Hz$) & ($\rm \mu Hz$) & \\ 
   \hline 
   12 & 1 & 1269.15 & 1.01 & 0.87 \\ 
   13 & 1 & 1359.15 & 0.74 & 0.93 \\ 
   14 & 1 & 1449.84 & 0.78 & 1.34 \\ 
   15 & 1 & 1543.24 & 0.94 & 0.8 \\ 
   16 & 0 & 1591.46 & 0.44 & $>6$ \\ 
   16 & 1 & 1636.19 & 0.55 & $>6$ \\ 
   16 & 2 & 1677.91 & 0.86 & -0.29 \\ 
   17 & 0 & 1685.43 & 0.63 & 1.11 \\ 
   17 & 1 & 1724.59 & 0.66 & 1.69 \\ 
   17 & 2 & 1768.02 & 0.87 & -0.1 \\ 
   18 & 0 & 1775.08 & 0.44 & 2.21 \\ 
   18 & 1 & 1817.74 & 0.59 & $>6$ \\ 
   18 & 2 & 1859.56 & 0.87 & 0.25 \\ 
   19 & 0 & 1867.46 & 0.62 & 1.85 \\ 
   19 & 1 & 1908.34 & 0.7 & 1.6 \\ 
   20 & 0 & 1959.85 & 0.61 & 1.31 \\ 
   20 & 1 & 1999.73 & 0.66 & $>6$ \\ 
   20 & 2 & 2045.68 & 0.97 & 0.03 \\ 
   21 & 0 & 2051.33 & 0.79 & 1.83 \\ 
   21 & 1 & 2095.43 & 0.84 & 1.88 \\ 
   22 & 0 & 2145.72 & 0.8 & 1.53 \\ 
   22 & 1 & 2190.19 & 0.84 & 2.78 \\ 
   23 & 1 & 2283.89 & 0.85 & 1.14 \\ 
   24 & 1 & 2373.99 & 0.79 & 0.91 \\ 
   25 & 1 & 2467.99 & 0.92 & 1.3 \\ 
   26 & 1 & 2563.4 & 0.86 & 1.66 \\ 
   \hline 
   \end{tabular} 
   \label{tab::3425851}\end{table} 
\begin{table} 
   \caption{Ratios for KIC 3425851.} 
   \begin{tabular}{cccc} 
   \hline 
   \hline 
   Ratio type & $n$ & Ratio & 68\% credible interval \\ 
   \hline 
   $r_{10}$ & 16 & 0.03 & 0.006 \\ 
   $r_{01}$ & 17 & 0.051 & 0.007 \\ 
   $r_{10}$ & 17 & 0.056 & 0.007 \\ 
   $r_{01}$ & 18 & 0.045 & 0.006 \\ 
   $r_{10}$ & 18 & 0.042 & 0.007 \\ 
   $r_{01}$ & 19 & 0.048 & 0.007 \\ 
   $r_{10}$ & 19 & 0.056 & 0.007 \\ 
   $r_{01}$ & 20 & 0.062 & 0.008 \\ 
   $r_{10}$ & 20 & 0.057 & 0.008 \\ 
   $r_{01}$ & 21 & 0.043 & 0.009 \\ 
   $r_{10}$ & 21 & 0.034 & 0.009 \\ 
   $r_{02}$ & 17 & 0.084 & 0.012 \\ 
   $r_{02}$ & 18 & 0.076 & 0.01 \\ 
   $r_{02}$ & 19 & 0.087 & 0.011 \\ 
   $r_{02}$ & 21 & 0.059 & 0.013 \\ 
   \hline 
   \end{tabular} 
   \label{tab::rat_3425851}\end{table} 
\clearpage
\subsubsection{3544595}
\begin{figure}
   \includegraphics[width=80mm]{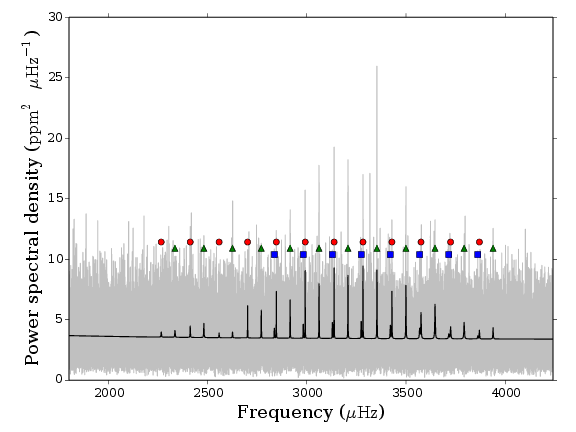}
   \includegraphics[width=80mm]{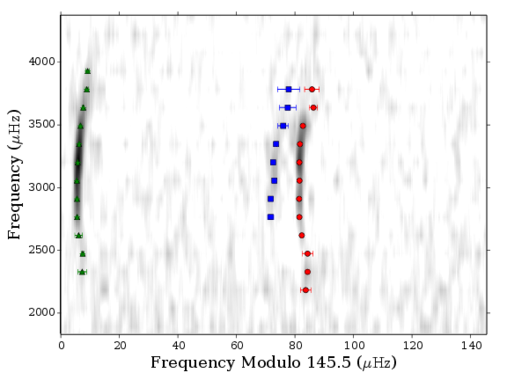}
   \caption{Power spectrum and echelle diagram for KIC 3544595.  Top: Power spectrum with data in grey smoothed over $3 \; \rm \mu Hz$ and best model in black.  Bottom: Echelle diagram with power in grey-scale.  Both: Mode frequencies are marked as: radial modes with red circles; dipole modes with green diamonds; quadrapole modes with blue squares; and octopole modes with yellow pentagons.}   \label{fig::3544595}\end{figure}
\begin{figure}
   \includegraphics[width=80mm]{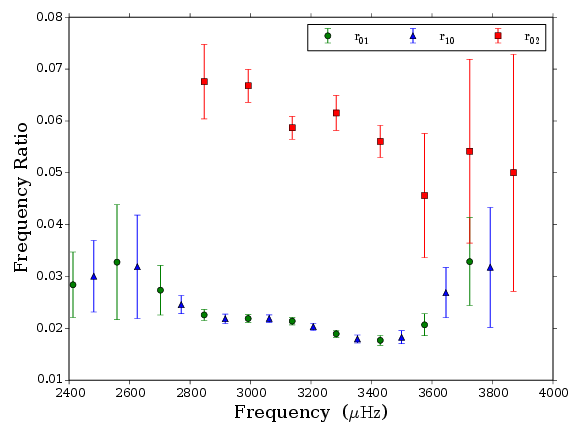}
   \caption{Ratios and $67 \%$ confidence intervals as a function of frequency for KIC 3544595.}   \label{fig::rat_3544595}\end{figure}
\begin{figure}
   \includegraphics[width=80mm]{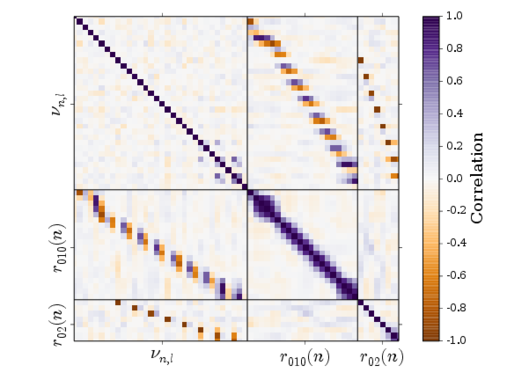}
   \caption{Correlation matrix of all frequencies and ratios for KIC 3544595.  The grid represents the matrix and hence the identity elements are all correlation 1.0.  The matrix is constructed so that frequencies and ratios are grouped separately.  If each matrix element is labelled $[i,j]$ then the first set of $i$ contains the mode frequencies, the second set contains the $r_{010}$, and the final set the $r_{02}$.}   \label{fig::cor_3544595}\end{figure}
\clearpage
\begin{table}
   \caption{Mode frequencies and statistics for KIC 3544595.}
   \begin{tabular}{ccccc}
   \hline 
   \hline 
   $n$ & $l$ & Frequency & 68\% credible & $ \ln{K}$  \\ 
 & & ($\rm \mu Hz$) & ($\rm \mu Hz$) & \\ 
   \hline 
   15 & 0 & 2266.65 & 1.78 & 0.99 \\ 
   15 & 1 & 2335.79 & 1.5 & 0.76 \\ 
   16 & 0 & 2412.81 & 0.53 & 2.13 \\ 
   16 & 1 & 2481.48 & 0.52 & 1.41 \\ 
   17 & 0 & 2558.34 & 1.79 & 0.6 \\ 
   17 & 1 & 2625.64 & 1.31 & 1.09 \\ 
   18 & 0 & 2701.9 & 0.19 & $>6$ \\ 
   18 & 1 & 2770.65 & 0.18 & $>6$ \\ 
   18 & 2 & 2836.77 & 1.05 & 0.35 \\ 
   19 & 0 & 2846.59 & 0.14 & 2.57 \\ 
   19 & 1 & 2916.15 & 0.12 & $>6$ \\ 
   19 & 2 & 2982.35 & 0.45 & 1.38 \\ 
   20 & 0 & 2992.05 & 0.11 & 3.59 \\ 
   20 & 1 & 3061.64 & 0.09 & $>6$ \\ 
   20 & 2 & 3129.13 & 0.29 & $>6$ \\ 
   21 & 0 & 3137.69 & 0.12 & $>6$ \\ 
   21 & 1 & 3207.46 & 0.09 & $>6$ \\ 
   21 & 2 & 3274.2 & 0.48 & $>6$ \\ 
   22 & 0 & 3283.18 & 0.09 & $>6$ \\ 
   22 & 1 & 3353.43 & 0.12 & $>6$ \\ 
   22 & 2 & 3420.75 & 0.45 & 1.7 \\ 
   23 & 0 & 3428.94 & 0.13 & 4.14 \\ 
   23 & 1 & 3499.47 & 0.15 & $>6$ \\ 
   23 & 2 & 3568.74 & 1.8 & 1.12 \\ 
   24 & 0 & 3575.44 & 0.24 & 4.2 \\ 
   24 & 1 & 3645.94 & 0.29 & $>6$ \\ 
   24 & 2 & 3715.8 & 2.78 & -0.01 \\ 
   25 & 0 & 3724.58 & 1.38 & 1.89 \\ 
   25 & 1 & 3792.67 & 0.48 & 2.92 \\ 
   25 & 2 & 3861.66 & 3.85 & -0.23 \\ 
   26 & 0 & 3869.64 & 2.52 & 1.41 \\ 
   26 & 1 & 3938.49 & 0.53 & 1.93 \\ 
   \hline 
   \end{tabular} 
   \label{tab::3544595}\end{table} 
\begin{table} 
   \caption{Ratios for KIC 3544595.} 
   \begin{tabular}{cccc} 
   \hline 
   \hline 
   Ratio type & $n$ & Ratio & 68\% credible interval \\ 
   \hline 
   $r_{01}$ & 16 & 0.028 & 0.006 \\ 
   $r_{10}$ & 16 & 0.03 & 0.007 \\ 
   $r_{01}$ & 17 & 0.033 & 0.011 \\ 
   $r_{10}$ & 17 & 0.032 & 0.01 \\ 
   $r_{01}$ & 18 & 0.027 & 0.005 \\ 
   $r_{10}$ & 18 & 0.025 & 0.002 \\ 
   $r_{01}$ & 19 & 0.023 & 0.001 \\ 
   $r_{10}$ & 19 & 0.022 & 0.001 \\ 
   $r_{01}$ & 20 & 0.022 & 0.001 \\ 
   $r_{10}$ & 20 & 0.022 & 0.001 \\ 
   $r_{01}$ & 21 & 0.021 & 0.001 \\ 
   $r_{10}$ & 21 & 0.02 & 0.001 \\ 
   $r_{01}$ & 22 & 0.019 & 0.001 \\ 
   $r_{10}$ & 22 & 0.018 & 0.001 \\ 
   $r_{01}$ & 23 & 0.018 & 0.001 \\ 
   $r_{10}$ & 23 & 0.018 & 0.001 \\ 
   $r_{01}$ & 24 & 0.021 & 0.002 \\ 
   $r_{10}$ & 24 & 0.027 & 0.005 \\ 
   $r_{01}$ & 25 & 0.033 & 0.008 \\ 
   $r_{10}$ & 25 & 0.032 & 0.012 \\ 
   $r_{02}$ & 19 & 0.068 & 0.007 \\ 
   $r_{02}$ & 20 & 0.067 & 0.003 \\ 
   $r_{02}$ & 21 & 0.059 & 0.002 \\ 
   $r_{02}$ & 22 & 0.062 & 0.003 \\ 
   $r_{02}$ & 23 & 0.056 & 0.003 \\ 
   $r_{02}$ & 24 & 0.046 & 0.012 \\ 
   $r_{02}$ & 25 & 0.054 & 0.018 \\ 
   $r_{02}$ & 26 & 0.05 & 0.023 \\ 
   \hline 
   \end{tabular} 
   \label{tab::rat_3544595}\end{table} 
\clearpage
\subsubsection{3632418}
\begin{figure}
   \includegraphics[width=80mm]{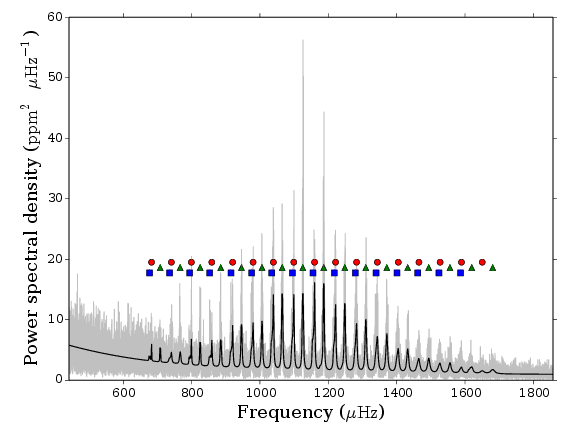}
   \includegraphics[width=80mm]{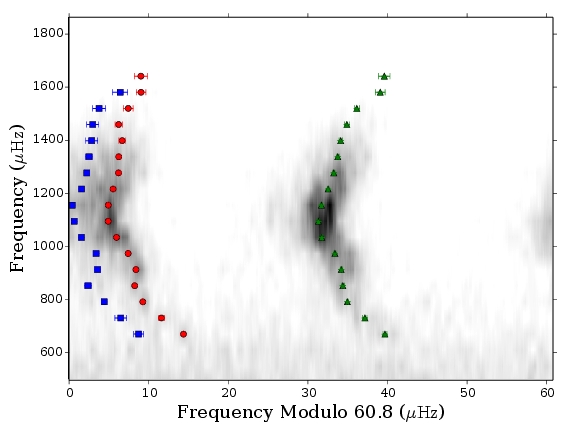}
   \caption{Power spectrum and echelle diagram for KIC 3632418.  Top: Power spectrum with data in grey smoothed over $3 \; \rm \mu Hz$ and best model in black.  Bottom: Echelle diagram with power in grey-scale.  Both: Mode frequencies are marked as: radial modes with red circles; dipole modes with green diamonds; quadrapole modes with blue squares; and octopole modes with yellow pentagons.}   \label{fig::3632418}\end{figure}
\begin{figure}
   \includegraphics[width=80mm]{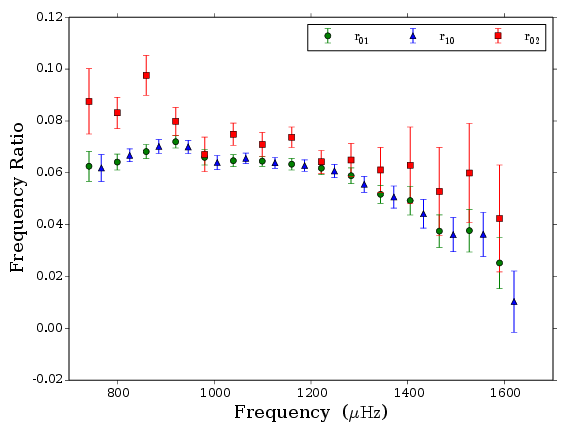}
   \caption{Ratios and $67 \%$ confidence intervals as a function of frequency for KIC 3632418.}   \label{fig::rat_3632418}\end{figure}
\begin{figure}
   \includegraphics[width=80mm]{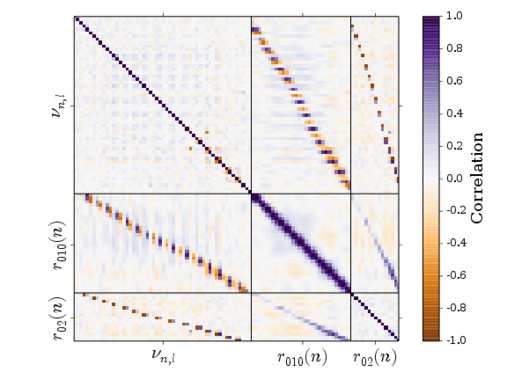}
   \caption{Correlation matrix of all frequencies and ratios for KIC 3632418.  The grid represents the matrix and hence the identity elements are all correlation 1.0.  The matrix is constructed so that frequencies and ratios are grouped separately.  If each matrix element is labelled $[i,j]$ then the first set of $i$ contains the mode frequencies, the second set contains the $r_{010}$, and the final set the $r_{02}$.}   \label{fig::cor_3632418}\end{figure}
\clearpage
\begin{table}
   \caption{Mode frequencies and statistics for KIC 3632418.}
   \begin{tabular}{ccccc}
   \hline 
   \hline 
   $n$ & $l$ & Frequency & 68\% credible & $ \ln{K}$  \\ 
 & & ($\rm \mu Hz$) & ($\rm \mu Hz$) & \\ 
   \hline 
   8 & 2 & 677.43 & 0.6 & 0.21 \\ 
   9 & 0 & 683.07 & 0.13 & 2.79 \\ 
   9 & 1 & 708.38 & 0.25 & $>6$ \\ 
   9 & 2 & 735.98 & 0.73 & 0.43 \\ 
   10 & 0 & 741.09 & 0.36 & 3.14 \\ 
   10 & 1 & 766.65 & 0.26 & $>6$ \\
\hline 
   \end{tabular} 
   \label{tab::3632418}\end{table} 
\begin{table} 
   \caption{Ratios for KIC 3632418.} 
   \begin{tabular}{cccc} 
   \hline 
   \hline 
   Ratio type & $n$ & Ratio & 68\% credible interval \\ 
   \hline 
   $r_{01}$ & 10 & 0.063 & 0.006 \\ 
   $r_{10}$ & 10 & 0.062 & 0.005 \\ 
   $r_{01}$ & 11 & 0.064 & 0.003 \\ 
   $r_{10}$ & 11 & 0.067 & 0.002 \\ 
   $r_{01}$ & 12 & 0.068 & 0.003 \\ 
   $r_{10}$ & 12 & 0.07 & 0.003 \\ 
   $r_{01}$ & 13 & 0.072 & 0.002 \\ 
   $r_{10}$ & 13 & 0.07 & 0.003 \\ 
   $r_{01}$ & 14 & 0.066 & 0.003 \\ 
   $r_{10}$ & 14 & 0.064 & 0.003 \\ 
   $r_{01}$ & 15 & 0.065 & 0.002 \\ 
   $r_{10}$ & 15 & 0.066 & 0.002 \\ 
   $r_{01}$ & 16 & 0.064 & 0.002 \\ 
   $r_{10}$ & 16 & 0.064 & 0.002 \\ 
   $r_{01}$ & 17 & 0.063 & 0.002 \\ 
   $r_{10}$ & 17 & 0.063 & 0.002 \\ 
   $r_{01}$ & 18 & 0.062 & 0.002 \\ 
   $r_{10}$ & 18 & 0.061 & 0.003 \\ 
   $r_{01}$ & 19 & 0.059 & 0.003 \\ 
   $r_{10}$ & 19 & 0.056 & 0.003 \\ 
   $r_{01}$ & 20 & 0.052 & 0.003 \\ 
   $r_{10}$ & 20 & 0.051 & 0.004 \\ 
   $r_{01}$ & 21 & 0.049 & 0.006 \\ 
   $r_{10}$ & 21 & 0.044 & 0.006 \\ 
   $r_{01}$ & 22 & 0.038 & 0.006 \\ 
   $r_{10}$ & 22 & 0.036 & 0.007 \\ 
   $r_{01}$ & 23 & 0.038 & 0.008 \\ 
   $r_{10}$ & 23 & 0.036 & 0.009 \\ 
   $r_{01}$ & 24 & 0.025 & 0.01 \\ 
   $r_{10}$ & 24 & 0.01 & 0.012 \\ 
   $r_{02}$ & 10 & 0.088 & 0.013 \\ 
   $r_{02}$ & 11 & 0.083 & 0.006 \\ 
   $r_{02}$ & 12 & 0.098 & 0.008 \\ 
   $r_{02}$ & 13 & 0.08 & 0.005 \\ 
   $r_{02}$ & 14 & 0.067 & 0.007 \\ 
   $r_{02}$ & 15 & 0.075 & 0.004 \\ 
   $r_{02}$ & 16 & 0.071 & 0.005 \\ 
   $r_{02}$ & 17 & 0.074 & 0.004 \\ 
   $r_{02}$ & 18 & 0.064 & 0.004 \\ 
   $r_{02}$ & 19 & 0.065 & 0.006 \\ 
   $r_{02}$ & 20 & 0.061 & 0.009 \\ 
   $r_{02}$ & 21 & 0.063 & 0.015 \\ 
   $r_{02}$ & 22 & 0.053 & 0.017 \\ 
   $r_{02}$ & 23 & 0.06 & 0.019 \\ 
   $r_{02}$ & 24 & 0.042 & 0.021 \\ 
   \hline 
   \end{tabular} 
   \label{tab::rat_3632418}\end{table} 
\clearpage
\subsubsection{4141376}
\begin{figure}
   \includegraphics[width=80mm]{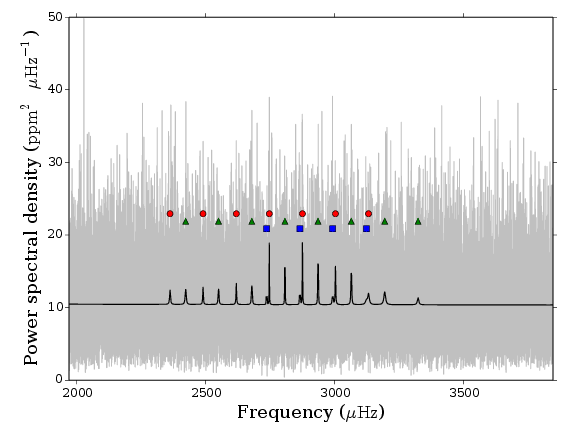}
   \includegraphics[width=80mm]{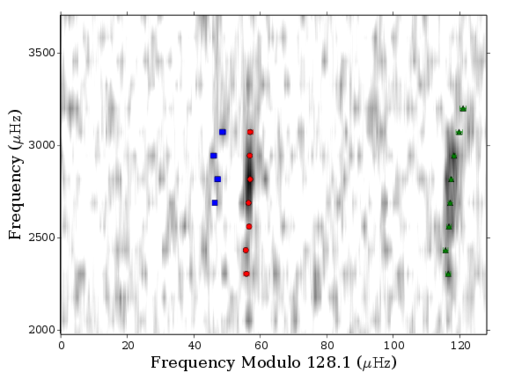}
   \caption{Power spectrum and echelle diagram for KIC 4141376.  Top: Power spectrum with data in grey smoothed over $3 \; \rm \mu Hz$ and best model in black.  Bottom: Echelle diagram with power in grey-scale.  Both: Mode frequencies are marked as: radial modes with red circles; dipole modes with green diamonds; quadrapole modes with blue squares; and octopole modes with yellow pentagons.}   \label{fig::4141376}\end{figure}
\begin{figure}
   \includegraphics[width=80mm]{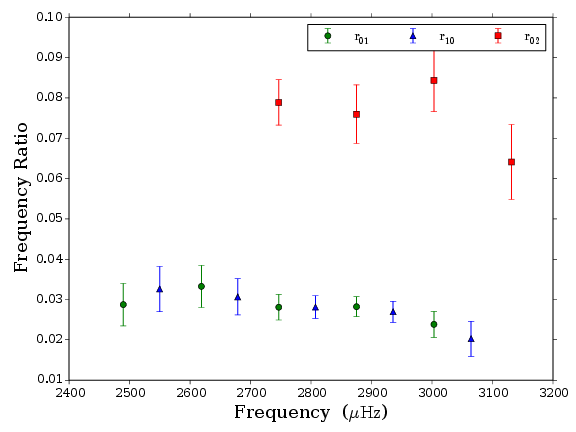}
   \caption{Ratios and $67 \%$ confidence intervals as a function of frequency for KIC 4141376.}   \label{fig::rat_4141376}\end{figure}
\begin{figure}
   \includegraphics[width=80mm]{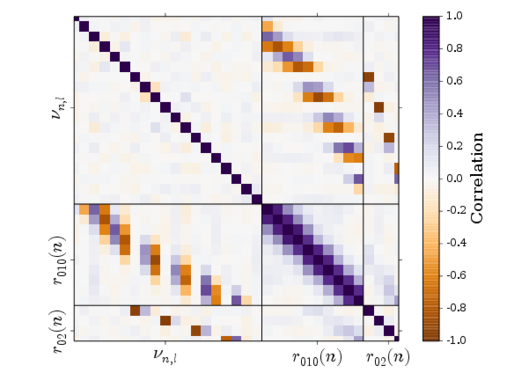}
   \caption{Correlation matrix of all frequencies and ratios for KIC 4141376.  The grid represents the matrix and hence the identity elements are all correlation 1.0.  The matrix is constructed so that frequencies and ratios are grouped separately.  If each matrix element is labelled $[i,j]$ then the first set of $i$ contains the mode frequencies, the second set contains the $r_{010}$, and the final set the $r_{02}$.}   \label{fig::cor_4141376}\end{figure}
\clearpage
\begin{table}
   \caption{Mode frequencies and statistics for KIC 4141376.}
   \begin{tabular}{ccccc}
   \hline 
   \hline 
   $n$ & $l$ & Frequency & 68\% credible & $ \ln{K}$  \\ 
 & & ($\rm \mu Hz$) & ($\rm \mu Hz$) & \\ 
   \hline 
   17 & 0 & 2361.69 & 0.82 & 0.8 \\ 
   17 & 1 & 2422.4 & 0.7 & $>6$ \\ 
   18 & 0 & 2489.59 & 0.53 & 1.54 \\ 
   18 & 1 & 2549.67 & 0.74 & 1.33 \\ 
   19 & 0 & 2618.68 & 0.57 & 0.81 \\ 
   19 & 1 & 2678.8 & 0.6 & 2.35 \\ 
   19 & 2 & 2736.49 & 0.69 & -0.07 \\ 
   20 & 0 & 2746.62 & 0.2 & 2.67 \\ 
   20 & 1 & 2807.28 & 0.41 & 2.27 \\ 
   20 & 2 & 2865.46 & 0.92 & 0.17 \\ 
   21 & 0 & 2875.22 & 0.22 & 3.02 \\ 
   21 & 1 & 2935.7 & 0.28 & 2.52 \\ 
   21 & 2 & 2992.34 & 0.94 & -0.42 \\ 
   22 & 0 & 3003.2 & 0.38 & 2.27 \\ 
   22 & 1 & 3064.66 & 0.44 & $>6$ \\ 
   22 & 2 & 3123.12 & 0.92 & -0.32 \\ 
   23 & 0 & 3131.43 & 0.78 & 1.23 \\ 
   23 & 1 & 3194.28 & 0.96 & 1.67 \\ 
   24 & 1 & 3323.58 & 0.94 & 0.92 \\ 
   \hline 
   \end{tabular} 
   \label{tab::4141376}\end{table} 
\begin{table} 
   \caption{Ratios for KIC 4141376.} 
   \begin{tabular}{cccc} 
   \hline 
   \hline 
   Ratio type & $n$ & Ratio & 68\% credible interval \\ 
   \hline 
   $r_{01}$ & 18 & 0.029 & 0.005 \\ 
   $r_{10}$ & 18 & 0.033 & 0.006 \\ 
   $r_{01}$ & 19 & 0.033 & 0.005 \\ 
   $r_{10}$ & 19 & 0.031 & 0.004 \\ 
   $r_{01}$ & 20 & 0.028 & 0.003 \\ 
   $r_{10}$ & 20 & 0.028 & 0.003 \\ 
   $r_{01}$ & 21 & 0.028 & 0.002 \\ 
   $r_{10}$ & 21 & 0.027 & 0.003 \\ 
   $r_{01}$ & 22 & 0.024 & 0.003 \\ 
   $r_{10}$ & 22 & 0.02 & 0.004 \\ 
   $r_{02}$ & 20 & 0.079 & 0.006 \\ 
   $r_{02}$ & 21 & 0.076 & 0.007 \\ 
   $r_{02}$ & 22 & 0.084 & 0.008 \\ 
   $r_{02}$ & 23 & 0.064 & 0.009 \\ 
   \hline 
   \end{tabular} 
   \label{tab::rat_4141376}\end{table} 
\clearpage
\subsubsection{4143755}
\begin{figure}
   \includegraphics[width=80mm]{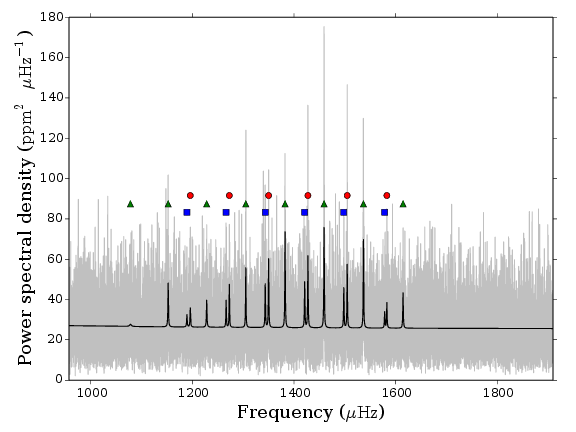}
   \includegraphics[width=80mm]{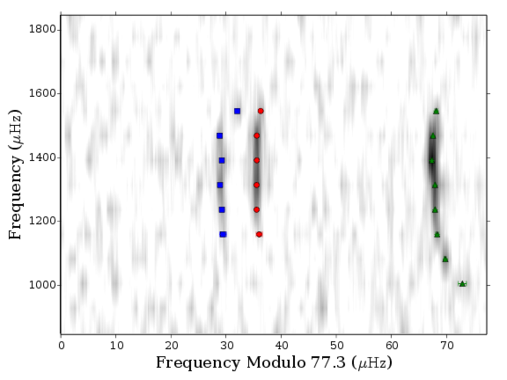}
   \caption{Power spectrum and echelle diagram for KIC 4143755.  Top: Power spectrum with data in grey smoothed over $3 \; \rm \mu Hz$ and best model in black.  Bottom: Echelle diagram with power in grey-scale.  Both: Mode frequencies are marked as: radial modes with red circles; dipole modes with green diamonds; quadrapole modes with blue squares; and octopole modes with yellow pentagons.}   \label{fig::4143755}\end{figure}
\begin{figure}
   \includegraphics[width=80mm]{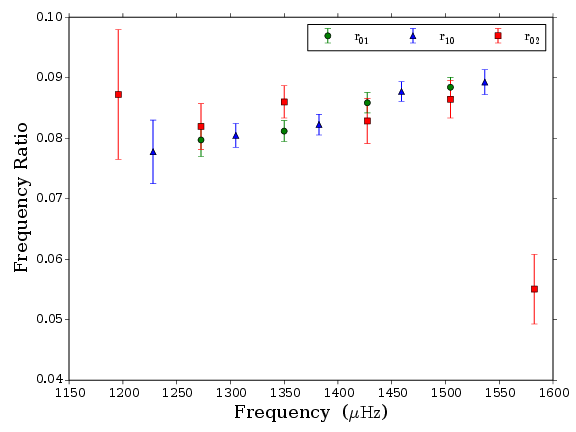}
   \caption{Ratios and $67 \%$ confidence intervals as a function of frequency for KIC 4143755.}   \label{fig::rat_4143755}\end{figure}
\begin{figure}
   \includegraphics[width=80mm]{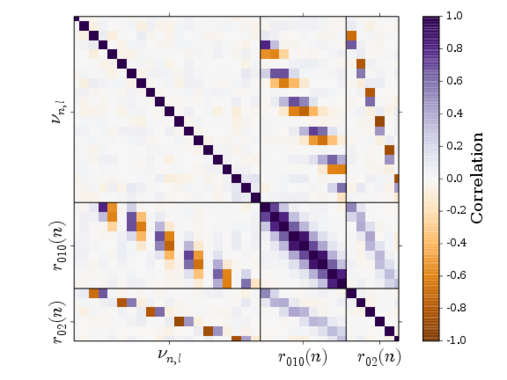}
   \caption{Correlation matrix of all frequencies and ratios for KIC 4143755.  The grid represents the matrix and hence the identity elements are all correlation 1.0.  The matrix is constructed so that frequencies and ratios are grouped separately.  If each matrix element is labelled $[i,j]$ then the first set of $i$ contains the mode frequencies, the second set contains the $r_{010}$, and the final set the $r_{02}$.}   \label{fig::cor_4143755}\end{figure}
\clearpage
\begin{table}
   \caption{Mode frequencies and statistics for KIC 4143755.}
   \begin{tabular}{ccccc}
   \hline 
   \hline 
   $n$ & $l$ & Frequency & 68\% credible & $ \ln{K}$  \\ 
 & & ($\rm \mu Hz$) & ($\rm \mu Hz$) & \\ 
   \hline 
   13 & 1 & 1078.06 & 0.73 & 1.27 \\ 
   14 & 1 & 1152.29 & 0.17 & $>6$ \\ 
   14 & 2 & 1189.21 & 0.63 & -0.41 \\ 
   15 & 0 & 1195.81 & 0.56 & 0.86 \\ 
   15 & 1 & 1228.12 & 0.29 & $>6$ \\ 
   15 & 2 & 1266.36 & 0.23 & $>6$ \\ 
   16 & 0 & 1272.67 & 0.17 & $>6$ \\ 
   16 & 1 & 1305.06 & 0.12 & $>6$ \\ 
   16 & 2 & 1343.34 & 0.16 & $>6$ \\ 
   17 & 0 & 1349.99 & 0.13 & $>6$ \\ 
   17 & 1 & 1382.37 & 0.13 & $>6$ \\ 
   17 & 2 & 1420.98 & 0.25 & $>6$ \\ 
   18 & 0 & 1427.34 & 0.13 & $>6$ \\ 
   18 & 1 & 1459.11 & 0.11 & $>6$ \\ 
   18 & 2 & 1497.93 & 0.22 & $>6$ \\ 
   19 & 0 & 1504.65 & 0.12 & $>6$ \\ 
   19 & 1 & 1536.65 & 0.14 & $>6$ \\ 
   19 & 2 & 1578.42 & 0.4 & $>6$ \\ 
   20 & 0 & 1582.68 & 0.24 & $>6$ \\ 
   20 & 1 & 1614.54 & 0.25 & $>6$ \\ 
   \hline 
   \end{tabular} 
   \label{tab::4143755}\end{table} 
\begin{table} 
   \caption{Ratios for KIC 4143755.} 
   \begin{tabular}{cccc} 
   \hline 
   \hline 
   Ratio type & $n$ & Ratio & 68\% credible interval \\ 
   \hline 
   $r_{10}$ & 15 & 0.078 & 0.005 \\ 
   $r_{01}$ & 16 & 0.08 & 0.003 \\ 
   $r_{10}$ & 16 & 0.08 & 0.002 \\ 
   $r_{01}$ & 17 & 0.081 & 0.002 \\ 
   $r_{10}$ & 17 & 0.082 & 0.002 \\ 
   $r_{01}$ & 18 & 0.086 & 0.002 \\ 
   $r_{10}$ & 18 & 0.088 & 0.002 \\ 
   $r_{01}$ & 19 & 0.088 & 0.002 \\ 
   $r_{10}$ & 19 & 0.089 & 0.002 \\ 
   $r_{02}$ & 15 & 0.087 & 0.011 \\ 
   $r_{02}$ & 16 & 0.082 & 0.004 \\ 
   $r_{02}$ & 17 & 0.086 & 0.003 \\ 
   $r_{02}$ & 18 & 0.083 & 0.004 \\ 
   $r_{02}$ & 19 & 0.086 & 0.003 \\ 
   $r_{02}$ & 20 & 0.055 & 0.006 \\ 
   \hline 
   \end{tabular} 
   \label{tab::rat_4143755}\end{table} 
\clearpage
\subsubsection{4349452}
\begin{figure}
   \includegraphics[width=80mm]{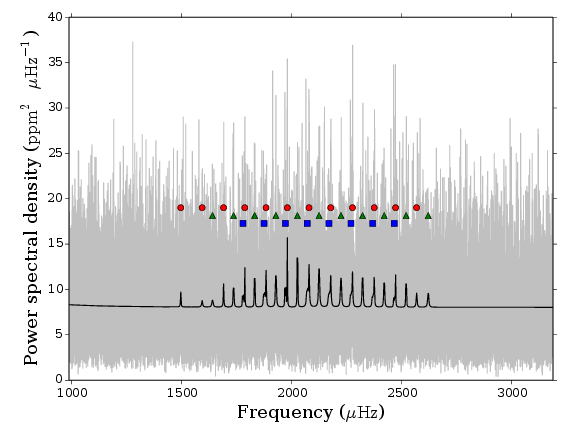}
   \includegraphics[width=80mm]{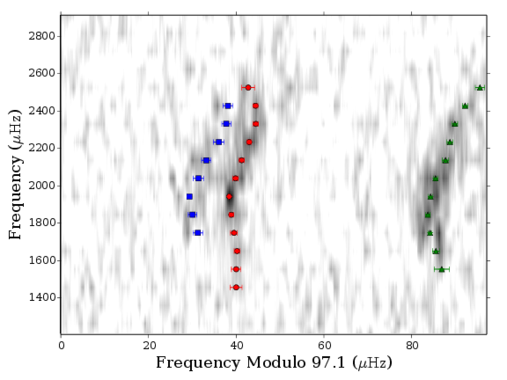}
   \caption{Power spectrum and echelle diagram for KIC 4349452.  Top: Power spectrum with data in grey smoothed over $3 \; \rm \mu Hz$ and best model in black.  Bottom: Echelle diagram with power in grey-scale.  Both: Mode frequencies are marked as: radial modes with red circles; dipole modes with green diamonds; quadrapole modes with blue squares; and octopole modes with yellow pentagons.}   \label{fig::4349452}\end{figure}
\begin{figure}
   \includegraphics[width=80mm]{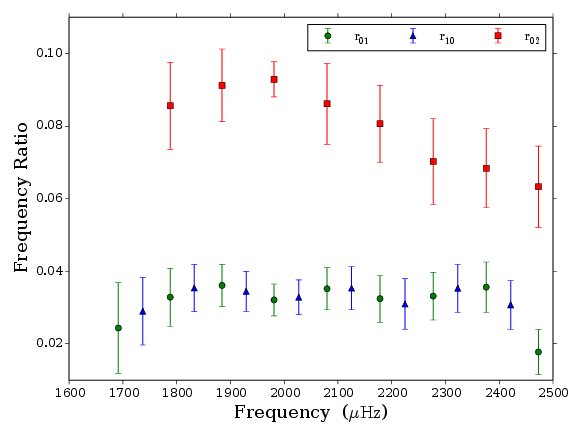}
   \caption{Ratios and $67 \%$ confidence intervals as a function of frequency for KIC 4349452.}   \label{fig::rat_4349452}\end{figure}
\begin{figure}
   \includegraphics[width=80mm]{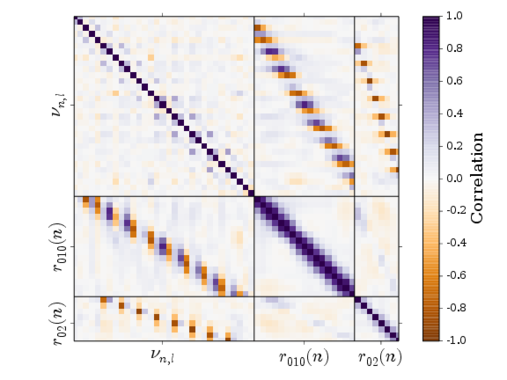}
   \caption{Correlation matrix of all frequencies and ratios for KIC 4349452.  The grid represents the matrix and hence the identity elements are all correlation 1.0.  The matrix is constructed so that frequencies and ratios are grouped separately.  If each matrix element is labelled $[i,j]$ then the first set of $i$ contains the mode frequencies, the second set contains the $r_{010}$, and the final set the $r_{02}$.}   \label{fig::cor_4349452}\end{figure}
\clearpage
\begin{table}
   \caption{Mode frequencies and statistics for KIC 4349452.}
   \begin{tabular}{ccccc}
   \hline 
   \hline 
   $n$ & $l$ & Frequency & 68\% credible & $ \ln{K}$  \\ 
 & & ($\rm \mu Hz$) & ($\rm \mu Hz$) & \\ 
   \hline 
   15 & 0 & 1497.1 & 1.34 & 0.85 \\ 
   16 & 0 & 1594.24 & 1.09 & 1.0 \\ 
   16 & 1 & 1641.14 & 1.75 & 1.02 \\ 
   17 & 0 & 1691.59 & 0.66 & 2.07 \\ 
   17 & 1 & 1736.93 & 0.81 & $>6$ \\ 
   17 & 2 & 1779.8 & 1.13 & 0.14 \\ 
   18 & 0 & 1788.03 & 0.79 & 1.81 \\ 
   18 & 1 & 1832.76 & 0.41 & 3.69 \\ 
   18 & 2 & 1875.72 & 0.98 & 1.68 \\ 
   19 & 0 & 1884.54 & 0.52 & 3.41 \\ 
   19 & 1 & 1929.37 & 0.56 & $>6$ \\ 
   19 & 2 & 1972.12 & 0.46 & 1.53 \\ 
   20 & 0 & 1981.2 & 0.18 & 3.72 \\ 
   20 & 1 & 2027.14 & 0.47 & $>6$ \\ 
   20 & 2 & 2071.34 & 1.21 & 1.46 \\ 
   21 & 0 & 2079.8 & 0.56 & 3.49 \\ 
   21 & 1 & 2125.42 & 0.46 & $>6$ \\ 
   21 & 2 & 2170.27 & 1.08 & 1.79 \\ 
   22 & 0 & 2178.33 & 0.56 & 3.49 \\ 
   22 & 1 & 2224.84 & 0.72 & $>6$ \\ 
   22 & 2 & 2270.29 & 1.21 & 1.26 \\ 
   23 & 0 & 2277.2 & 0.44 & 3.77 \\ 
   23 & 1 & 2322.98 & 0.68 & 2.89 \\ 
   23 & 2 & 2369.11 & 1.02 & 0.52 \\ 
   24 & 0 & 2375.86 & 0.63 & 2.8 \\ 
   24 & 1 & 2421.25 & 0.62 & 3.06 \\ 
   24 & 2 & 2466.63 & 1.12 & 0.2 \\ 
   25 & 0 & 2472.92 & 0.52 & 2.61 \\ 
   25 & 1 & 2520.72 & 0.56 & 1.94 \\ 
   26 & 0 & 2568.4 & 1.45 & 1.81 \\ 
   26 & 1 & 2621.25 & 1.06 & 2.0 \\ 
   \hline 
   \end{tabular} 
   \label{tab::4349452}\end{table} 
\begin{table} 
   \caption{Ratios for KIC 4349452.} 
   \begin{tabular}{cccc} 
   \hline 
   \hline 
   Ratio type & $n$ & Ratio & 68\% credible interval \\ 
   \hline 
   $r_{01}$ & 17 & 0.024 & 0.012 \\ 
   $r_{10}$ & 17 & 0.029 & 0.009 \\ 
   $r_{01}$ & 18 & 0.033 & 0.008 \\ 
   $r_{10}$ & 18 & 0.035 & 0.006 \\ 
   $r_{01}$ & 19 & 0.036 & 0.006 \\ 
   $r_{10}$ & 19 & 0.034 & 0.006 \\ 
   $r_{01}$ & 20 & 0.032 & 0.004 \\ 
   $r_{10}$ & 20 & 0.033 & 0.005 \\ 
   $r_{01}$ & 21 & 0.035 & 0.006 \\ 
   $r_{10}$ & 21 & 0.035 & 0.006 \\ 
   $r_{01}$ & 22 & 0.032 & 0.006 \\ 
   $r_{10}$ & 22 & 0.031 & 0.007 \\ 
   $r_{01}$ & 23 & 0.033 & 0.007 \\ 
   $r_{10}$ & 23 & 0.035 & 0.007 \\ 
   $r_{01}$ & 24 & 0.036 & 0.007 \\ 
   $r_{10}$ & 24 & 0.031 & 0.007 \\ 
   $r_{01}$ & 25 & 0.018 & 0.006 \\ 
   $r_{02}$ & 18 & 0.086 & 0.012 \\ 
   $r_{02}$ & 19 & 0.091 & 0.01 \\ 
   $r_{02}$ & 20 & 0.093 & 0.005 \\ 
   $r_{02}$ & 21 & 0.086 & 0.011 \\ 
   $r_{02}$ & 22 & 0.081 & 0.011 \\ 
   $r_{02}$ & 23 & 0.07 & 0.012 \\ 
   $r_{02}$ & 24 & 0.068 & 0.011 \\ 
   $r_{02}$ & 25 & 0.063 & 0.011 \\ 
   \hline 
   \end{tabular} 
   \label{tab::rat_4349452}\end{table} 
\clearpage
\subsubsection{4914423}
\begin{figure}
   \includegraphics[width=80mm]{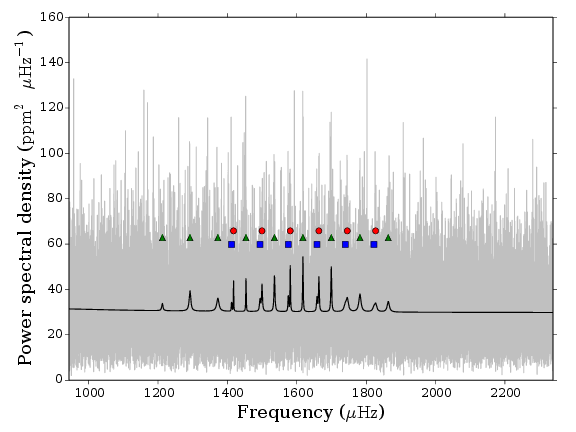}
   \includegraphics[width=80mm]{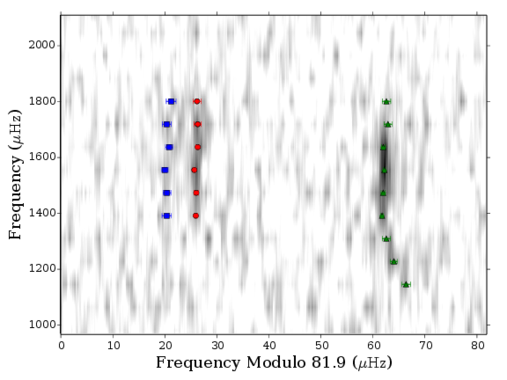}
   \caption{Power spectrum and echelle diagram for KIC 4914423.  Top: Power spectrum with data in grey smoothed over $3 \; \rm \mu Hz$ and best model in black.  Bottom: Echelle diagram with power in grey-scale.  Both: Mode frequencies are marked as: radial modes with red circles; dipole modes with green diamonds; quadrapole modes with blue squares; and octopole modes with yellow pentagons.}   \label{fig::4914423}\end{figure}
\begin{figure}
   \includegraphics[width=80mm]{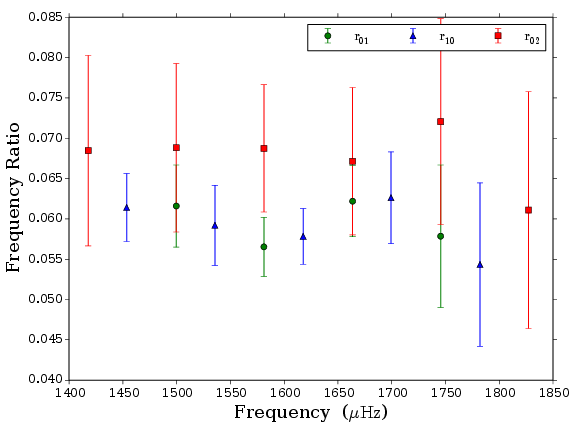}
   \caption{Ratios and $67 \%$ confidence intervals as a function of frequency for KIC 4914423.}   \label{fig::rat_4914423}\end{figure}
\begin{figure}
   \includegraphics[width=80mm]{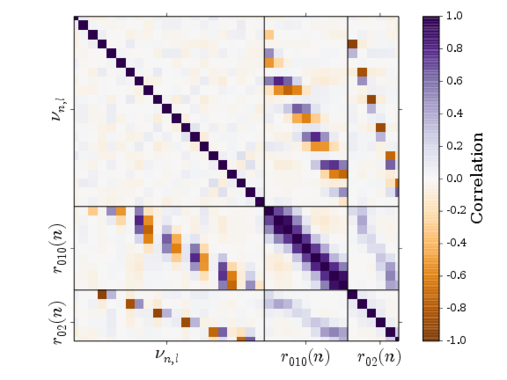}
   \caption{Correlation matrix of all frequencies and ratios for KIC 4914423.  The grid represents the matrix and hence the identity elements are all correlation 1.0.  The matrix is constructed so that frequencies and ratios are grouped separately.  If each matrix element is labelled $[i,j]$ then the first set of $i$ contains the mode frequencies, the second set contains the $r_{010}$, and the final set the $r_{02}$.}   \label{fig::cor_4914423}\end{figure}
\clearpage
\begin{table}
   \caption{Mode frequencies and statistics for KIC 4914423.}
   \begin{tabular}{ccccc}
   \hline 
   \hline 
   $n$ & $l$ & Frequency & 68\% credible & $ \ln{K}$  \\ 
 & & ($\rm \mu Hz$) & ($\rm \mu Hz$) & \\ 
   \hline 
   13 & 1 & 1212.51 & 0.82 & 1.07 \\ 
   14 & 1 & 1292.05 & 0.66 & $>6$ \\ 
   15 & 1 & 1372.45 & 0.8 & 1.53 \\ 
   15 & 2 & 1412.11 & 0.9 & -0.14 \\ 
   16 & 0 & 1417.69 & 0.33 & 1.57 \\ 
   16 & 1 & 1453.49 & 0.27 & $>6$ \\ 
   16 & 2 & 1494.01 & 0.75 & -0.18 \\ 
   17 & 0 & 1499.65 & 0.45 & 1.97 \\ 
   17 & 1 & 1535.58 & 0.38 & $>6$ \\ 
   17 & 2 & 1575.49 & 0.61 & 0.87 \\ 
   18 & 0 & 1581.14 & 0.26 & 2.74 \\ 
   18 & 1 & 1617.68 & 0.22 & $>6$ \\ 
   18 & 2 & 1658.17 & 0.7 & 0.49 \\ 
   19 & 0 & 1663.67 & 0.35 & 2.7 \\ 
   19 & 1 & 1699.33 & 0.3 & $>6$ \\ 
   19 & 2 & 1739.58 & 0.8 & 0.27 \\ 
   20 & 0 & 1745.54 & 0.71 & 2.67 \\ 
   20 & 1 & 1782.11 & 0.79 & 1.15 \\ 
   20 & 2 & 1822.29 & 0.94 & -0.01 \\ 
   21 & 0 & 1827.27 & 0.78 & 1.92 \\ 
   21 & 1 & 1863.68 & 0.83 & 0.87 \\ 
   \hline 
   \end{tabular} 
   \label{tab::4914423}\end{table} 
\begin{table} 
   \caption{Ratios for KIC 4914423.} 
   \begin{tabular}{cccc} 
   \hline 
   \hline 
   Ratio type & $n$ & Ratio & 68\% credible interval \\ 
   \hline 
   $r_{10}$ & 16 & 0.061 & 0.004 \\ 
   $r_{01}$ & 17 & 0.062 & 0.005 \\ 
   $r_{10}$ & 17 & 0.059 & 0.005 \\ 
   $r_{01}$ & 18 & 0.057 & 0.004 \\ 
   $r_{10}$ & 18 & 0.058 & 0.003 \\ 
   $r_{01}$ & 19 & 0.062 & 0.004 \\ 
   $r_{10}$ & 19 & 0.063 & 0.006 \\ 
   $r_{01}$ & 20 & 0.058 & 0.009 \\ 
   $r_{10}$ & 20 & 0.054 & 0.01 \\ 
   $r_{02}$ & 16 & 0.068 & 0.012 \\ 
   $r_{02}$ & 17 & 0.069 & 0.01 \\ 
   $r_{02}$ & 18 & 0.069 & 0.008 \\ 
   $r_{02}$ & 19 & 0.067 & 0.009 \\ 
   $r_{02}$ & 20 & 0.072 & 0.013 \\ 
   $r_{02}$ & 21 & 0.061 & 0.015 \\ 
   \hline 
   \end{tabular} 
   \label{tab::rat_4914423}\end{table} 
\clearpage
\subsubsection{5094751}
\begin{figure}
   \includegraphics[width=80mm]{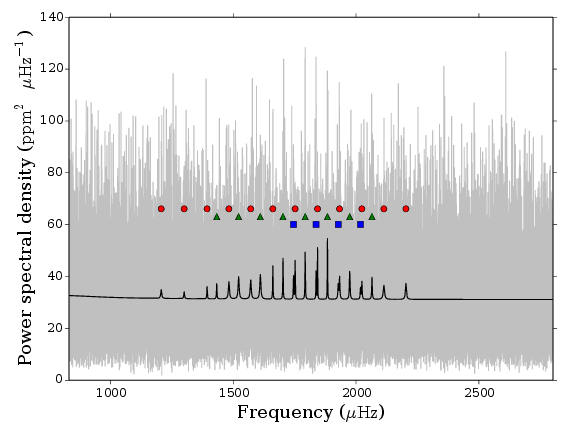}
   \includegraphics[width=80mm]{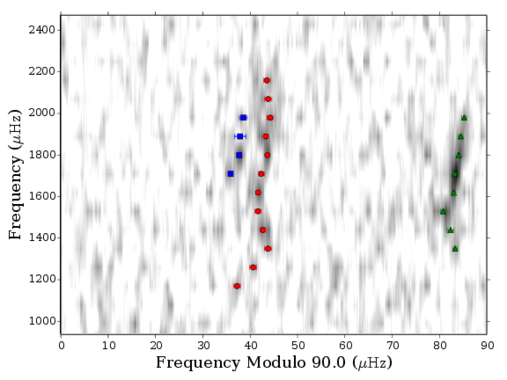}
   \caption{Power spectrum and echelle diagram for KIC 5094751.  Top: Power spectrum with data in grey smoothed over $3 \; \rm \mu Hz$ and best model in black.  Bottom: Echelle diagram with power in grey-scale.  Both: Mode frequencies are marked as: radial modes with red circles; dipole modes with green diamonds; quadrapole modes with blue squares; and octopole modes with yellow pentagons.}   \label{fig::5094751}\end{figure}
\begin{figure}
   \includegraphics[width=80mm]{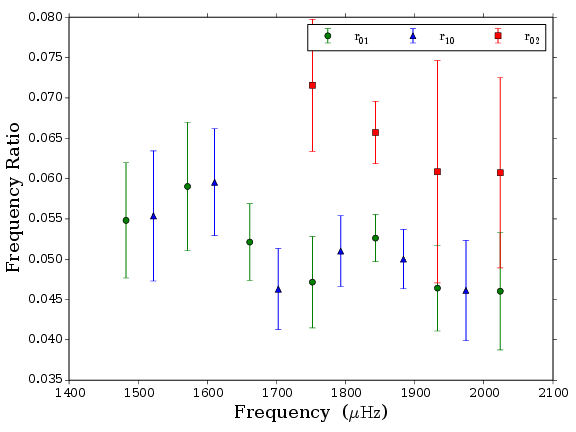}
   \caption{Ratios and $67 \%$ confidence intervals as a function of frequency for KIC 5094751.}   \label{fig::rat_5094751}\end{figure}
\begin{figure}
   \includegraphics[width=80mm]{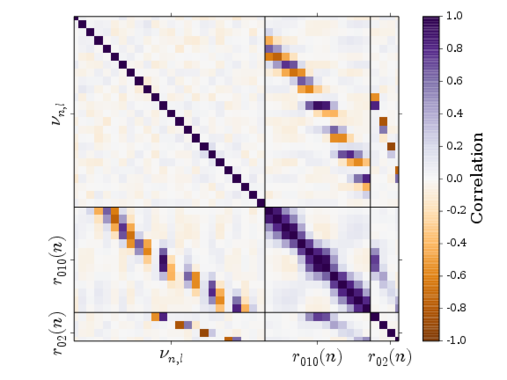}
   \caption{Correlation matrix of all frequencies and ratios for KIC 5094751.  The grid represents the matrix and hence the identity elements are all correlation 1.0.  The matrix is constructed so that frequencies and ratios are grouped separately.  If each matrix element is labelled $[i,j]$ then the first set of $i$ contains the mode frequencies, the second set contains the $r_{010}$, and the final set the $r_{02}$.}   \label{fig::cor_5094751}\end{figure}
\clearpage
\begin{table}
   \caption{Mode frequencies and statistics for KIC 5094751.}
   \begin{tabular}{ccccc}
   \hline 
   \hline 
   $n$ & $l$ & Frequency & 68\% credible & $ \ln{K}$  \\ 
 & & ($\rm \mu Hz$) & ($\rm \mu Hz$) & \\ 
   \hline 
   12 & 0 & 1206.97 & 0.71 & 1.01 \\ 
   13 & 0 & 1300.35 & 0.71 & 0.94 \\ 
   14 & 0 & 1393.45 & 0.8 & 1.18 \\ 
   14 & 1 & 1433.01 & 0.48 & 1.3 \\ 
   15 & 0 & 1482.37 & 0.54 & 2.36 \\ 
   15 & 1 & 1522.0 & 0.73 & 1.44 \\ 
   16 & 0 & 1571.32 & 0.72 & 0.8 \\ 
   16 & 1 & 1610.44 & 0.53 & $>6$ \\ 
   17 & 0 & 1661.3 & 0.36 & $>6$ \\ 
   17 & 1 & 1702.63 & 0.33 & $>6$ \\ 
   17 & 2 & 1745.48 & 0.41 & 0.52 \\ 
   18 & 0 & 1751.95 & 0.62 & 1.87 \\ 
   18 & 1 & 1792.86 & 0.21 & $>6$ \\ 
   18 & 2 & 1837.27 & 0.28 & $>6$ \\ 
   19 & 0 & 1843.23 & 0.21 & $>6$ \\ 
   19 & 1 & 1883.65 & 0.26 & $>6$ \\ 
   19 & 2 & 1927.46 & 1.2 & -0.29 \\ 
   20 & 0 & 1932.88 & 0.54 & 3.05 \\ 
   20 & 1 & 1974.06 & 0.47 & 2.42 \\ 
   20 & 2 & 2018.12 & 0.84 & 0.09 \\ 
   21 & 0 & 2023.76 & 0.7 & 2.27 \\ 
   21 & 1 & 2064.76 & 0.48 & 1.25 \\ 
   22 & 0 & 2113.34 & 0.79 & 2.08 \\ 
   23 & 0 & 2203.04 & 0.72 & 1.78 \\ 
   \hline 
   \end{tabular} 
   \label{tab::5094751}\end{table} 
\begin{table} 
   \caption{Ratios for KIC 5094751.} 
   \begin{tabular}{cccc} 
   \hline 
   \hline 
   Ratio type & $n$ & Ratio & 68\% credible interval \\ 
   \hline 
   $r_{01}$ & 15 & 0.055 & 0.007 \\ 
   $r_{10}$ & 15 & 0.055 & 0.008 \\ 
   $r_{01}$ & 16 & 0.059 & 0.008 \\ 
   $r_{10}$ & 16 & 0.06 & 0.007 \\ 
   $r_{01}$ & 17 & 0.052 & 0.005 \\ 
   $r_{10}$ & 17 & 0.046 & 0.005 \\ 
   $r_{01}$ & 18 & 0.047 & 0.006 \\ 
   $r_{10}$ & 18 & 0.051 & 0.004 \\ 
   $r_{01}$ & 19 & 0.053 & 0.003 \\ 
   $r_{10}$ & 19 & 0.05 & 0.004 \\ 
   $r_{01}$ & 20 & 0.046 & 0.005 \\ 
   $r_{10}$ & 20 & 0.046 & 0.006 \\ 
   $r_{01}$ & 21 & 0.046 & 0.007 \\ 
   $r_{02}$ & 18 & 0.072 & 0.008 \\ 
   $r_{02}$ & 19 & 0.066 & 0.004 \\ 
   $r_{02}$ & 20 & 0.061 & 0.014 \\ 
   $r_{02}$ & 21 & 0.061 & 0.012 \\ 
   \hline 
   \end{tabular} 
   \label{tab::rat_5094751}\end{table} 
\clearpage
\subsubsection{5866724}
\begin{figure}
   \includegraphics[width=80mm]{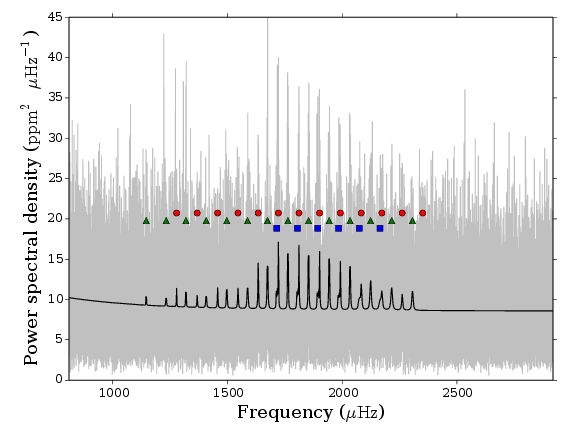}
   \includegraphics[width=80mm]{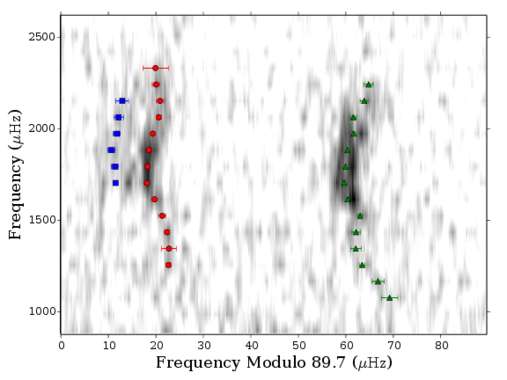}
   \caption{Power spectrum and echelle diagram for KIC 5866724.  Top: Power spectrum with data in grey smoothed over $3 \; \rm \mu Hz$ and best model in black.  Bottom: Echelle diagram with power in grey-scale.  Both: Mode frequencies are marked as: radial modes with red circles; dipole modes with green diamonds; quadrapole modes with blue squares; and octopole modes with yellow pentagons.}   \label{fig::5866724}\end{figure}
\begin{figure}
   \includegraphics[width=80mm]{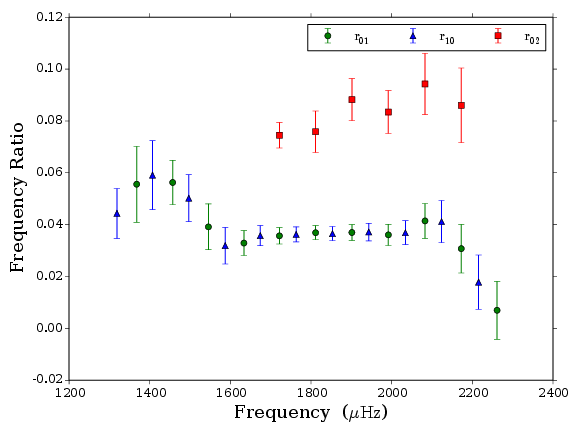}
   \caption{Ratios and $67 \%$ confidence intervals as a function of frequency for KIC 5866724.}   \label{fig::rat_5866724}\end{figure}
\begin{figure}
   \includegraphics[width=80mm]{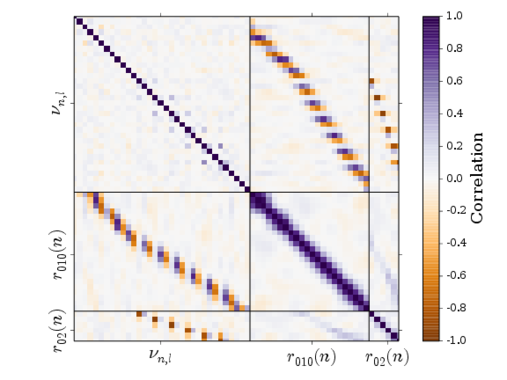}
   \caption{Correlation matrix of all frequencies and ratios for KIC 5866724.  The grid represents the matrix and hence the identity elements are all correlation 1.0.  The matrix is constructed so that frequencies and ratios are grouped separately.  If each matrix element is labelled $[i,j]$ then the first set of $i$ contains the mode frequencies, the second set contains the $r_{010}$, and the final set the $r_{02}$.}   \label{fig::cor_5866724}\end{figure}
\clearpage
\begin{table}
   \caption{Mode frequencies and statistics for KIC 5866724.}
   \begin{tabular}{ccccc}
   \hline 
   \hline 
   $n$ & $l$ & Frequency & 68\% credible & $ \ln{K}$  \\ 
 & & ($\rm \mu Hz$) & ($\rm \mu Hz$) & \\ 
   \hline 
   12 & 1 & 1145.06 & 1.74 & 1.17 \\ 
   13 & 1 & 1232.26 & 1.34 & 1.16 \\ 
   14 & 0 & 1277.81 & 0.54 & 1.25 \\ 
   14 & 1 & 1318.57 & 0.52 & 1.6 \\ 
   15 & 0 & 1367.58 & 1.52 & 1.44 \\ 
   15 & 1 & 1406.91 & 1.1 & 1.17 \\ 
   16 & 0 & 1456.82 & 0.5 & 1.82 \\ 
   16 & 1 & 1496.62 & 0.86 & 1.58 \\ 
   17 & 0 & 1545.47 & 0.85 & 0.96 \\ 
   17 & 1 & 1587.13 & 0.47 & 3.03 \\ 
   18 & 0 & 1633.51 & 0.4 & 3.25 \\ 
   18 & 1 & 1674.16 & 0.31 & $>6$ \\ 
   18 & 2 & 1714.95 & 0.42 & 2.34 \\ 
   19 & 0 & 1721.55 & 0.23 & 4.09 \\ 
   19 & 1 & 1763.06 & 0.25 & $>6$ \\ 
   19 & 2 & 1804.5 & 0.72 & 1.52 \\ 
   20 & 0 & 1811.32 & 0.23 & 3.94 \\ 
   20 & 1 & 1853.02 & 0.2 & $>6$ \\ 
   20 & 2 & 1893.38 & 0.72 & 2.16 \\ 
   21 & 0 & 1901.3 & 0.27 & 4.32 \\ 
   21 & 1 & 1943.05 & 0.27 & $>6$ \\ 
   21 & 2 & 1984.2 & 0.74 & 1.21 \\ 
   22 & 0 & 1991.77 & 0.42 & 3.85 \\ 
   22 & 1 & 2034.04 & 0.28 & $>6$ \\ 
   22 & 2 & 2074.21 & 1.02 & 1.27 \\ 
   23 & 0 & 2082.68 & 0.61 & 3.24 \\ 
   23 & 1 & 2123.62 & 0.66 & 3.67 \\ 
   23 & 2 & 2164.64 & 1.31 & 0.46 \\ 
   24 & 0 & 2172.57 & 0.81 & 3.0 \\ 
   24 & 1 & 2215.57 & 0.87 & $>6$ \\ 
   25 & 0 & 2261.41 & 0.82 & $>6$ \\ 
   25 & 1 & 2306.14 & 0.96 & 2.26 \\ 
   26 & 0 & 2350.94 & 2.69 & 2.3 \\ 
   \hline 
   \end{tabular} 
   \label{tab::5866724}\end{table} 
\begin{table} 
   \caption{Ratios for KIC 5866724.} 
   \begin{tabular}{cccc} 
   \hline 
   \hline 
   Ratio type & $n$ & Ratio & 68\% credible interval \\ 
   \hline 
   $r_{10}$ & 14 & 0.044 & 0.01 \\ 
   $r_{01}$ & 15 & 0.056 & 0.015 \\ 
   $r_{10}$ & 15 & 0.059 & 0.013 \\ 
   $r_{01}$ & 16 & 0.056 & 0.009 \\ 
   $r_{10}$ & 16 & 0.05 & 0.009 \\ 
   $r_{01}$ & 17 & 0.039 & 0.009 \\ 
   $r_{10}$ & 17 & 0.032 & 0.007 \\ 
   $r_{01}$ & 18 & 0.033 & 0.005 \\ 
   $r_{10}$ & 18 & 0.036 & 0.004 \\ 
   $r_{01}$ & 19 & 0.036 & 0.003 \\ 
   $r_{10}$ & 19 & 0.036 & 0.003 \\ 
   $r_{01}$ & 20 & 0.037 & 0.003 \\ 
   $r_{10}$ & 20 & 0.037 & 0.003 \\ 
   $r_{01}$ & 21 & 0.037 & 0.003 \\ 
   $r_{10}$ & 21 & 0.037 & 0.003 \\ 
   $r_{01}$ & 22 & 0.036 & 0.004 \\ 
   $r_{10}$ & 22 & 0.037 & 0.005 \\ 
   $r_{01}$ & 23 & 0.041 & 0.007 \\ 
   $r_{10}$ & 23 & 0.041 & 0.008 \\ 
   $r_{01}$ & 24 & 0.031 & 0.009 \\ 
   $r_{10}$ & 24 & 0.018 & 0.01 \\ 
   $r_{01}$ & 25 & 0.007 & 0.011 \\ 
   $r_{02}$ & 19 & 0.075 & 0.005 \\ 
   $r_{02}$ & 20 & 0.076 & 0.008 \\ 
   $r_{02}$ & 21 & 0.088 & 0.008 \\ 
   $r_{02}$ & 22 & 0.083 & 0.008 \\ 
   $r_{02}$ & 23 & 0.094 & 0.012 \\ 
   $r_{02}$ & 24 & 0.086 & 0.014 \\ 
   \hline 
   \end{tabular} 
   \label{tab::rat_5866724}\end{table} 
\clearpage
\subsubsection{6196457}
\begin{figure}
   \includegraphics[width=80mm]{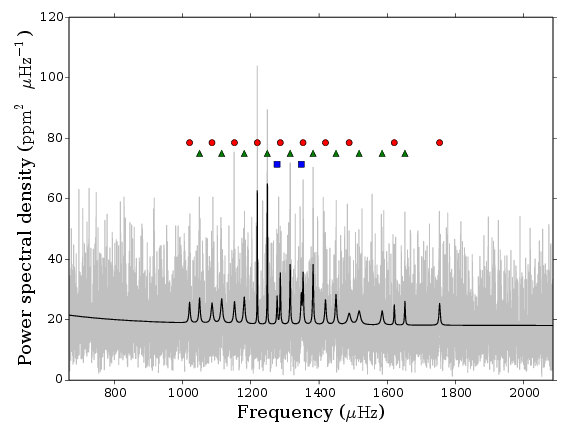}
   \includegraphics[width=80mm]{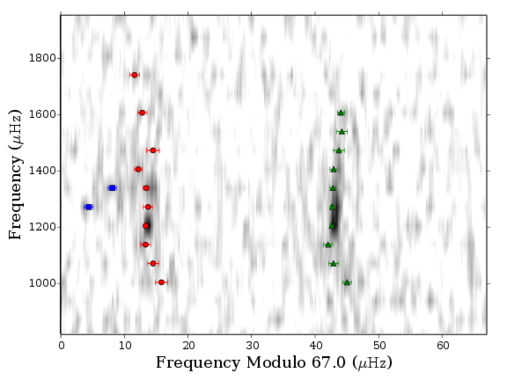}
   \caption{Power spectrum and echelle diagram for KIC 6196457.  Top: Power spectrum with data in grey smoothed over $3 \; \rm \mu Hz$ and best model in black.  Bottom: Echelle diagram with power in grey-scale.  Both: Mode frequencies are marked as: radial modes with red circles; dipole modes with green diamonds; quadrapole modes with blue squares; and octopole modes with yellow pentagons.}   \label{fig::6196457}\end{figure}
\begin{figure}
   \includegraphics[width=80mm]{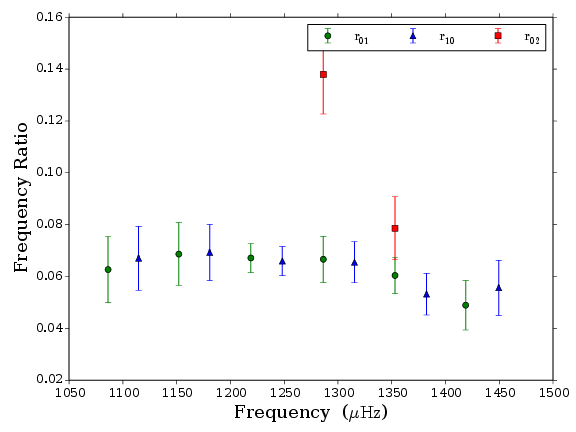}
   \caption{Ratios and $67 \%$ confidence intervals as a function of frequency for KIC 6196457.}   \label{fig::rat_6196457}\end{figure}
\begin{figure}
   \includegraphics[width=80mm]{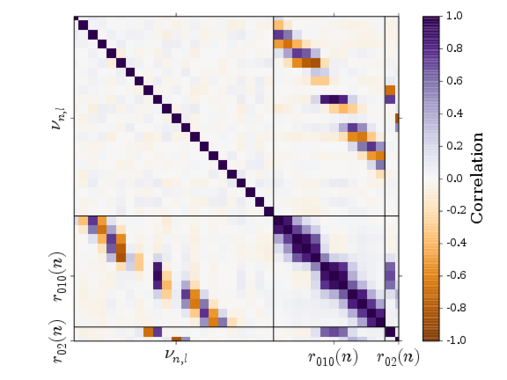}
   \caption{Correlation matrix of all frequencies and ratios for KIC 6196457.  The grid represents the matrix and hence the identity elements are all correlation 1.0.  The matrix is constructed so that frequencies and ratios are grouped separately.  If each matrix element is labelled $[i,j]$ then the first set of $i$ contains the mode frequencies, the second set contains the $r_{010}$, and the final set the $r_{02}$.}   \label{fig::cor_6196457}\end{figure}
\clearpage
\begin{table}
   \caption{Mode frequencies and statistics for KIC 6196457.}
   \begin{tabular}{ccccc}
   \hline 
   \hline 
   $n$ & $l$ & Frequency & 68\% credible & $ \ln{K}$  \\ 
 & & ($\rm \mu Hz$) & ($\rm \mu Hz$) & \\ 
   \hline 
   14 & 0 & 1020.58 & 0.94 & 0.79 \\ 
   14 & 1 & 1049.78 & 0.7 & 1.58 \\ 
   15 & 0 & 1086.24 & 0.86 & 0.96 \\ 
   15 & 1 & 1114.61 & 0.73 & 1.57 \\ 
   16 & 0 & 1152.05 & 0.76 & $>6$ \\ 
   16 & 1 & 1180.75 & 0.71 & 1.39 \\ 
   17 & 0 & 1219.04 & 0.16 & $>6$ \\ 
   17 & 1 & 1248.33 & 0.17 & $>6$ \\ 
   17 & 2 & 1277.08 & 0.69 & -0.25 \\ 
   18 & 0 & 1286.41 & 0.76 & 1.12 \\ 
   18 & 1 & 1315.37 & 0.29 & $>6$ \\ 
   18 & 2 & 1347.8 & 0.68 & $>6$ \\ 
   19 & 0 & 1353.11 & 0.47 & $>6$ \\ 
   19 & 1 & 1382.46 & 0.44 & $>6$ \\ 
   20 & 0 & 1418.82 & 0.67 & $>6$ \\ 
   20 & 1 & 1449.55 & 0.59 & 1.41 \\ 
   21 & 0 & 1488.14 & 1.0 & 0.93 \\ 
   21 & 1 & 1517.37 & 0.9 & 1.11 \\ 
   22 & 1 & 1584.82 & 0.81 & $>6$ \\ 
   23 & 0 & 1620.42 & 0.74 & 1.42 \\ 
   23 & 1 & 1651.65 & 0.57 & 0.97 \\ 
   25 & 0 & 1753.17 & 0.77 & 1.18 \\ 
   \hline 
   \end{tabular} 
   \label{tab::6196457}\end{table} 
\begin{table} 
   \caption{Ratios for KIC 6196457.} 
   \begin{tabular}{cccc} 
   \hline 
   \hline 
   Ratio type & $n$ & Ratio & 68\% credible interval \\ 
   \hline 
   $r_{01}$ & 15 & 0.063 & 0.013 \\ 
   $r_{10}$ & 15 & 0.067 & 0.012 \\ 
   $r_{01}$ & 16 & 0.069 & 0.012 \\ 
   $r_{10}$ & 16 & 0.069 & 0.011 \\ 
   $r_{01}$ & 17 & 0.067 & 0.006 \\ 
   $r_{10}$ & 17 & 0.066 & 0.006 \\ 
   $r_{01}$ & 18 & 0.067 & 0.009 \\ 
   $r_{10}$ & 18 & 0.065 & 0.008 \\ 
   $r_{01}$ & 19 & 0.06 & 0.007 \\ 
   $r_{10}$ & 19 & 0.053 & 0.008 \\ 
   $r_{01}$ & 20 & 0.049 & 0.01 \\ 
   $r_{10}$ & 20 & 0.056 & 0.011 \\ 
   $r_{02}$ & 18 & 0.138 & 0.015 \\ 
   $r_{02}$ & 19 & 0.079 & 0.012 \\ 
   \hline 
   \end{tabular} 
   \label{tab::rat_6196457}\end{table} 
\clearpage
\subsubsection{6278762}
\begin{figure}
   \includegraphics[width=80mm]{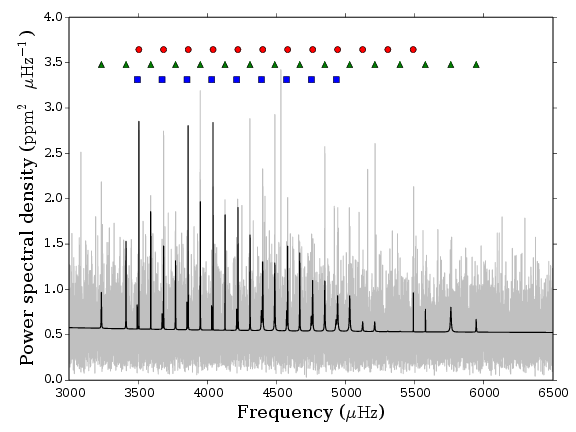}
   \includegraphics[width=80mm]{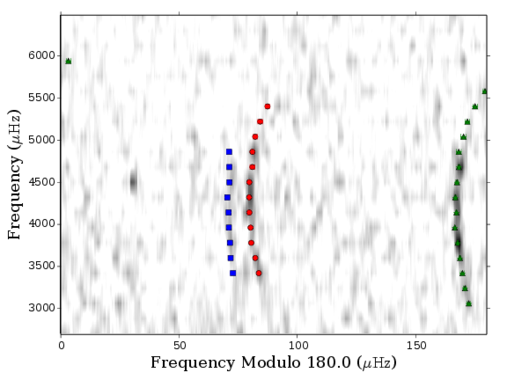}
   \caption{Power spectrum and echelle diagram for KIC 6278762.  Top: Power spectrum with data in grey smoothed over $3 \; \rm \mu Hz$ and best model in black.  Bottom: Echelle diagram with power in grey-scale.  Both: Mode frequencies are marked as: radial modes with red circles; dipole modes with green diamonds; quadrapole modes with blue squares; and octopole modes with yellow pentagons.}   \label{fig::6278762}\end{figure}
\begin{figure}
   \includegraphics[width=80mm]{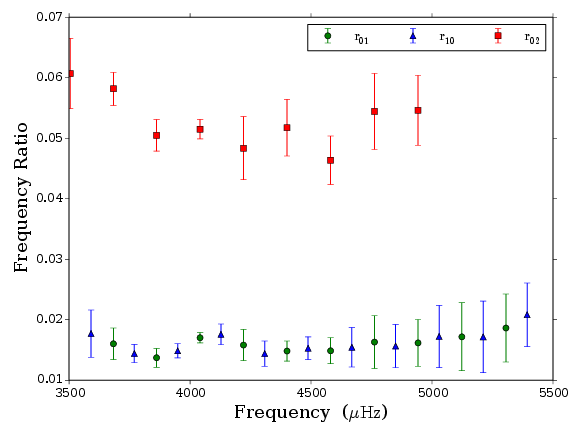}
   \caption{Ratios and $67 \%$ confidence intervals as a function of frequency for KIC 6278762.}   \label{fig::rat_6278762}\end{figure}
\begin{figure}
   \includegraphics[width=80mm]{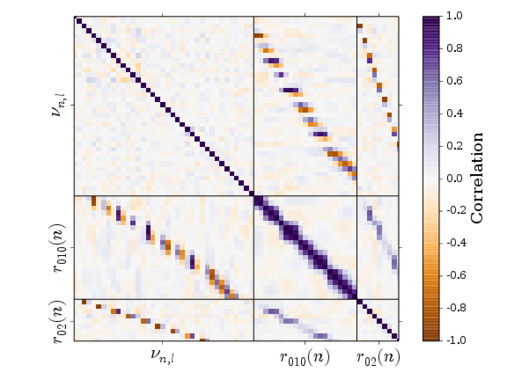}
   \caption{Correlation matrix of all frequencies and ratios for KIC 6278762.  The grid represents the matrix and hence the identity elements are all correlation 1.0.  The matrix is constructed so that frequencies and ratios are grouped separately.  If each matrix element is labelled $[i,j]$ then the first set of $i$ contains the mode frequencies, the second set contains the $r_{010}$, and the final set the $r_{02}$.}   \label{fig::cor_6278762}\end{figure}
\clearpage
\begin{table}
   \caption{Mode frequencies and statistics for KIC 6278762.}
   \begin{tabular}{ccccc}
   \hline 
   \hline 
   $n$ & $l$ & Frequency & 68\% credible & $ \ln{K}$  \\ 
 & & ($\rm \mu Hz$) & ($\rm \mu Hz$) & \\ 
   \hline 
   16 & 1 & 3233.26 & 0.7 & 1.7 \\ 
   17 & 1 & 3411.7 & 0.52 & $>6$ \\ 
   17 & 2 & 3493.67 & 1.03 & 0.1 \\ 
   18 & 0 & 3504.57 & 0.34 & 2.69 \\ 
   18 & 1 & 3590.67 & 0.88 & $>6$ \\ 
   18 & 2 & 3672.71 & 0.48 & 0.21 \\ 
   19 & 0 & 3683.12 & 0.14 & 2.47 \\ 
   19 & 1 & 3769.67 & 0.25 & $>6$ \\ 
   19 & 2 & 3852.45 & 0.31 & 0.56 \\ 
   20 & 0 & 3861.5 & 0.32 & 1.81 \\ 
   20 & 1 & 3948.73 & 0.13 & $>6$ \\ 
   20 & 2 & 4032.07 & 0.27 & 0.34 \\ 
   21 & 0 & 4041.25 & 0.15 & 1.91 \\ 
   21 & 1 & 4127.62 & 0.09 & $>6$ \\ 
   21 & 2 & 4212.04 & 0.75 & 0.68 \\ 
   22 & 0 & 4220.76 & 0.61 & 2.96 \\ 
   22 & 1 & 4308.33 & 0.24 & $>6$ \\ 
   22 & 2 & 4391.47 & 0.8 & 0.06 \\ 
   23 & 0 & 4400.74 & 0.25 & 2.82 \\ 
   23 & 1 & 4487.89 & 0.36 & $>6$ \\ 
   23 & 2 & 4572.47 & 0.68 & 0.69 \\ 
   24 & 0 & 4580.84 & 0.31 & 2.85 \\ 
   24 & 1 & 4668.67 & 0.49 & 1.52 \\ 
   24 & 2 & 4752.43 & 0.82 & -0.04 \\ 
   25 & 0 & 4762.2 & 0.99 & 1.32 \\ 
   25 & 1 & 4849.51 & 0.38 & $>6$ \\ 
   25 & 2 & 4932.4 & 0.91 & -0.84 \\ 
   26 & 0 & 4942.25 & 0.53 & 2.74 \\ 
   26 & 1 & 5029.49 & 1.05 & $>6$ \\ 
   27 & 0 & 5123.48 & 0.92 & 1.29 \\ 
   27 & 1 & 5211.57 & 1.05 & 0.94 \\ 
   28 & 0 & 5305.62 & 0.97 & 0.73 \\ 
   28 & 1 & 5393.24 & 0.85 & 0.72 \\ 
   29 & 0 & 5488.79 & 1.04 & 0.68 \\ 
   29 & 1 & 5576.51 & 0.91 & 0.68 \\ 
   30 & 1 & 5760.77 & 0.77 & 1.32 \\ 
   31 & 1 & 5944.74 & 0.93 & 1.07 \\ 
   \hline 
   \end{tabular} 
   \label{tab::6278762}\end{table} 
\begin{table} 
   \caption{Ratios for KIC 6278762.} 
   \begin{tabular}{cccc} 
   \hline 
   \hline 
   Ratio type & $n$ & Ratio & 68\% credible interval \\ 
   \hline 
   $r_{10}$ & 18 & 0.018 & 0.004 \\ 
   $r_{01}$ & 19 & 0.016 & 0.003 \\ 
   $r_{10}$ & 19 & 0.014 & 0.002 \\ 
   $r_{01}$ & 20 & 0.014 & 0.002 \\ 
   $r_{10}$ & 20 & 0.015 & 0.001 \\ 
   $r_{01}$ & 21 & 0.017 & 0.001 \\ 
   $r_{10}$ & 21 & 0.018 & 0.002 \\ 
   $r_{01}$ & 22 & 0.016 & 0.003 \\ 
   $r_{10}$ & 22 & 0.014 & 0.002 \\ 
   $r_{01}$ & 23 & 0.015 & 0.002 \\ 
   $r_{10}$ & 23 & 0.015 & 0.002 \\ 
   $r_{01}$ & 24 & 0.015 & 0.002 \\ 
   $r_{10}$ & 24 & 0.015 & 0.003 \\ 
   $r_{01}$ & 25 & 0.016 & 0.004 \\ 
   $r_{10}$ & 25 & 0.016 & 0.004 \\ 
   $r_{01}$ & 26 & 0.016 & 0.004 \\ 
   $r_{10}$ & 26 & 0.017 & 0.005 \\ 
   $r_{01}$ & 27 & 0.017 & 0.006 \\ 
   $r_{10}$ & 27 & 0.017 & 0.006 \\ 
   $r_{01}$ & 28 & 0.019 & 0.006 \\ 
   $r_{10}$ & 28 & 0.021 & 0.005 \\ 
   $r_{02}$ & 18 & 0.061 & 0.006 \\ 
   $r_{02}$ & 19 & 0.058 & 0.003 \\ 
   $r_{02}$ & 20 & 0.05 & 0.003 \\ 
   $r_{02}$ & 21 & 0.052 & 0.002 \\ 
   $r_{02}$ & 22 & 0.048 & 0.005 \\ 
   $r_{02}$ & 23 & 0.052 & 0.005 \\ 
   $r_{02}$ & 24 & 0.046 & 0.004 \\ 
   $r_{02}$ & 25 & 0.054 & 0.006 \\ 
   $r_{02}$ & 26 & 0.055 & 0.006 \\ 
   \hline 
   \end{tabular} 
   \label{tab::rat_6278762}\end{table} 
\clearpage
\subsubsection{6521045}
\begin{figure}
   \includegraphics[width=80mm]{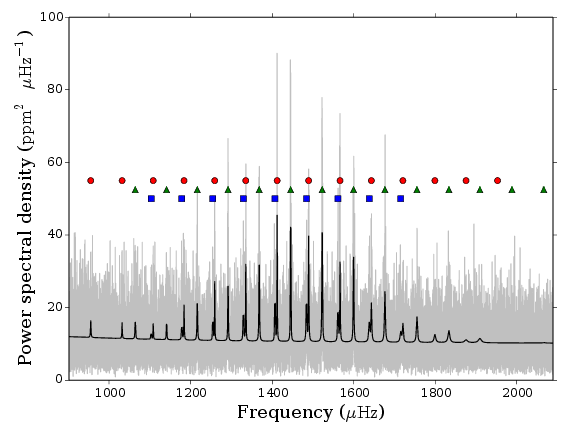}
   \includegraphics[width=80mm]{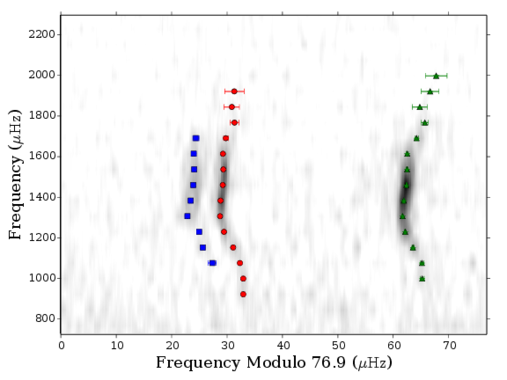}
   \caption{Power spectrum and echelle diagram for KIC 6521045.  Top: Power spectrum with data in grey smoothed over $3 \; \rm \mu Hz$ and best model in black.  Bottom: Echelle diagram with power in grey-scale.  Both: Mode frequencies are marked as: radial modes with red circles; dipole modes with green diamonds; quadrapole modes with blue squares; and octopole modes with yellow pentagons.}   \label{fig::6521045}\end{figure}
\begin{figure}
   \includegraphics[width=80mm]{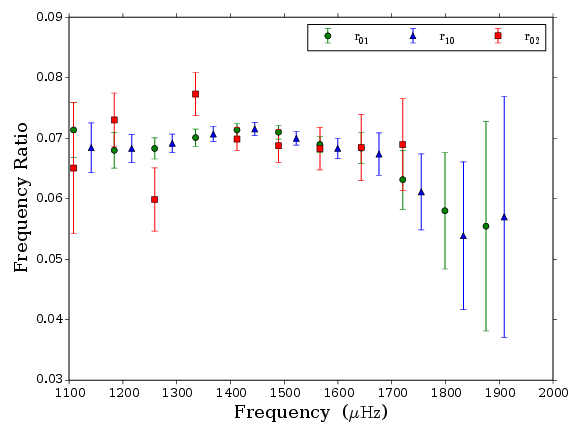}
   \caption{Ratios and $67 \%$ confidence intervals as a function of frequency for KIC 6521045.}   \label{fig::rat_6521045}\end{figure}
\begin{figure}
   \includegraphics[width=80mm]{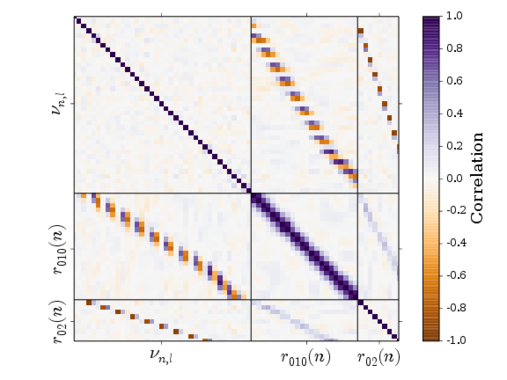}
   \caption{Correlation matrix of all frequencies and ratios for KIC 6521045.  The grid represents the matrix and hence the identity elements are all correlation 1.0.  The matrix is constructed so that frequencies and ratios are grouped separately.  If each matrix element is labelled $[i,j]$ then the first set of $i$ contains the mode frequencies, the second set contains the $r_{010}$, and the final set the $r_{02}$.}   \label{fig::cor_6521045}\end{figure}
\clearpage
\begin{table}
   \caption{Mode frequencies and statistics for KIC 6521045.}
   \begin{tabular}{ccccc}
   \hline 
   \hline 
   $n$ & $l$ & Frequency & 68\% credible & $ \ln{K}$  \\ 
 & & ($\rm \mu Hz$) & ($\rm \mu Hz$) & \\ 
   \hline 
   12 & 0 & 955.19 & 0.32 & $>6$ \\ 
   13 & 0 & 1032.04 & 0.3 & $>6$ \\ 
   13 & 1 & 1064.34 & 0.36 & 2.38 \\ 
   13 & 2 & 1103.31 & 0.75 & -0.15 \\ 
   14 & 0 & 1108.3 & 0.34 & 1.8 \\ 
   14 & 1 & 1141.16 & 0.31 & 1.68 \\ 
   14 & 2 & 1178.46 & 0.28 & 1.16 \\ 
   15 & 0 & 1183.94 & 0.2 & 3.15 \\ 
   15 & 1 & 1216.38 & 0.16 & $>6$ \\ 
   15 & 2 & 1254.63 & 0.38 & 0.36 \\ 
   16 & 0 & 1259.14 & 0.12 & 3.58 \\ 
   16 & 1 & 1291.84 & 0.11 & $>6$ \\ 
   16 & 2 & 1329.39 & 0.26 & $>6$ \\ 
   17 & 0 & 1335.29 & 0.1 & $>6$ \\ 
   17 & 1 & 1368.23 & 0.1 & $>6$ \\ 
   17 & 2 & 1406.84 & 0.14 & $>6$ \\ 
   18 & 0 & 1412.23 & 0.07 & $>6$ \\ 
   18 & 1 & 1445.3 & 0.08 & $>6$ \\ 
   18 & 2 & 1484.17 & 0.19 & $>6$ \\ 
   19 & 0 & 1489.48 & 0.09 & $>6$ \\ 
   19 & 1 & 1522.62 & 0.08 & $>6$ \\ 
   19 & 2 & 1561.19 & 0.26 & $>6$ \\ 
   20 & 0 & 1566.44 & 0.09 & $>6$ \\ 
   20 & 1 & 1599.6 & 0.11 & $>6$ \\ 
   20 & 2 & 1637.96 & 0.4 & $>6$ \\ 
   21 & 0 & 1643.23 & 0.17 & $>6$ \\ 
   21 & 1 & 1676.49 & 0.21 & $>6$ \\ 
   21 & 2 & 1715.2 & 0.56 & $>6$ \\ 
   22 & 0 & 1720.6 & 0.41 & $>6$ \\ 
   22 & 1 & 1755.02 & 0.31 & $>6$ \\ 
   23 & 0 & 1799.02 & 0.8 & 2.2 \\ 
   23 & 1 & 1833.38 & 0.62 & 2.62 \\ 
   24 & 0 & 1875.39 & 1.38 & 1.48 \\ 
   24 & 1 & 1909.32 & 1.36 & 0.78 \\ 
   25 & 0 & 1952.7 & 1.75 & 0.84 \\ 
   25 & 1 & 1988.05 & 1.59 & 0.68 \\ 
   26 & 1 & 2066.0 & 1.96 & 0.67 \\ 
   \hline 
   \end{tabular} 
   \label{tab::6521045}\end{table} 
\begin{table} 
   \caption{Ratios for KIC 6521045.} 
   \begin{tabular}{cccc} 
   \hline 
   \hline 
   Ratio type & $n$ & Ratio & 68\% credible interval \\ 
   \hline 
   $r_{01}$ & 14 & 0.071 & 0.005 \\ 
   $r_{10}$ & 14 & 0.068 & 0.004 \\ 
   $r_{01}$ & 15 & 0.068 & 0.003 \\ 
   $r_{10}$ & 15 & 0.068 & 0.002 \\ 
   $r_{01}$ & 16 & 0.068 & 0.002 \\ 
   $r_{10}$ & 16 & 0.069 & 0.002 \\ 
   $r_{01}$ & 17 & 0.07 & 0.001 \\ 
   $r_{10}$ & 17 & 0.071 & 0.001 \\ 
   $r_{01}$ & 18 & 0.071 & 0.001 \\ 
   $r_{10}$ & 18 & 0.072 & 0.001 \\ 
   $r_{01}$ & 19 & 0.071 & 0.001 \\ 
   $r_{10}$ & 19 & 0.07 & 0.001 \\ 
   $r_{01}$ & 20 & 0.069 & 0.001 \\ 
   $r_{10}$ & 20 & 0.068 & 0.002 \\ 
   $r_{01}$ & 21 & 0.068 & 0.003 \\ 
   $r_{10}$ & 21 & 0.067 & 0.004 \\ 
   $r_{01}$ & 22 & 0.063 & 0.005 \\ 
   $r_{10}$ & 22 & 0.061 & 0.006 \\ 
   $r_{01}$ & 23 & 0.058 & 0.01 \\ 
   $r_{10}$ & 23 & 0.054 & 0.012 \\ 
   $r_{01}$ & 24 & 0.055 & 0.017 \\ 
   $r_{10}$ & 24 & 0.057 & 0.02 \\ 
   $r_{02}$ & 14 & 0.065 & 0.011 \\ 
   $r_{02}$ & 15 & 0.073 & 0.004 \\ 
   $r_{02}$ & 16 & 0.06 & 0.005 \\ 
   $r_{02}$ & 17 & 0.077 & 0.004 \\ 
   $r_{02}$ & 18 & 0.07 & 0.002 \\ 
   $r_{02}$ & 19 & 0.069 & 0.003 \\ 
   $r_{02}$ & 20 & 0.068 & 0.003 \\ 
   $r_{02}$ & 21 & 0.068 & 0.005 \\ 
   $r_{02}$ & 22 & 0.069 & 0.008 \\ 
   \hline 
   \end{tabular} 
   \label{tab::rat_6521045}\end{table} 
\clearpage
\subsubsection{7199397}
\begin{figure}
   \includegraphics[width=80mm]{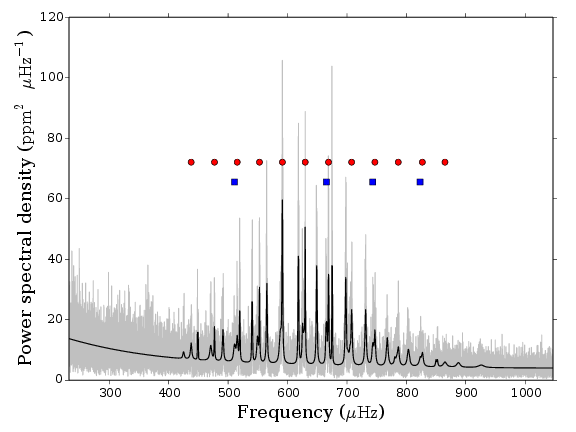}
   \includegraphics[width=80mm]{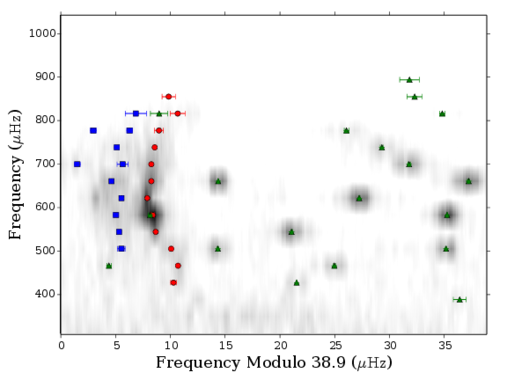}
   \caption{Power spectrum and echelle diagram for 7199397.}\end{figure}
\clearpage
\begin{table}
   \caption{Frequencies for 7199397}
   \begin{tabular}{cccc}
   \hline 
   \hline 
   $n$ & $l$ & Frequency ($\rm \mu Hz$) & 68\% credible ($\rm \mu Hz$) \\ 
   \hline 
   8(01) & 1 & 425.05 & 0.58 \\ 
   10 & 0 & 437.8 & 0.26 \\ 
   9(01) & 1 & 449.04 & 0.06 \\ 
   11 & 0 & 477.07 & 0.11 \\ 
   10(01) & 1 & 491.34 & 0.13 \\ 
   10(02) & 1 & 470.77 & 0.21 \\ 
   11 & 2 & 510.77 & 0.34 \\ 
   12 & 0 & 515.31 & 0.19 \\ 
   11(01) & 1 & 540.4 & 0.06 \\ 
   11(02) & 1 & 519.58 & 0.1 \\ 
   12(02) & 2 & 549.42 & 0.16 \\ 
   13 & 0 & 552.77 & 0.07 \\ 
   12(01) & 1 & 565.16 & 0.08 \\ 
   13(01) & 2 & 588.0 & 0.26 \\ 
   14 & 0 & 591.35 & 0.17 \\ 
   13(01) & 1 & 618.23 & 0.06 \\ 
   13(02) & 1 & 591.13 & 0.15 \\ 
   14(02) & 2 & 627.38 & 0.12 \\ 
   15 & 0 & 629.72 & 0.06 \\ 
   14(01) & 1 & 649.07 & 0.07 \\ 
   15 & 2 & 665.34 & 0.11 \\ 
   16 & 0 & 668.96 & 0.07 \\ 
   15(02) & 1 & 675.06 & 0.04 \\ 
   16(01) & 2 & 705.21 & 0.51 \\ 
   16(02) & 2 & 701.07 & 0.3 \\ 
   17 & 0 & 707.83 & 0.13 \\ 
   16(01) & 1 & 731.35 & 0.11 \\ 
   16(02) & 1 & 697.92 & 0.11 \\ 
   17 & 2 & 743.55 & 0.22 \\ 
   18 & 0 & 747.0 & 0.14 \\ 
   17(01) & 1 & 767.73 & 0.15 \\ 
   19 & 0 & 786.26 & 0.4 \\ 
   18(02) & 2 & 783.58 & 0.28 \\ 
   18(01) & 2 & 780.26 & 0.29 \\ 
   18(01) & 1 & 803.36 & 0.2 \\ 
   20 & 0 & 826.83 & 0.69 \\ 
   19(01) & 1 & 850.97 & 0.24 \\ 
   19(02) & 1 & 825.14 & 0.79 \\ 
   20 & 2 & 823.01 & 0.97 \\ 
   21 & 0 & 864.87 & 0.61 \\ 
   20(01) & 1 & 887.32 & 0.7 \\ 
   20(02) & 1 & 925.71 & 0.89 \\ 
   \hline 
   \end{tabular} 
\end{table} 
\clearpage
\subsubsection{7670943}
\begin{figure}
   \includegraphics[width=80mm]{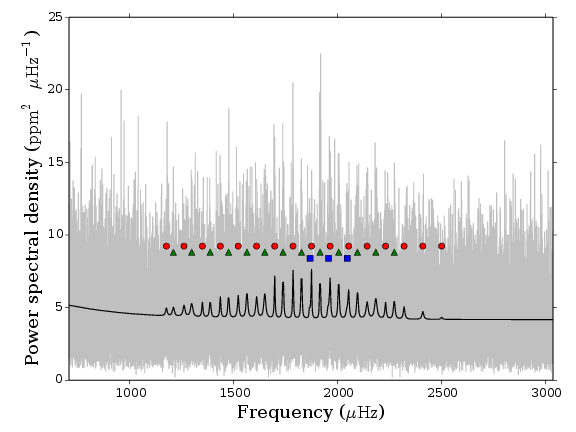}
   \includegraphics[width=80mm]{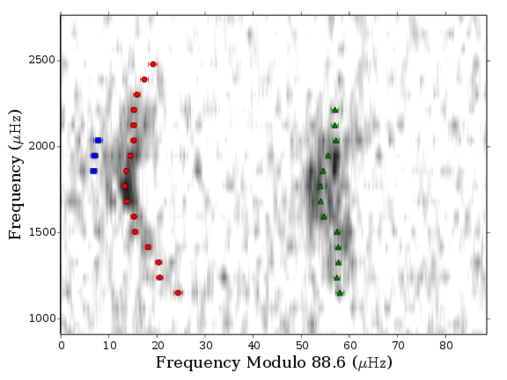}
   \caption{Power spectrum and echelle diagram for KIC 7670943.  Top: Power spectrum with data in grey smoothed over $3 \; \rm \mu Hz$ and best model in black.  Bottom: Echelle diagram with power in grey-scale.  Both: Mode frequencies are marked as: radial modes with red circles; dipole modes with green diamonds; quadrapole modes with blue squares; and octopole modes with yellow pentagons.}   \label{fig::7670943}\end{figure}
\begin{figure}
   \includegraphics[width=80mm]{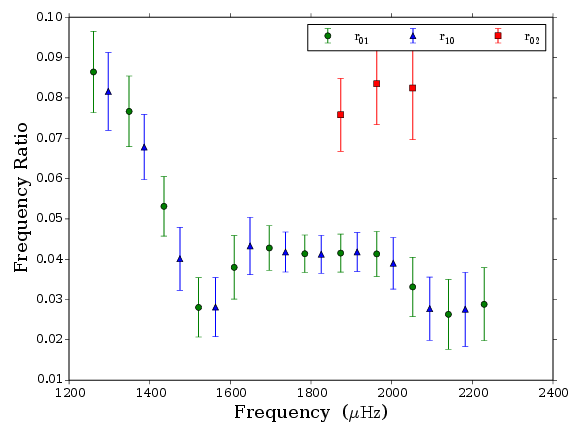}
   \caption{Ratios and $67 \%$ confidence intervals as a function of frequency for KIC 7670943.}   \label{fig::rat_7670943}\end{figure}
\begin{figure}
   \includegraphics[width=80mm]{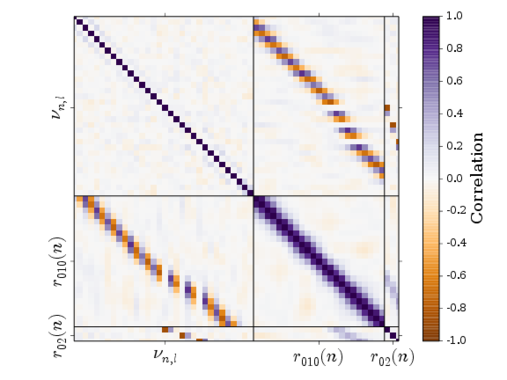}
   \caption{Correlation matrix of all frequencies and ratios for KIC 7670943.  The grid represents the matrix and hence the identity elements are all correlation 1.0.  The matrix is constructed so that frequencies and ratios are grouped separately.  If each matrix element is labelled $[i,j]$ then the first set of $i$ contains the mode frequencies, the second set contains the $r_{010}$, and the final set the $r_{02}$.}   \label{fig::cor_7670943}\end{figure}
\clearpage
\begin{table}
   \caption{Mode frequencies and statistics for KIC 7670943.}
   \begin{tabular}{ccccc}
   \hline 
   \hline 
   $n$ & $l$ & Frequency & 68\% credible & $ \ln{K}$  \\ 
 & & ($\rm \mu Hz$) & ($\rm \mu Hz$) & \\ 
   \hline 
   12 & 0 & 1175.7 & 0.9 & 1.03 \\ 
   12 & 1 & 1209.37 & 0.87 & 1.16 \\ 
   13 & 0 & 1260.45 & 0.78 & 0.86 \\ 
   13 & 1 & 1297.3 & 0.7 & 1.2 \\ 
   14 & 0 & 1348.81 & 0.76 & 0.8 \\ 
   14 & 1 & 1386.18 & 0.61 & 2.55 \\ 
   15 & 0 & 1435.14 & 0.55 & 2.17 \\ 
   15 & 1 & 1474.69 & 0.64 & 2.74 \\ 
   16 & 0 & 1521.05 & 0.63 & 2.67 \\ 
   16 & 1 & 1563.03 & 0.59 & 3.2 \\ 
   17 & 0 & 1609.37 & 0.73 & 1.58 \\ 
   17 & 1 & 1648.85 & 0.59 & $>6$ \\ 
   18 & 0 & 1696.44 & 0.39 & $>6$ \\ 
   18 & 1 & 1736.74 & 0.46 & 3.92 \\ 
   19 & 0 & 1784.58 & 0.35 & $>6$ \\ 
   19 & 1 & 1825.25 & 0.43 & $>6$ \\ 
   19 & 2 & 1866.69 & 0.72 & 1.26 \\ 
   20 & 0 & 1873.44 & 0.41 & 3.77 \\ 
   20 & 1 & 1914.37 & 0.34 & $>6$ \\ 
   20 & 2 & 1955.41 & 0.82 & 0.67 \\ 
   21 & 0 & 1962.91 & 0.48 & 4.04 \\ 
   21 & 1 & 2003.97 & 0.49 & $>6$ \\ 
   21 & 2 & 2044.7 & 0.92 & 1.4 \\ 
   22 & 0 & 2052.12 & 0.7 & 4.01 \\ 
   22 & 1 & 2094.2 & 0.61 & 4.08 \\ 
   23 & 0 & 2140.68 & 0.68 & 3.13 \\ 
   23 & 1 & 2182.5 & 0.78 & 1.53 \\ 
   24 & 0 & 2229.3 & 0.65 & 2.66 \\ 
   24 & 1 & 2271.12 & 0.74 & 1.72 \\ 
   25 & 0 & 2318.5 & 0.76 & 1.72 \\ 
   26 & 0 & 2408.59 & 0.84 & 0.9 \\ 
   27 & 0 & 2498.96 & 0.93 & 0.67 \\ 
   \hline 
   \end{tabular} 
   \label{tab::7670943}\end{table} 
\begin{table} 
   \caption{Ratios for KIC 7670943.} 
   \begin{tabular}{cccc} 
   \hline 
   \hline 
   Ratio type & $n$ & Ratio & 68\% credible interval \\ 
   \hline 
   $r_{01}$ & 13 & 0.086 & 0.01 \\ 
   $r_{10}$ & 13 & 0.082 & 0.01 \\ 
   $r_{01}$ & 14 & 0.077 & 0.009 \\ 
   $r_{10}$ & 14 & 0.068 & 0.008 \\ 
   $r_{01}$ & 15 & 0.053 & 0.007 \\ 
   $r_{10}$ & 15 & 0.04 & 0.008 \\ 
   $r_{01}$ & 16 & 0.028 & 0.007 \\ 
   $r_{10}$ & 16 & 0.028 & 0.007 \\ 
   $r_{01}$ & 17 & 0.038 & 0.008 \\ 
   $r_{10}$ & 17 & 0.043 & 0.007 \\ 
   $r_{01}$ & 18 & 0.043 & 0.006 \\ 
   $r_{10}$ & 18 & 0.042 & 0.005 \\ 
   $r_{01}$ & 19 & 0.041 & 0.005 \\ 
   $r_{10}$ & 19 & 0.041 & 0.005 \\ 
   $r_{01}$ & 20 & 0.042 & 0.005 \\ 
   $r_{10}$ & 20 & 0.042 & 0.005 \\ 
   $r_{01}$ & 21 & 0.041 & 0.006 \\ 
   $r_{10}$ & 21 & 0.039 & 0.006 \\ 
   $r_{01}$ & 22 & 0.033 & 0.007 \\ 
   $r_{10}$ & 22 & 0.028 & 0.008 \\ 
   $r_{01}$ & 23 & 0.026 & 0.009 \\ 
   $r_{10}$ & 23 & 0.028 & 0.009 \\ 
   $r_{01}$ & 24 & 0.029 & 0.009 \\ 
   $r_{02}$ & 20 & 0.076 & 0.009 \\ 
   $r_{02}$ & 21 & 0.084 & 0.01 \\ 
   $r_{02}$ & 22 & 0.082 & 0.013 \\ 
   \hline 
   \end{tabular} 
   \label{tab::rat_7670943}\end{table} 
\clearpage
\subsubsection{8077137}
\begin{figure}
   \includegraphics[width=80mm]{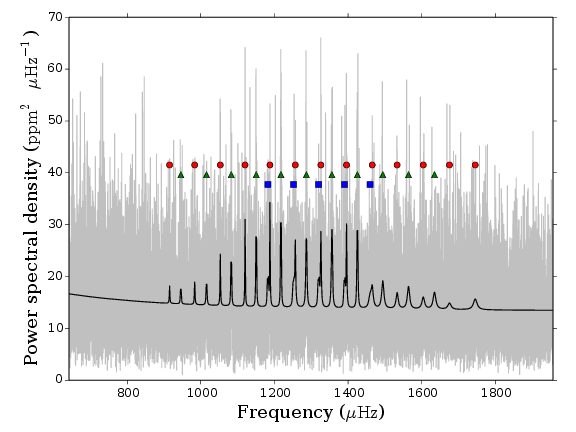}
   \includegraphics[width=80mm]{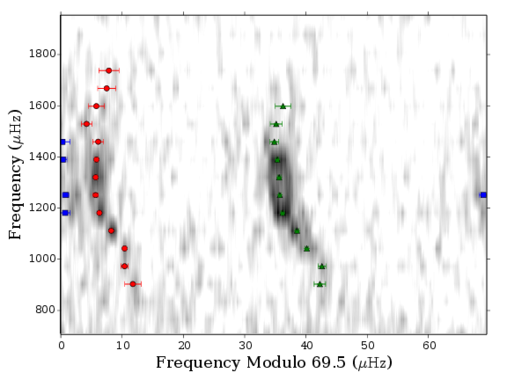}
   \caption{Power spectrum and echelle diagram for KIC 8077137.  Top: Power spectrum with data in grey smoothed over $3 \; \rm \mu Hz$ and best model in black.  Bottom: Echelle diagram with power in grey-scale.  Both: Mode frequencies are marked as: radial modes with red circles; dipole modes with green diamonds; quadrapole modes with blue squares; and octopole modes with yellow pentagons.}   \label{fig::8077137}\end{figure}
\begin{figure}
   \includegraphics[width=80mm]{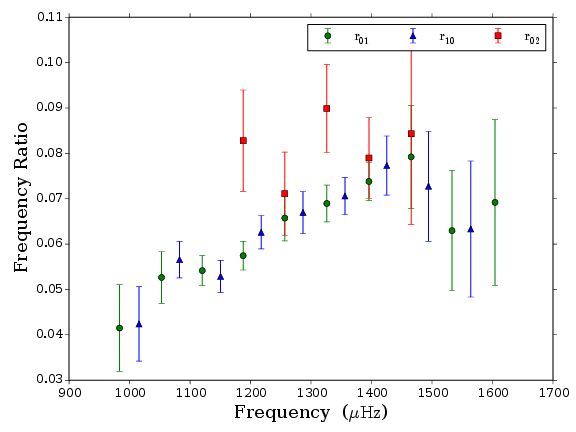}
   \caption{Ratios and $67 \%$ confidence intervals as a function of frequency for KIC 8077137.}   \label{fig::rat_8077137}\end{figure}
\begin{figure}
   \includegraphics[width=80mm]{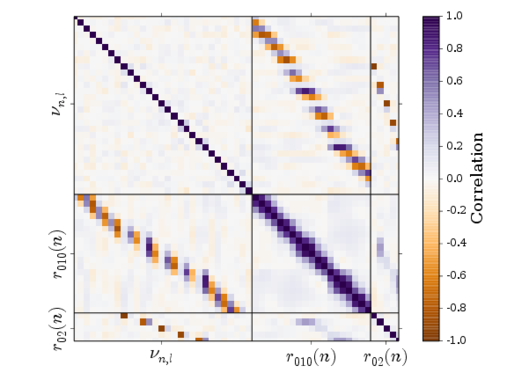}
   \caption{Correlation matrix of all frequencies and ratios for KIC 8077137.  The grid represents the matrix and hence the identity elements are all correlation 1.0.  The matrix is constructed so that frequencies and ratios are grouped separately.  If each matrix element is labelled $[i,j]$ then the first set of $i$ contains the mode frequencies, the second set contains the $r_{010}$, and the final set the $r_{02}$.}   \label{fig::cor_8077137}\end{figure}
\clearpage
\begin{table}
   \caption{Mode frequencies and statistics for KIC 8077137.}
   \begin{tabular}{ccccc}
   \hline 
   \hline 
   $n$ & $l$ & Frequency & 68\% credible & $ \ln{K}$  \\ 
 & & ($\rm \mu Hz$) & ($\rm \mu Hz$) & \\ 
   \hline 
   13 & 0 & 915.2 & 1.33 & $>6$ \\ 
   13 & 1 & 945.72 & 0.93 & 1.24 \\ 
   14 & 0 & 983.35 & 0.55 & 1.6 \\ 
   14 & 1 & 1015.56 & 0.64 & $>6$ \\ 
   15 & 0 & 1052.85 & 0.25 & $>6$ \\ 
   15 & 1 & 1082.59 & 0.28 & $>6$ \\ 
   16 & 0 & 1120.16 & 0.16 & $>6$ \\ 
   16 & 1 & 1150.48 & 0.29 & $>6$ \\ 
   16 & 2 & 1182.15 & 0.76 & $>6$ \\ 
   17 & 0 & 1187.73 & 0.16 & $>6$ \\ 
   17 & 1 & 1217.61 & 0.19 & $>6$ \\ 
   17 & 2 & 1251.73 & 0.51 & 2.02 \\ 
   18 & 0 & 1256.61 & 0.4 & 3.72 \\ 
   18 & 1 & 1286.67 & 0.26 & $>6$ \\ 
   18 & 2 & 1319.88 & 0.67 & 2.48 \\ 
   19 & 0 & 1326.11 & 0.24 & 4.4 \\ 
   19 & 1 & 1356.05 & 0.28 & $>6$ \\ 
   19 & 2 & 1390.26 & 0.65 & 1.52 \\ 
   20 & 0 & 1395.74 & 0.28 & 4.06 \\ 
   20 & 1 & 1425.27 & 0.2 & $>6$ \\ 
   20 & 2 & 1459.69 & 1.23 & 0.27 \\ 
   21 & 0 & 1465.51 & 0.89 & 2.73 \\ 
   21 & 1 & 1494.26 & 0.71 & 2.59 \\ 
   22 & 0 & 1533.12 & 0.86 & $>6$ \\ 
   22 & 1 & 1564.06 & 0.98 & 2.13 \\ 
   23 & 0 & 1604.21 & 1.28 & 2.73 \\ 
   23 & 1 & 1634.67 & 1.26 & 1.43 \\ 
   24 & 0 & 1675.39 & 1.46 & 0.93 \\ 
   25 & 0 & 1745.24 & 1.66 & 1.14 \\ 
   \hline 
   \end{tabular} 
   \label{tab::8077137}\end{table} 
\begin{table} 
   \caption{Ratios for KIC 8077137.} 
   \begin{tabular}{cccc} 
   \hline 
   \hline 
   Ratio type & $n$ & Ratio & 68\% credible interval \\ 
   \hline 
   $r_{01}$ & 14 & 0.041 & 0.01 \\ 
   $r_{10}$ & 14 & 0.042 & 0.008 \\ 
   $r_{01}$ & 15 & 0.053 & 0.006 \\ 
   $r_{10}$ & 15 & 0.057 & 0.004 \\ 
   $r_{01}$ & 16 & 0.054 & 0.003 \\ 
   $r_{10}$ & 16 & 0.053 & 0.004 \\ 
   $r_{01}$ & 17 & 0.057 & 0.003 \\ 
   $r_{10}$ & 17 & 0.063 & 0.004 \\ 
   $r_{01}$ & 18 & 0.066 & 0.005 \\ 
   $r_{10}$ & 18 & 0.067 & 0.005 \\ 
   $r_{01}$ & 19 & 0.069 & 0.004 \\ 
   $r_{10}$ & 19 & 0.071 & 0.004 \\ 
   $r_{01}$ & 20 & 0.074 & 0.004 \\ 
   $r_{10}$ & 20 & 0.077 & 0.007 \\ 
   $r_{01}$ & 21 & 0.079 & 0.011 \\ 
   $r_{10}$ & 21 & 0.073 & 0.012 \\ 
   $r_{01}$ & 22 & 0.063 & 0.013 \\ 
   $r_{10}$ & 22 & 0.063 & 0.015 \\ 
   $r_{01}$ & 23 & 0.069 & 0.018 \\ 
   $r_{02}$ & 17 & 0.083 & 0.011 \\ 
   $r_{02}$ & 18 & 0.071 & 0.009 \\ 
   $r_{02}$ & 19 & 0.09 & 0.01 \\ 
   $r_{02}$ & 20 & 0.079 & 0.009 \\ 
   $r_{02}$ & 21 & 0.084 & 0.02 \\ 
   \hline 
   \end{tabular} 
   \label{tab::rat_8077137}\end{table} 
\clearpage
\subsubsection{8292840}
\begin{figure}
   \includegraphics[width=80mm]{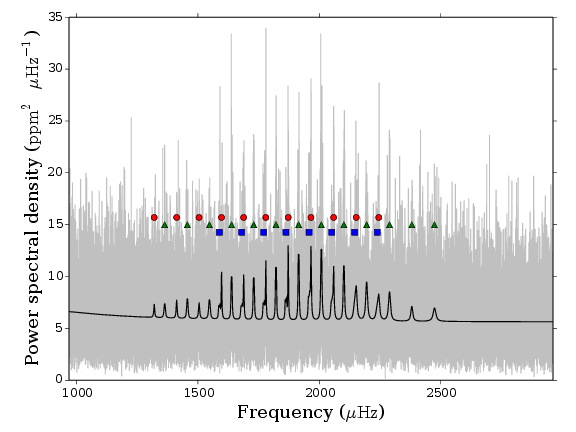}
   \includegraphics[width=80mm]{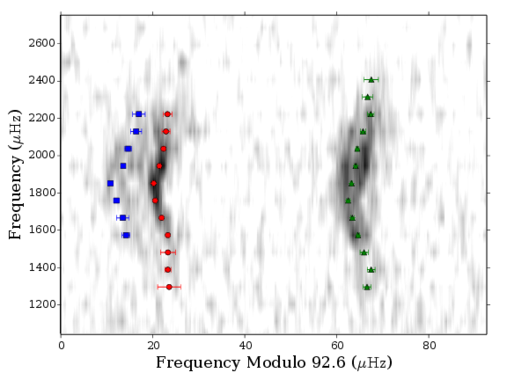}
   \caption{Power spectrum and echelle diagram for KIC 8292840.  Top: Power spectrum with data in grey smoothed over $3 \; \rm \mu Hz$ and best model in black.  Bottom: Echelle diagram with power in grey-scale.  Both: Mode frequencies are marked as: radial modes with red circles; dipole modes with green diamonds; quadrapole modes with blue squares; and octopole modes with yellow pentagons.}   \label{fig::8292840}\end{figure}
\begin{figure}
   \includegraphics[width=80mm]{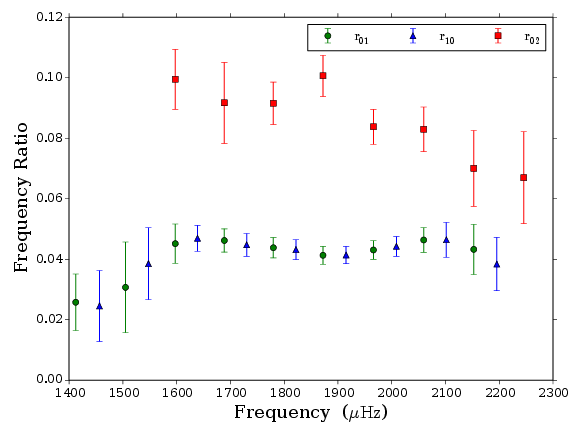}
   \caption{Ratios and $67 \%$ confidence intervals as a function of frequency for KIC 8292840.}   \label{fig::rat_8292840}\end{figure}
\begin{figure}
   \includegraphics[width=80mm]{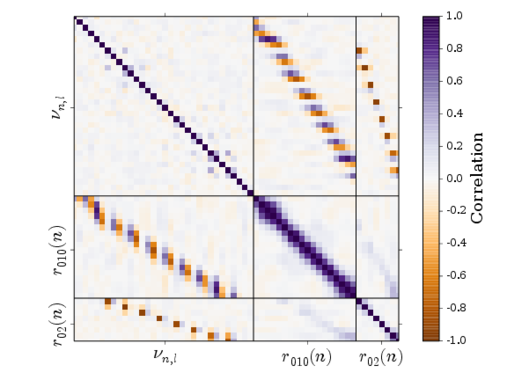}
   \caption{Correlation matrix of all frequencies and ratios for KIC 8292840.  The grid represents the matrix and hence the identity elements are all correlation 1.0.  The matrix is constructed so that frequencies and ratios are grouped separately.  If each matrix element is labelled $[i,j]$ then the first set of $i$ contains the mode frequencies, the second set contains the $r_{010}$, and the final set the $r_{02}$.}   \label{fig::cor_8292840}\end{figure}
\clearpage
\begin{table}
   \caption{Mode frequencies and statistics for KIC 8292840.}
   \begin{tabular}{ccccc}
   \hline 
   \hline 
   $n$ & $l$ & Frequency & 68\% credible & $ \ln{K}$  \\ 
 & & ($\rm \mu Hz$) & ($\rm \mu Hz$) & \\ 
   \hline 
   14 & 0 & 1319.97 & 2.51 & 0.75 \\ 
   14 & 1 & 1363.0 & 0.83 & 1.95 \\ 
   15 & 0 & 1412.29 & 0.68 & 1.85 \\ 
   15 & 1 & 1456.49 & 0.85 & 2.84 \\ 
   16 & 0 & 1504.9 & 1.61 & 1.44 \\ 
   16 & 1 & 1547.56 & 0.87 & 1.46 \\ 
   16 & 2 & 1588.39 & 0.86 & 1.34 \\ 
   17 & 0 & 1597.49 & 0.41 & 3.03 \\ 
   17 & 1 & 1638.84 & 0.3 & $>6$ \\ 
   17 & 2 & 1680.31 & 1.32 & 0.1 \\ 
   18 & 0 & 1688.72 & 0.34 & 3.72 \\ 
   18 & 1 & 1730.21 & 0.36 & $>6$ \\ 
   18 & 2 & 1771.53 & 0.62 & 2.45 \\ 
   19 & 0 & 1779.92 & 0.23 & 4.28 \\ 
   19 & 1 & 1821.88 & 0.33 & $>6$ \\ 
   19 & 2 & 1862.82 & 0.61 & 2.44 \\ 
   20 & 0 & 1872.23 & 0.2 & 4.6 \\ 
   20 & 1 & 1915.25 & 0.26 & $>6$ \\ 
   20 & 2 & 1958.19 & 0.52 & 2.93 \\ 
   21 & 0 & 1966.04 & 0.26 & 4.71 \\ 
   21 & 1 & 2008.73 & 0.27 & $>6$ \\ 
   21 & 2 & 2051.84 & 0.71 & 2.65 \\ 
   22 & 0 & 2059.56 & 0.34 & 4.46 \\ 
   22 & 1 & 2101.69 & 0.34 & $>6$ \\ 
   22 & 2 & 2146.2 & 1.2 & 0.15 \\ 
   23 & 0 & 2152.69 & 0.86 & 4.17 \\ 
   23 & 1 & 2195.53 & 0.52 & 4.19 \\ 
   23 & 2 & 2239.42 & 1.35 & 0.56 \\ 
   24 & 0 & 2245.63 & 0.92 & 3.65 \\ 
   24 & 1 & 2289.83 & 0.71 & $>6$ \\ 
   25 & 1 & 2381.74 & 1.12 & 2.7 \\ 
   26 & 1 & 2475.14 & 1.54 & 1.39 \\ 
   \hline 
   \end{tabular} 
   \label{tab::8292840}\end{table} 
\begin{table} 
   \caption{Ratios for KIC 8292840.} 
   \begin{tabular}{cccc} 
   \hline 
   \hline 
   Ratio type & $n$ & Ratio & 68\% credible interval \\ 
   \hline 
   $r_{01}$ & 15 & 0.026 & 0.009 \\ 
   $r_{10}$ & 15 & 0.025 & 0.012 \\ 
   $r_{01}$ & 16 & 0.031 & 0.015 \\ 
   $r_{10}$ & 16 & 0.038 & 0.012 \\ 
   $r_{01}$ & 17 & 0.045 & 0.006 \\ 
   $r_{10}$ & 17 & 0.047 & 0.004 \\ 
   $r_{01}$ & 18 & 0.046 & 0.004 \\ 
   $r_{10}$ & 18 & 0.045 & 0.004 \\ 
   $r_{01}$ & 19 & 0.044 & 0.003 \\ 
   $r_{10}$ & 19 & 0.043 & 0.003 \\ 
   $r_{01}$ & 20 & 0.041 & 0.003 \\ 
   $r_{10}$ & 20 & 0.041 & 0.003 \\ 
   $r_{01}$ & 21 & 0.043 & 0.003 \\ 
   $r_{10}$ & 21 & 0.044 & 0.003 \\ 
   $r_{01}$ & 22 & 0.046 & 0.004 \\ 
   $r_{10}$ & 22 & 0.046 & 0.006 \\ 
   $r_{01}$ & 23 & 0.043 & 0.008 \\ 
   $r_{10}$ & 23 & 0.038 & 0.009 \\ 
   $r_{02}$ & 17 & 0.1 & 0.01 \\ 
   $r_{02}$ & 18 & 0.092 & 0.013 \\ 
   $r_{02}$ & 19 & 0.092 & 0.007 \\ 
   $r_{02}$ & 20 & 0.101 & 0.007 \\ 
   $r_{02}$ & 21 & 0.084 & 0.006 \\ 
   $r_{02}$ & 22 & 0.083 & 0.007 \\ 
   $r_{02}$ & 23 & 0.07 & 0.013 \\ 
   $r_{02}$ & 24 & 0.067 & 0.015 \\ 
   \hline 
   \end{tabular} 
   \label{tab::rat_8292840}\end{table} 
\clearpage
\subsubsection{8349582}
\begin{figure}
   \includegraphics[width=80mm]{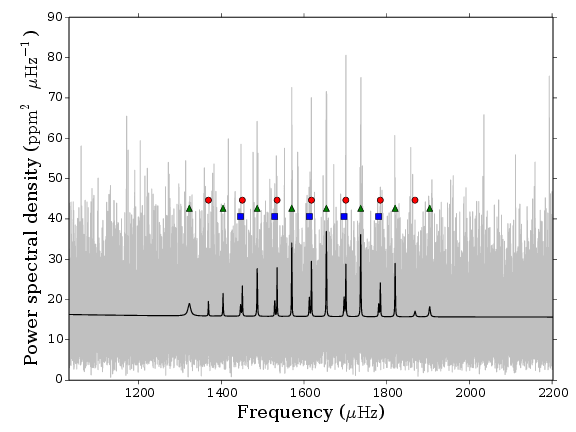}
   \includegraphics[width=80mm]{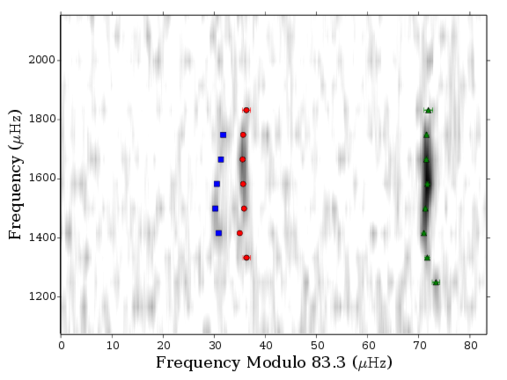}
   \caption{Power spectrum and echelle diagram for KIC 8349582.  Top: Power spectrum with data in grey smoothed over $3 \; \rm \mu Hz$ and best model in black.  Bottom: Echelle diagram with power in grey-scale.  Both: Mode frequencies are marked as: radial modes with red circles; dipole modes with green diamonds; quadrapole modes with blue squares; and octopole modes with yellow pentagons.}   \label{fig::8349582}\end{figure}
\begin{figure}
   \includegraphics[width=80mm]{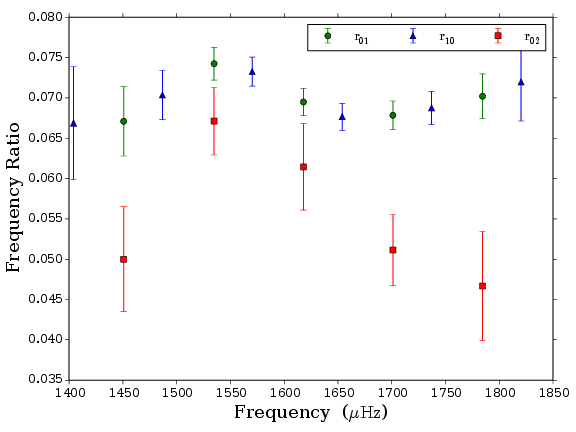}
   \caption{Ratios and $67 \%$ confidence intervals as a function of frequency for KIC 8349582.}   \label{fig::rat_8349582}\end{figure}
\begin{figure}
   \includegraphics[width=80mm]{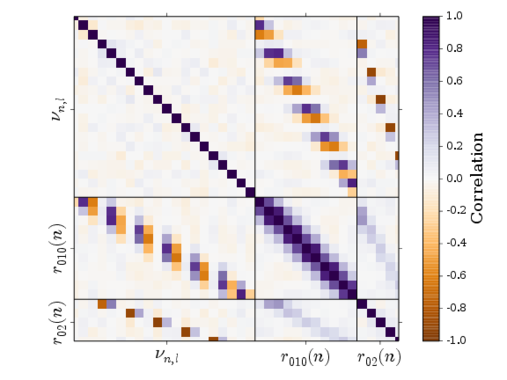}
   \caption{Correlation matrix of all frequencies and ratios for KIC 8349582.  The grid represents the matrix and hence the identity elements are all correlation 1.0.  The matrix is constructed so that frequencies and ratios are grouped separately.  If each matrix element is labelled $[i,j]$ then the first set of $i$ contains the mode frequencies, the second set contains the $r_{010}$, and the final set the $r_{02}$.}   \label{fig::cor_8349582}\end{figure}
\clearpage
\begin{table}
   \caption{Mode frequencies and statistics for KIC 8349582.}
   \begin{tabular}{ccccc}
   \hline 
   \hline 
   $n$ & $l$ & Frequency & 68\% credible & $ \ln{K}$  \\ 
 & & ($\rm \mu Hz$) & ($\rm \mu Hz$) & \\ 
   \hline 
   14 & 1 & 1322.53 & 0.74 & $>6$ \\ 
   15 & 0 & 1368.77 & 0.71 & 1.26 \\ 
   15 & 1 & 1404.12 & 0.42 & 1.42 \\ 
   15 & 2 & 1446.6 & 0.45 & 0.57 \\ 
   16 & 0 & 1450.72 & 0.36 & 1.84 \\ 
   16 & 1 & 1486.74 & 0.18 & $>6$ \\ 
   16 & 2 & 1529.23 & 0.32 & 0.87 \\ 
   17 & 0 & 1534.84 & 0.16 & 2.33 \\ 
   17 & 1 & 1570.3 & 0.12 & $>6$ \\ 
   17 & 2 & 1612.81 & 0.42 & 0.19 \\ 
   18 & 0 & 1617.95 & 0.14 & 2.86 \\ 
   18 & 1 & 1653.98 & 0.12 & $>6$ \\ 
   18 & 2 & 1696.88 & 0.34 & 0.49 \\ 
   19 & 0 & 1701.13 & 0.14 & 3.3 \\ 
   19 & 1 & 1737.04 & 0.14 & $>6$ \\ 
   19 & 2 & 1780.62 & 0.53 & 0.22 \\ 
   20 & 0 & 1784.5 & 0.23 & 2.94 \\ 
   20 & 1 & 1820.37 & 0.18 & $>6$ \\ 
   21 & 0 & 1868.42 & 0.8 & 1.14 \\ 
   21 & 1 & 1904.01 & 0.84 & 1.53 \\ 
   \hline 
   \end{tabular} 
   \label{tab::8349582}\end{table} 
\begin{table} 
   \caption{Ratios for KIC 8349582.} 
   \begin{tabular}{cccc} 
   \hline 
   \hline 
   Ratio type & $n$ & Ratio & 68\% credible interval \\ 
   \hline 
   $r_{10}$ & 15 & 0.067 & 0.007 \\ 
   $r_{01}$ & 16 & 0.067 & 0.004 \\ 
   $r_{10}$ & 16 & 0.07 & 0.003 \\ 
   $r_{01}$ & 17 & 0.074 & 0.002 \\ 
   $r_{10}$ & 17 & 0.073 & 0.002 \\ 
   $r_{01}$ & 18 & 0.069 & 0.002 \\ 
   $r_{10}$ & 18 & 0.068 & 0.002 \\ 
   $r_{01}$ & 19 & 0.068 & 0.002 \\ 
   $r_{10}$ & 19 & 0.069 & 0.002 \\ 
   $r_{01}$ & 20 & 0.07 & 0.003 \\ 
   $r_{10}$ & 20 & 0.072 & 0.005 \\ 
   $r_{02}$ & 16 & 0.05 & 0.007 \\ 
   $r_{02}$ & 17 & 0.067 & 0.004 \\ 
   $r_{02}$ & 18 & 0.061 & 0.005 \\ 
   $r_{02}$ & 19 & 0.051 & 0.004 \\ 
   $r_{02}$ & 20 & 0.047 & 0.007 \\ 
   \hline 
   \end{tabular} 
   \label{tab::rat_8349582}\end{table} 
\clearpage
\subsubsection{8478994}
\begin{figure}
   \includegraphics[width=80mm]{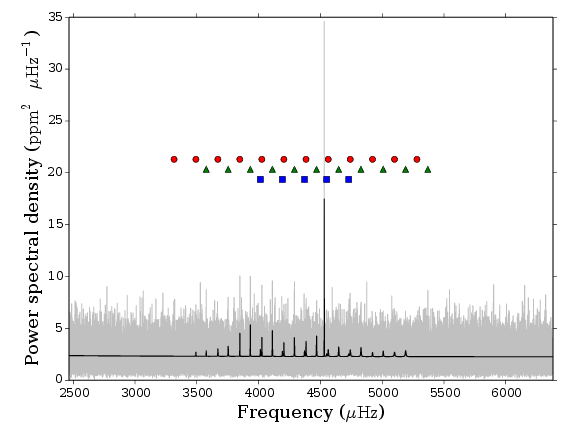}
   \includegraphics[width=80mm]{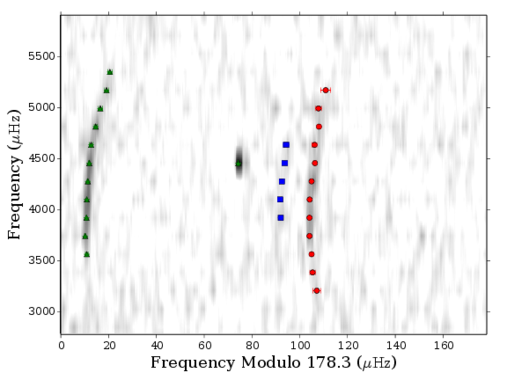}
   \caption{Power spectrum and echelle diagram for KIC 8478994.  Top: Power spectrum with data in grey smoothed over $3 \; \rm \mu Hz$ and best model in black.  Bottom: Echelle diagram with power in grey-scale.  Both: Mode frequencies are marked as: radial modes with red circles; dipole modes with green diamonds; quadrapole modes with blue squares; and octopole modes with yellow pentagons.}   \label{fig::8478994}\end{figure}
\begin{figure}
   \includegraphics[width=80mm]{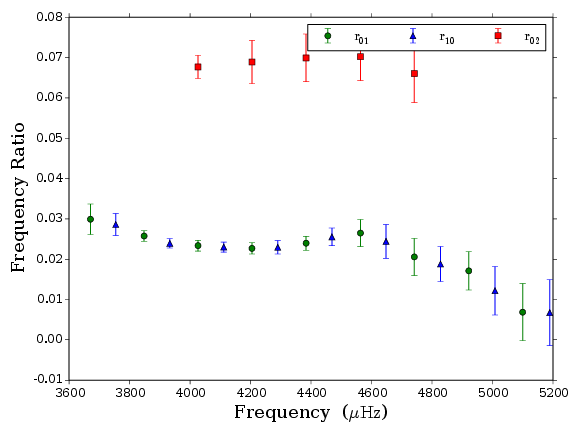}
   \caption{Ratios and $67 \%$ confidence intervals as a function of frequency for KIC 8478994.}   \label{fig::rat_8478994}\end{figure}
\begin{figure}
   \includegraphics[width=80mm]{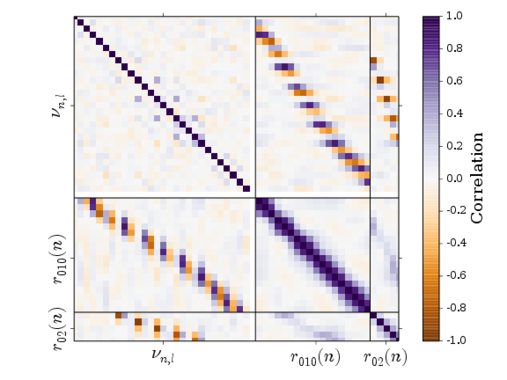}
   \caption{Correlation matrix of all frequencies and ratios for KIC 8478994.  The grid represents the matrix and hence the identity elements are all correlation 1.0.  The matrix is constructed so that frequencies and ratios are grouped separately.  If each matrix element is labelled $[i,j]$ then the first set of $i$ contains the mode frequencies, the second set contains the $r_{010}$, and the final set the $r_{02}$.}   \label{fig::cor_8478994}\end{figure}
\clearpage
\begin{table}
   \caption{Mode frequencies and statistics for KIC 8478994.}
   \begin{tabular}{ccccc}
   \hline 
   \hline 
   $n$ & $l$ & Frequency & 68\% credible & $ \ln{K}$  \\ 
 & & ($\rm \mu Hz$) & ($\rm \mu Hz$) & \\ 
   \hline 
   19 & 0 & 3316.21 & 1.61 & 1.0 \\ 
   20 & 0 & 3492.73 & 1.16 & 0.93 \\ 
   20 & 1 & 3576.49 & 0.55 & 1.39 \\ 
   21 & 0 & 3670.58 & 0.71 & 0.99 \\ 
   21 & 1 & 3754.12 & 0.34 & $>6$ \\ 
   22 & 0 & 3847.98 & 0.13 & $>6$ \\ 
   22 & 1 & 3932.91 & 0.15 & $>6$ \\ 
   22 & 2 & 4014.21 & 0.46 & 1.07 \\ 
   23 & 0 & 4026.26 & 0.27 & 2.97 \\ 
   23 & 1 & 4111.32 & 0.15 & $>6$ \\ 
   23 & 2 & 4192.3 & 0.92 & 0.1 \\ 
   24 & 0 & 4204.63 & 0.22 & 2.58 \\ 
   24 & 1 & 4290.03 & 0.29 & $>6$ \\ 
   24 & 2 & 4371.23 & 1.03 & 0.17 \\ 
   25 & 0 & 4383.7 & 0.33 & 3.24 \\ 
   25 & 1 & 4468.93 & 0.22 & $>6$ \\ 
   25 & 2 & 4550.79 & 1.03 & -0.02 \\ 
   26 & 0 & 4563.4 & 0.67 & 1.58 \\ 
   26 & 1 & 4647.97 & 0.61 & $>6$ \\ 
   26 & 2 & 4729.63 & 1.24 & -0.03 \\ 
   27 & 0 & 4741.53 & 0.95 & 2.21 \\ 
   27 & 1 & 4828.17 & 0.6 & $>6$ \\ 
   28 & 0 & 4921.64 & 0.82 & 1.43 \\ 
   28 & 1 & 5008.38 & 0.82 & 1.42 \\ 
   29 & 0 & 5099.72 & 1.33 & 2.66 \\ 
   29 & 1 & 5189.28 & 0.83 & 1.67 \\ 
   30 & 0 & 5281.02 & 2.15 & 0.87 \\ 
   30 & 1 & 5368.96 & 0.89 & 1.05 \\ 
   \hline 
   \end{tabular} 
   \label{tab::8478994}\end{table} 
\begin{table} 
   \caption{Ratios for KIC 8478994.} 
   \begin{tabular}{cccc} 
   \hline 
   \hline 
   Ratio type & $n$ & Ratio & 68\% credible interval \\ 
   \hline 
   $r_{01}$ & 21 & 0.03 & 0.004 \\ 
   $r_{10}$ & 21 & 0.029 & 0.003 \\ 
   $r_{01}$ & 22 & 0.026 & 0.001 \\ 
   $r_{10}$ & 22 & 0.024 & 0.001 \\ 
   $r_{01}$ & 23 & 0.023 & 0.001 \\ 
   $r_{10}$ & 23 & 0.023 & 0.001 \\ 
   $r_{01}$ & 24 & 0.023 & 0.001 \\ 
   $r_{10}$ & 24 & 0.023 & 0.002 \\ 
   $r_{01}$ & 25 & 0.024 & 0.002 \\ 
   $r_{10}$ & 25 & 0.026 & 0.002 \\ 
   $r_{01}$ & 26 & 0.026 & 0.003 \\ 
   $r_{10}$ & 26 & 0.024 & 0.004 \\ 
   $r_{01}$ & 27 & 0.021 & 0.005 \\ 
   $r_{10}$ & 27 & 0.019 & 0.004 \\ 
   $r_{01}$ & 28 & 0.017 & 0.005 \\ 
   $r_{10}$ & 28 & 0.012 & 0.006 \\ 
   $r_{01}$ & 29 & 0.007 & 0.007 \\ 
   $r_{10}$ & 29 & 0.007 & 0.008 \\ 
   $r_{02}$ & 23 & 0.068 & 0.003 \\ 
   $r_{02}$ & 24 & 0.069 & 0.005 \\ 
   $r_{02}$ & 25 & 0.07 & 0.006 \\ 
   $r_{02}$ & 26 & 0.07 & 0.006 \\ 
   $r_{02}$ & 27 & 0.066 & 0.007 \\ 
   \hline 
   \end{tabular} 
   \label{tab::rat_8478994}\end{table} 
\clearpage
\subsubsection{8494142}
\begin{figure}
   \includegraphics[width=80mm]{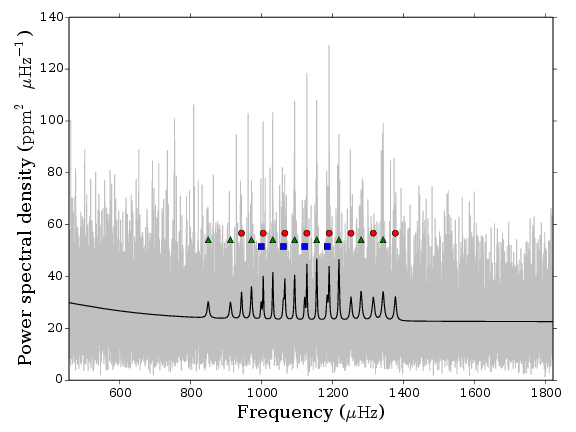}
   \includegraphics[width=80mm]{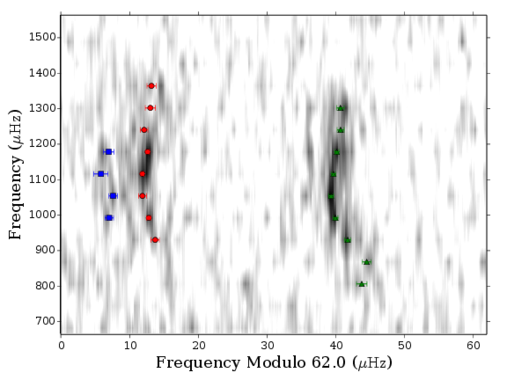}
   \caption{Power spectrum and echelle diagram for KIC 8494142.  Top: Power spectrum with data in grey smoothed over $3 \; \rm \mu Hz$ and best model in black.  Bottom: Echelle diagram with power in grey-scale.  Both: Mode frequencies are marked as: radial modes with red circles; dipole modes with green diamonds; quadrapole modes with blue squares; and octopole modes with yellow pentagons.}   \label{fig::8494142}\end{figure}
\begin{figure}
   \includegraphics[width=80mm]{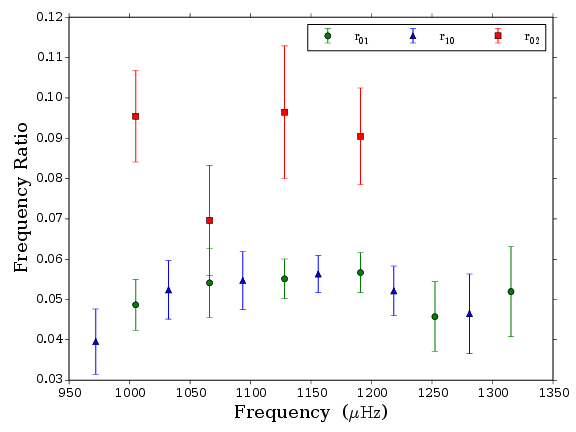}
   \caption{Ratios and $67 \%$ confidence intervals as a function of frequency for KIC 8494142.}   \label{fig::rat_8494142}\end{figure}
\begin{figure}
   \includegraphics[width=80mm]{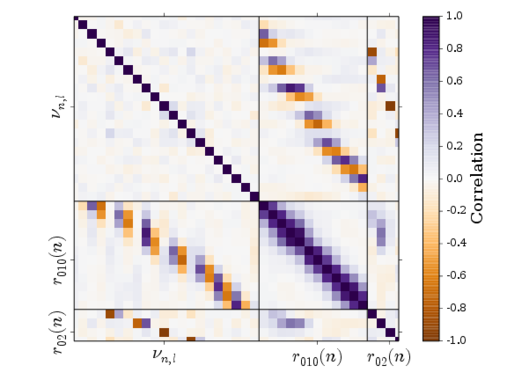}
   \caption{Correlation matrix of all frequencies and ratios for KIC 8494142.  The grid represents the matrix and hence the identity elements are all correlation 1.0.  The matrix is constructed so that frequencies and ratios are grouped separately.  If each matrix element is labelled $[i,j]$ then the first set of $i$ contains the mode frequencies, the second set contains the $r_{010}$, and the final set the $r_{02}$.}   \label{fig::cor_8494142}\end{figure}
\clearpage
\begin{table}
   \caption{Mode frequencies and statistics for KIC 8494142.}
   \begin{tabular}{ccccc}
   \hline 
   \hline 
   $n$ & $l$ & Frequency & 68\% credible & $ \ln{K}$  \\ 
 & & ($\rm \mu Hz$) & ($\rm \mu Hz$) & \\ 
   \hline 
   13 & 1 & 849.99 & 0.76 & 1.93 \\ 
   14 & 1 & 912.73 & 0.66 & $>6$ \\ 
   15 & 0 & 943.94 & 0.6 & 2.02 \\ 
   15 & 1 & 971.97 & 0.44 & $>6$ \\ 
   15 & 2 & 999.26 & 0.63 & 0.85 \\ 
   16 & 0 & 1005.02 & 0.28 & 3.01 \\ 
   16 & 1 & 1032.19 & 0.36 & $>6$ \\ 
   16 & 2 & 1061.83 & 0.6 & 1.63 \\ 
   17 & 0 & 1066.1 & 0.59 & 3.15 \\ 
   17 & 1 & 1093.57 & 0.36 & $>6$ \\ 
   17 & 2 & 1122.08 & 1.03 & 1.53 \\ 
   18 & 0 & 1128.12 & 0.24 & 3.93 \\ 
   18 & 1 & 1155.94 & 0.3 & $>6$ \\ 
   18 & 2 & 1185.22 & 0.73 & $>6$ \\ 
   19 & 0 & 1190.89 & 0.28 & $>6$ \\ 
   19 & 1 & 1218.43 & 0.29 & $>6$ \\ 
   20 & 0 & 1252.4 & 0.54 & 2.9 \\ 
   20 & 1 & 1281.0 & 0.51 & 2.47 \\ 
   21 & 0 & 1315.31 & 0.75 & 2.18 \\ 
   21 & 1 & 1342.99 & 0.55 & $>6$ \\ 
   22 & 0 & 1377.49 & 0.65 & 3.2 \\ 
   \hline 
   \end{tabular} 
   \label{tab::8494142}\end{table} 
\begin{table} 
   \caption{Ratios for KIC 8494142.} 
   \begin{tabular}{cccc} 
   \hline 
   \hline 
   Ratio type & $n$ & Ratio & 68\% credible interval \\ 
   \hline 
   $r_{10}$ & 15 & 0.04 & 0.008 \\ 
   $r_{01}$ & 16 & 0.049 & 0.006 \\ 
   $r_{10}$ & 16 & 0.052 & 0.007 \\ 
   $r_{01}$ & 17 & 0.054 & 0.009 \\ 
   $r_{10}$ & 17 & 0.055 & 0.007 \\ 
   $r_{01}$ & 18 & 0.055 & 0.005 \\ 
   $r_{10}$ & 18 & 0.056 & 0.005 \\ 
   $r_{01}$ & 19 & 0.057 & 0.005 \\ 
   $r_{10}$ & 19 & 0.052 & 0.006 \\ 
   $r_{01}$ & 20 & 0.046 & 0.009 \\ 
   $r_{10}$ & 20 & 0.046 & 0.01 \\ 
   $r_{01}$ & 21 & 0.052 & 0.011 \\ 
   $r_{02}$ & 16 & 0.095 & 0.011 \\ 
   $r_{02}$ & 17 & 0.07 & 0.014 \\ 
   $r_{02}$ & 18 & 0.096 & 0.016 \\ 
   $r_{02}$ & 19 & 0.09 & 0.012 \\ 
   \hline 
   \end{tabular} 
   \label{tab::rat_8494142}\end{table} 
\clearpage
\subsubsection{8554498}
\begin{figure}
   \includegraphics[width=80mm]{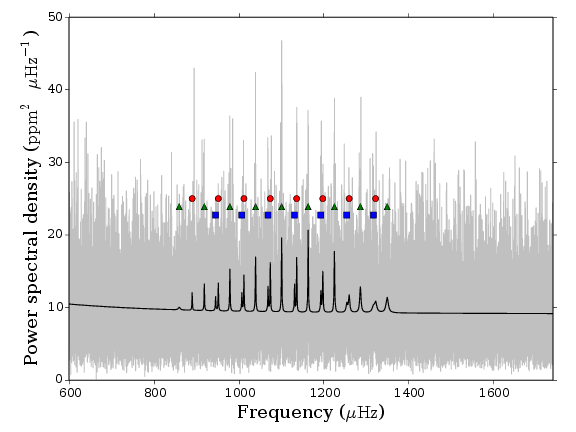}
   \includegraphics[width=80mm]{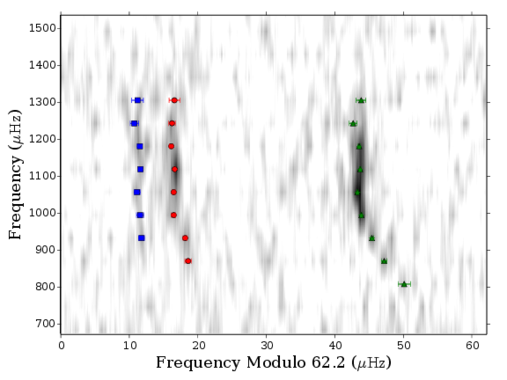}
   \caption{Power spectrum and echelle diagram for KIC 8554498.  Top: Power spectrum with data in grey smoothed over $3 \; \rm \mu Hz$ and best model in black.  Bottom: Echelle diagram with power in grey-scale.  Both: Mode frequencies are marked as: radial modes with red circles; dipole modes with green diamonds; quadrapole modes with blue squares; and octopole modes with yellow pentagons.}   \label{fig::8554498}\end{figure}
\begin{figure}
   \includegraphics[width=80mm]{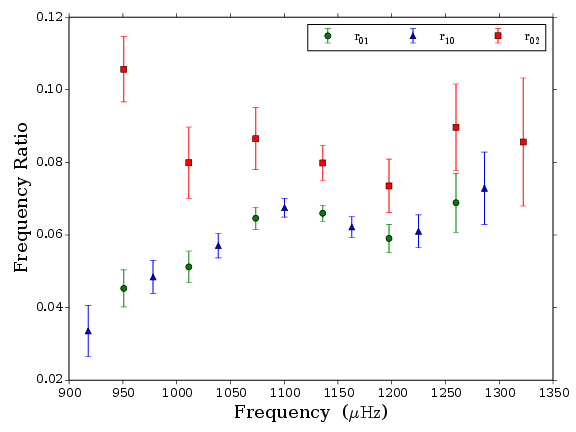}
   \caption{Ratios and $67 \%$ confidence intervals as a function of frequency for KIC 8554498.}   \label{fig::rat_8554498}\end{figure}
\begin{figure}
   \includegraphics[width=80mm]{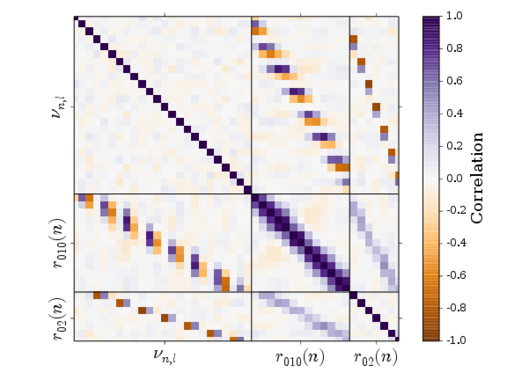}
   \caption{Correlation matrix of all frequencies and ratios for KIC 8554498.  The grid represents the matrix and hence the identity elements are all correlation 1.0.  The matrix is constructed so that frequencies and ratios are grouped separately.  If each matrix element is labelled $[i,j]$ then the first set of $i$ contains the mode frequencies, the second set contains the $r_{010}$, and the final set the $r_{02}$.}   \label{fig::cor_8554498}\end{figure}
\clearpage
\begin{table}
   \caption{Mode frequencies and statistics for KIC 8554498.}
   \begin{tabular}{ccccc}
   \hline 
   \hline 
   $n$ & $l$ & Frequency & 68\% credible & $ \ln{K}$  \\ 
 & & ($\rm \mu Hz$) & ($\rm \mu Hz$) & \\ 
   \hline 
   12 & 1 & 858.46 & 0.9 & 1.46148055213 \\ 
   13 & 0 & 889.1 & 0.47 & 1.25317679976 \\ 
   13 & 1 & 917.72 & 0.38 & 2.32886725415 \\ 
   13 & 2 & 944.46 & 0.44 & 0.768006877931 \\ 
   14 & 0 & 950.84 & 0.3 & 2.99602555427 \\ 
   14 & 1 & 978.11 & 0.2 & 2.89941107311 \\ 
   14 & 2 & 1006.44 & 0.52 & -0.172803622598 \\ 
   15 & 0 & 1011.34 & 0.33 & 3.13144959404 \\ 
   15 & 1 & 1038.75 & 0.14 & 3.93980605899 \\ 
   15 & 2 & 1068.19 & 0.47 & 0.531822771261 \\ 
   16 & 0 & 1073.52 & 0.23 & 3.31355771105 \\ 
   16 & 1 & 1100.35 & 0.12 & 4.28855218519 \\ 
   16 & 2 & 1130.85 & 0.27 & 1.09445490339 \\ 
   17 & 0 & 1135.85 & 0.13 & 4.09598958677 \\ 
   17 & 1 & 1162.92 & 0.13 & 4.20627269067 \\ 
   17 & 2 & 1192.94 & 0.38 & 2.03101625621 \\ 
   18 & 0 & 1197.53 & 0.29 & 3.88062619201 \\ 
   18 & 1 & 1224.96 & 0.16 & 4.12411412731 \\ 
   18 & 2 & 1254.31 & 0.59 & 1.03625592511 \\ 
   19 & 0 & 1259.82 & 0.47 & 3.26298520248 \\ 
   19 & 1 & 1286.26 & 0.57 & 2.63852220956 \\ 
   19 & 2 & 1316.96 & 0.85 & 0.546199079961 \\ 
   20 & 0 & 1322.34 & 0.78 & 2.58950211922 \\ 
   20 & 1 & 1349.61 & 0.71 & 2.20229920263 \\ 
   \hline 
   \end{tabular} 
   \label{tab::8554498}\end{table} 
\begin{table} 
   \caption{Ratios for KIC 8554498.} 
   \begin{tabular}{cccc} 
   \hline 
   \hline 
   Ratio type & $n$ & Ratio & 68\% credible interval \\ 
   \hline 
   $r_{10}$ & 13 & 0.034 & 0.007 \\ 
   $r_{01}$ & 14 & 0.045 & 0.005 \\ 
   $r_{10}$ & 14 & 0.048 & 0.004 \\ 
   $r_{01}$ & 15 & 0.051 & 0.004 \\ 
   $r_{10}$ & 15 & 0.057 & 0.003 \\ 
   $r_{01}$ & 16 & 0.065 & 0.003 \\ 
   $r_{10}$ & 16 & 0.068 & 0.003 \\ 
   $r_{01}$ & 17 & 0.066 & 0.002 \\ 
   $r_{10}$ & 17 & 0.062 & 0.003 \\ 
   $r_{01}$ & 18 & 0.059 & 0.004 \\ 
   $r_{10}$ & 18 & 0.061 & 0.005 \\ 
   $r_{01}$ & 19 & 0.069 & 0.008 \\ 
   $r_{10}$ & 19 & 0.073 & 0.01 \\ 
   $r_{02}$ & 14 & 0.106 & 0.009 \\ 
   $r_{02}$ & 15 & 0.08 & 0.01 \\ 
   $r_{02}$ & 16 & 0.087 & 0.009 \\ 
   $r_{02}$ & 17 & 0.08 & 0.005 \\ 
   $r_{02}$ & 18 & 0.074 & 0.007 \\ 
   $r_{02}$ & 19 & 0.09 & 0.012 \\ 
   $r_{02}$ & 20 & 0.086 & 0.018 \\ 
   \hline 
   \end{tabular} 
   \label{tab::rat_8554498}\end{table} 
\clearpage
\subsubsection{8684730}
\begin{figure}
   \includegraphics[width=80mm]{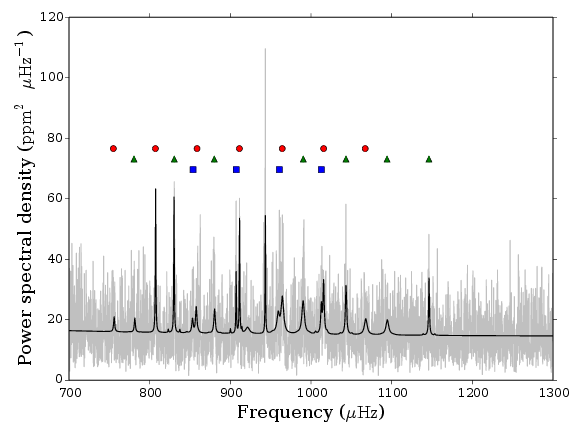}
   \includegraphics[width=80mm]{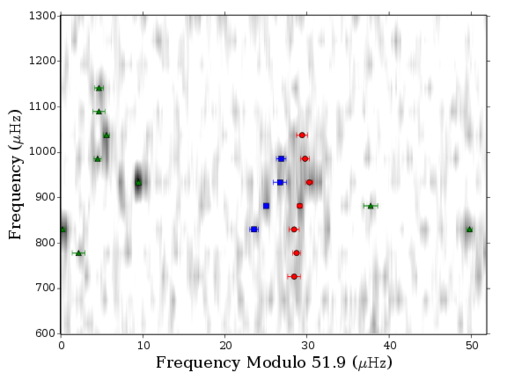}
   \caption{Power spectrum and echelle diagram for KIC 8684730.  Top: Power spectrum with data in grey smoothed over $3 \; \rm \mu Hz$ and best model in black.  Bottom: Echelle diagram with power in grey-scale.  Both: Mode frequencies are marked as: radial modes with red circles; dipole modes with green diamonds; quadrapole modes with blue squares; and octopole modes with yellow pentagons.}   \label{fig::8684730}\end{figure}
\clearpage
\begin{table}
   \caption{Mode frequencies and statistics for KIC 8684730.}
   \begin{tabular}{ccccc}
   \hline 
   \hline 
   $n$ & $l$ & Frequency & 68\% credible \\ 
 & & ($\rm \mu Hz$) & ($\rm \mu Hz$) \\ 
   \hline 
   13 & 0 & 754.85 & 0.78 \\ 
   13 & 1 & 780.48 & 0.78 \\ 
   14 & 0 & 807.04 & 0.49 \\ 
   14 & 1 & 830.41 & 0.22 \\ 
   14 & 2 & 853.74 & 0.55 \\ 
   15 & 0 & 858.63 & 0.63 \\ 
   15 & 1 & 880.01 & 0.35 \\ 
   15 & 2 & 907.1 & 0.29  \\ 
   16 & 0 & 911.19 & 0.34 \\ 
   16(01) & 1 & 919.83 & 0.86 \\
   16(02) & 1 & 943.45 & 0.18 \\
   16 & 2 & 960.68 & 0.81 \\ 
   17 & 0 & 964.26 & 0.41 \\ 
   17 & 1 & 990.35 & 0.43 \\ 
   17 & 2 & 1012.72 & 0.59 \\ 
   18 & 0 & 1015.62 & 0.53 \\ 
   18 & 1 & 1043.32 & 0.36 \\ 
   19 & 0 & 1067.15 & 0.68 \\ 
   19 & 1 & 1094.28 & 0.74 \\ 
   20 & 1 & 1146.17 & 0.56 \\ 
   \hline 
   \end{tabular} 
   \label{tab::8684730}\end{table} 
\clearpage
\subsubsection{8866102}
\begin{figure}
   \includegraphics[width=80mm]{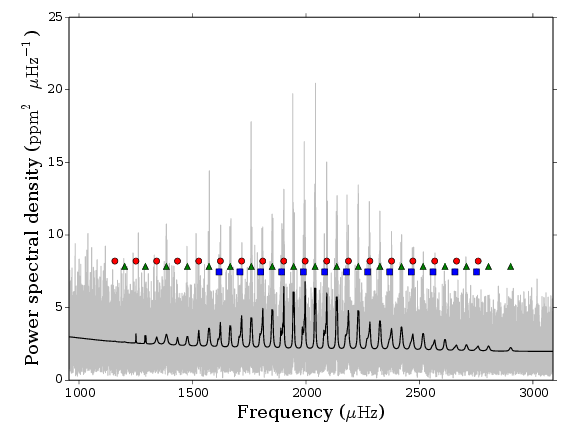}
   \includegraphics[width=80mm]{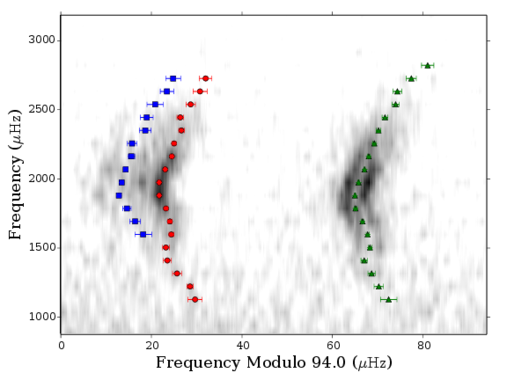}
   \caption{Power spectrum and echelle diagram for KIC 8866102.  Top: Power spectrum with data in grey smoothed over $3 \; \rm \mu Hz$ and best model in black.  Bottom: Echelle diagram with power in grey-scale.  Both: Mode frequencies are marked as: radial modes with red circles; dipole modes with green diamonds; quadrapole modes with blue squares; and octopole modes with yellow pentagons.}   \label{fig::8866102}\end{figure}
\begin{figure}
   \includegraphics[width=80mm]{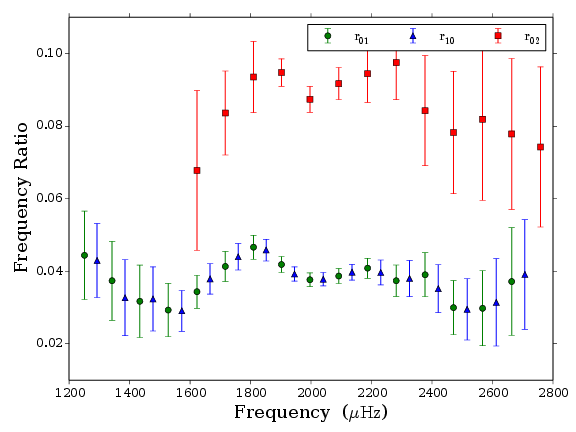}
   \caption{Ratios and $67 \%$ confidence intervals as a function of frequency for KIC 8866102.}   \label{fig::rat_8866102}\end{figure}
\begin{figure}
   \includegraphics[width=80mm]{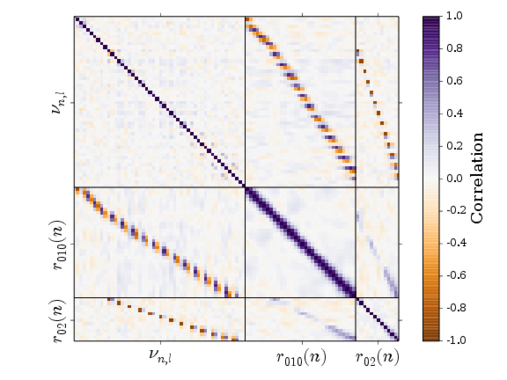}
   \caption{Correlation matrix of all frequencies and ratios for KIC 8866102.  The grid represents the matrix and hence the identity elements are all correlation 1.0.  The matrix is constructed so that frequencies and ratios are grouped separately.  If each matrix element is labelled $[i,j]$ then the first set of $i$ contains the mode frequencies, the second set contains the $r_{010}$, and the final set the $r_{02}$.}   \label{fig::cor_8866102}\end{figure}
\clearpage
\begin{table}
   \caption{Mode frequencies and statistics for KIC 8866102.}
   \begin{tabular}{ccccc}
   \hline 
   \hline 
   $n$ & $l$ & Frequency & 68\% credible & $ \ln{K}$  \\ 
 & & ($\rm \mu Hz$) & ($\rm \mu Hz$) & \\ 
   \hline 
   12 & 0 & 1157.94 & 1.54 & 1.31 \\ 
   12 & 1 & 1200.69 & 1.83 & 0.78 \\ 
   13 & 0 & 1250.85 & 0.64 & 1.67 \\ 
   13 & 1 & 1292.54 & 1.04 & 0.98 \\ 
   14 & 0 & 1342.01 & 0.99 & 1.97 \\ 
   14 & 1 & 1384.97 & 0.8 & 2.92 \\
\hline 
   \end{tabular} 
   \label{tab::8866102}\end{table} 
\begin{table} 
   \caption{Ratios for KIC 8866102.} 
   \begin{tabular}{cccc} 
   \hline 
   \hline 
   Ratio type & $n$ & Ratio & 68\% credible interval \\ 
   \hline 
   $r_{01}$ & 13 & 0.044 & 0.012 \\ 
   $r_{10}$ & 13 & 0.043 & 0.01 \\ 
   $r_{01}$ & 14 & 0.037 & 0.011 \\ 
   $r_{10}$ & 14 & 0.033 & 0.01 \\ 
   $r_{01}$ & 15 & 0.032 & 0.01 \\ 
   $r_{10}$ & 15 & 0.032 & 0.009 \\ 
   $r_{01}$ & 16 & 0.029 & 0.007 \\ 
   $r_{10}$ & 16 & 0.029 & 0.006 \\ 
   $r_{01}$ & 17 & 0.034 & 0.005 \\ 
   $r_{10}$ & 17 & 0.038 & 0.004 \\ 
   $r_{01}$ & 18 & 0.041 & 0.004 \\ 
   $r_{10}$ & 18 & 0.044 & 0.004 \\ 
   $r_{01}$ & 19 & 0.047 & 0.003 \\ 
   $r_{10}$ & 19 & 0.046 & 0.003 \\ 
   $r_{01}$ & 20 & 0.042 & 0.002 \\ 
   $r_{10}$ & 20 & 0.039 & 0.002 \\ 
   $r_{01}$ & 21 & 0.038 & 0.002 \\ 
   $r_{10}$ & 21 & 0.038 & 0.002 \\ 
   $r_{01}$ & 22 & 0.039 & 0.002 \\ 
   $r_{10}$ & 22 & 0.04 & 0.002 \\ 
   $r_{01}$ & 23 & 0.041 & 0.003 \\ 
   $r_{10}$ & 23 & 0.04 & 0.003 \\ 
   $r_{01}$ & 24 & 0.037 & 0.004 \\ 
   $r_{10}$ & 24 & 0.038 & 0.005 \\ 
   $r_{01}$ & 25 & 0.039 & 0.006 \\ 
   $r_{10}$ & 25 & 0.035 & 0.007 \\ 
   $r_{01}$ & 26 & 0.03 & 0.007 \\ 
   $r_{10}$ & 26 & 0.03 & 0.008 \\ 
   $r_{01}$ & 27 & 0.03 & 0.01 \\ 
   $r_{10}$ & 27 & 0.031 & 0.012 \\ 
   $r_{01}$ & 28 & 0.037 & 0.015 \\ 
   $r_{10}$ & 28 & 0.039 & 0.015 \\ 
   $r_{02}$ & 17 & 0.068 & 0.022 \\ 
   $r_{02}$ & 18 & 0.084 & 0.012 \\ 
   $r_{02}$ & 19 & 0.094 & 0.01 \\ 
   $r_{02}$ & 20 & 0.095 & 0.004 \\ 
   $r_{02}$ & 21 & 0.087 & 0.004 \\ 
   $r_{02}$ & 22 & 0.092 & 0.004 \\ 
   $r_{02}$ & 23 & 0.094 & 0.008 \\ 
   $r_{02}$ & 24 & 0.098 & 0.01 \\ 
   $r_{02}$ & 25 & 0.084 & 0.015 \\ 
   $r_{02}$ & 26 & 0.078 & 0.017 \\ 
   $r_{02}$ & 27 & 0.082 & 0.022 \\ 
   $r_{02}$ & 28 & 0.078 & 0.021 \\ 
   $r_{02}$ & 29 & 0.074 & 0.022 \\ 
   \hline 
   \end{tabular} 
   \label{tab::rat_8866102}\end{table} 
\clearpage
\subsubsection{9414417}
\begin{figure}
   \includegraphics[width=80mm]{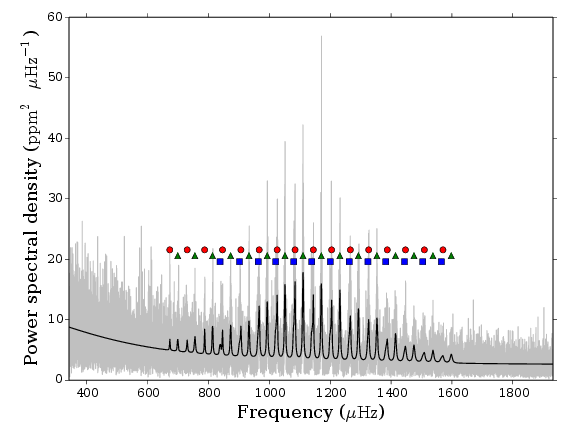}
   \includegraphics[width=80mm]{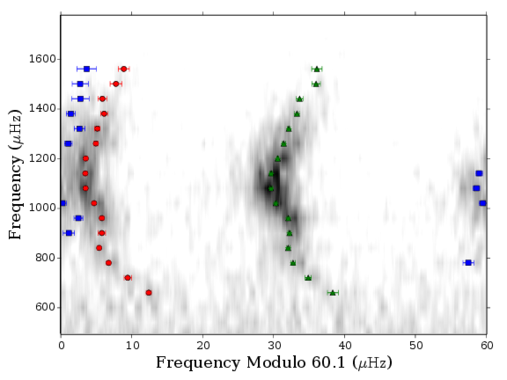}
   \caption{Power spectrum and echelle diagram for KIC 9414417.  Top: Power spectrum with data in grey smoothed over $3 \; \rm \mu Hz$ and best model in black.  Bottom: Echelle diagram with power in grey-scale.  Both: Mode frequencies are marked as: radial modes with red circles; dipole modes with green diamonds; quadrapole modes with blue squares; and octopole modes with yellow pentagons.}   \label{fig::9414417}\end{figure}
\begin{figure}
   \includegraphics[width=80mm]{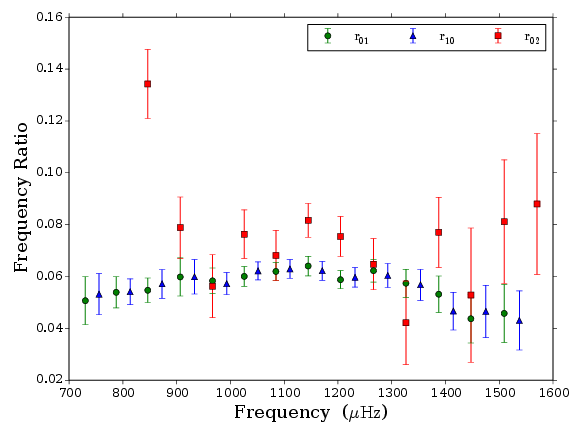}
   \caption{Ratios and $67 \%$ confidence intervals as a function of frequency for KIC 9414417.}   \label{fig::rat_9414417}\end{figure}
\begin{figure}
   \includegraphics[width=80mm]{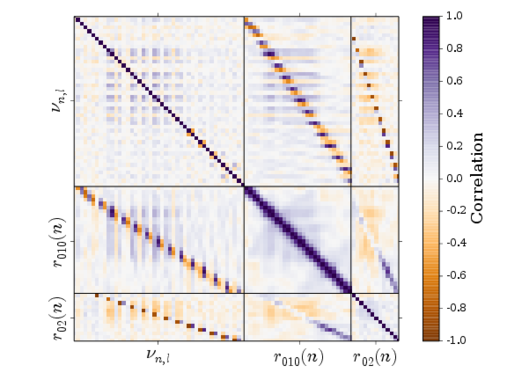}
   \caption{Correlation matrix of all frequencies and ratios for KIC 9414417.  The grid represents the matrix and hence the identity elements are all correlation 1.0.  The matrix is constructed so that frequencies and ratios are grouped separately.  If each matrix element is labelled $[i,j]$ then the first set of $i$ contains the mode frequencies, the second set contains the $r_{010}$, and the final set the $r_{02}$.}   \label{fig::cor_9414417}\end{figure}
\clearpage
\input{Results/9414417_freqs_table}
\input{Results/9414417_ratios_table}
\clearpage
\subsubsection{9592705}
\begin{figure}
   \includegraphics[width=80mm]{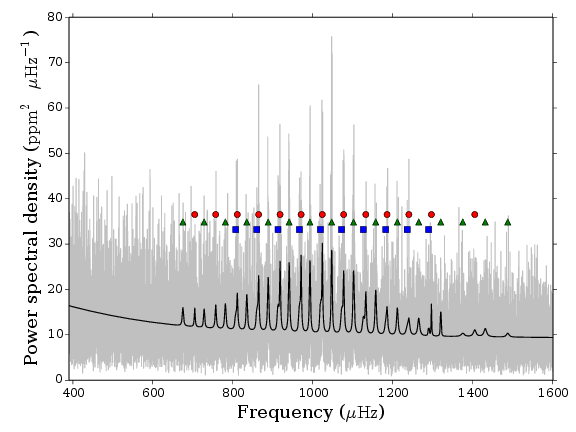}
   \includegraphics[width=80mm]{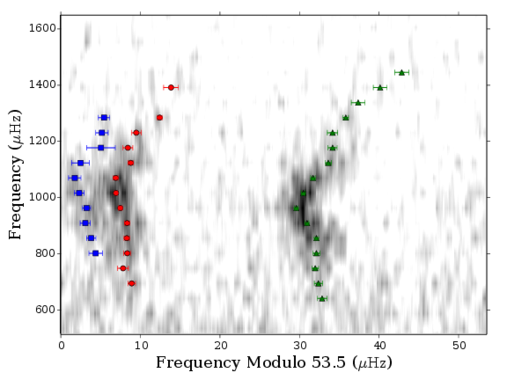}
   \caption{Power spectrum and echelle diagram for KIC 9592705.  Top: Power spectrum with data in grey smoothed over $3 \; \rm \mu Hz$ and best model in black.  Bottom: Echelle diagram with power in grey-scale.  Both: Mode frequencies are marked as: radial modes with red circles; dipole modes with green diamonds; quadrapole modes with blue squares; and octopole modes with yellow pentagons.}   \label{fig::9592705}\end{figure}
\begin{figure}
   \includegraphics[width=80mm]{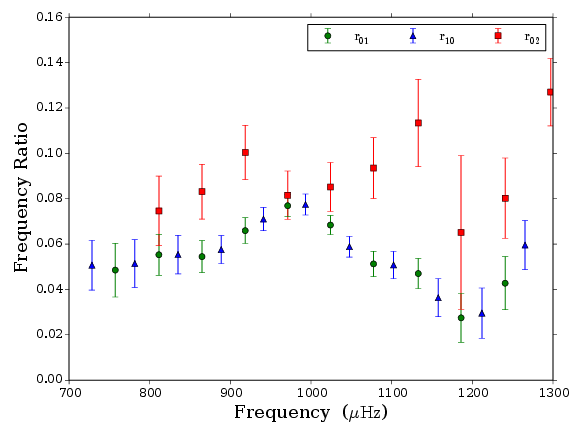}
   \caption{Ratios and $67 \%$ confidence intervals as a function of frequency for KIC 9592705.}   \label{fig::rat_9592705}\end{figure}
\begin{figure}
   \includegraphics[width=80mm]{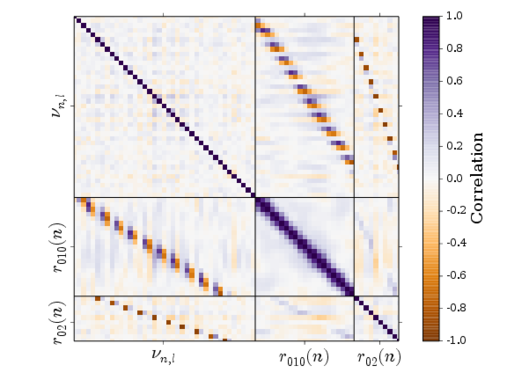}
   \caption{Correlation matrix of all frequencies and ratios for KIC 9592705.  The grid represents the matrix and hence the identity elements are all correlation 1.0.  The matrix is constructed so that frequencies and ratios are grouped separately.  If each matrix element is labelled $[i,j]$ then the first set of $i$ contains the mode frequencies, the second set contains the $r_{010}$, and the final set the $r_{02}$.}   \label{fig::cor_9592705}\end{figure}
\clearpage
\input{Results/9592705_freqs_table}
\input{Results/9592705_ratios_table}
\clearpage
\subsubsection{9955598}
\begin{figure}
   \includegraphics[width=80mm]{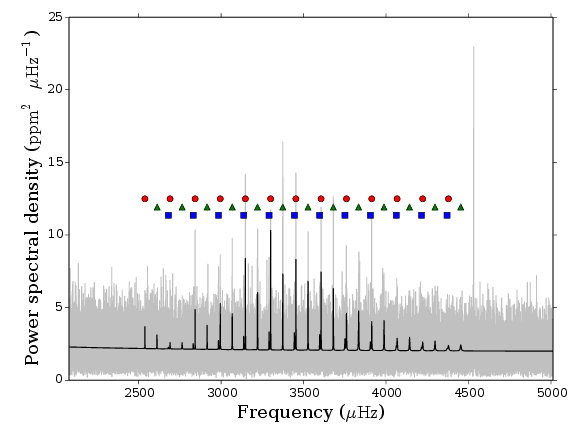}
   \includegraphics[width=80mm]{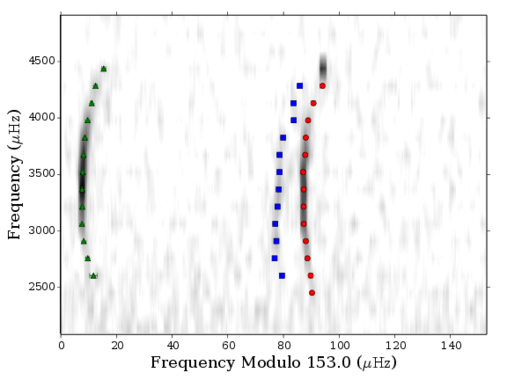}
   \caption{Power spectrum and echelle diagram for KIC 9955598.  Top: Power spectrum with data in grey smoothed over $3 \; \rm \mu Hz$ and best model in black.  Bottom: Echelle diagram with power in grey-scale.  Both: Mode frequencies are marked as: radial modes with red circles; dipole modes with green diamonds; quadrapole modes with blue squares; and octopole modes with yellow pentagons.}   \label{fig::9955598}\end{figure}
\begin{figure}
   \includegraphics[width=80mm]{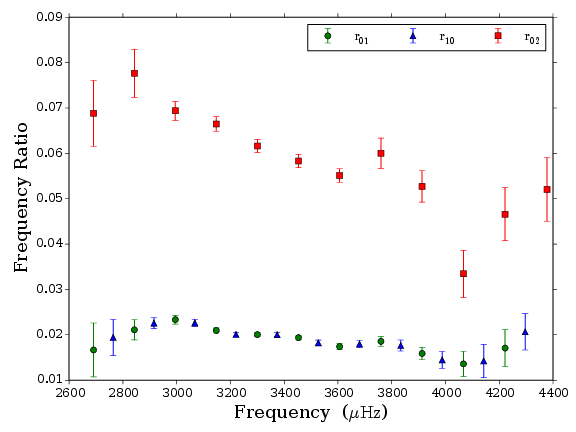}
   \caption{Ratios and $67 \%$ confidence intervals as a function of frequency for KIC 9955598.}   \label{fig::rat_9955598}\end{figure}
\begin{figure}
   \includegraphics[width=80mm]{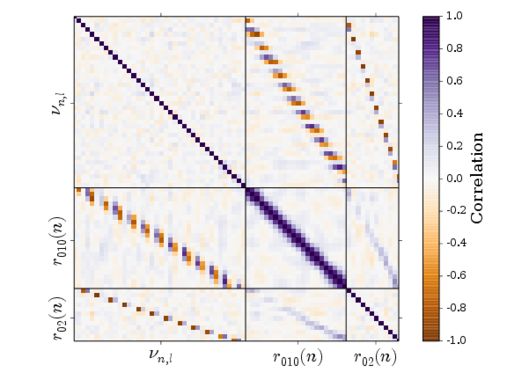}
   \caption{Correlation matrix of all frequencies and ratios for KIC 9955598.  The grid represents the matrix and hence the identity elements are all correlation 1.0.  The matrix is constructed so that frequencies and ratios are grouped separately.  If each matrix element is labelled $[i,j]$ then the first set of $i$ contains the mode frequencies, the second set contains the $r_{010}$, and the final set the $r_{02}$.}   \label{fig::cor_9955598}\end{figure}
\clearpage
\input{Results/9955598_freqs_table}
\input{Results/9955598_ratios_table}
\clearpage
\subsubsection{10514430}
\begin{figure}
   \includegraphics[width=80mm]{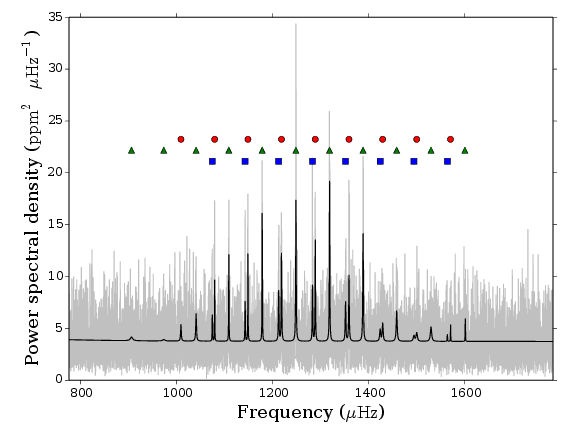}
   \includegraphics[width=80mm]{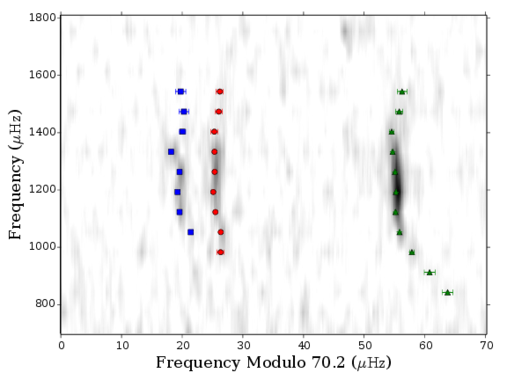}
   \caption{Power spectrum and echelle diagram for KIC 10514430.  Top: Power spectrum with data in grey smoothed over $3 \; \rm \mu Hz$ and best model in black.  Bottom: Echelle diagram with power in grey-scale.  Both: Mode frequencies are marked as: radial modes with red circles; dipole modes with green diamonds; quadrapole modes with blue squares; and octopole modes with yellow pentagons.}   \label{fig::10514430}\end{figure}
\begin{figure}
   \includegraphics[width=80mm]{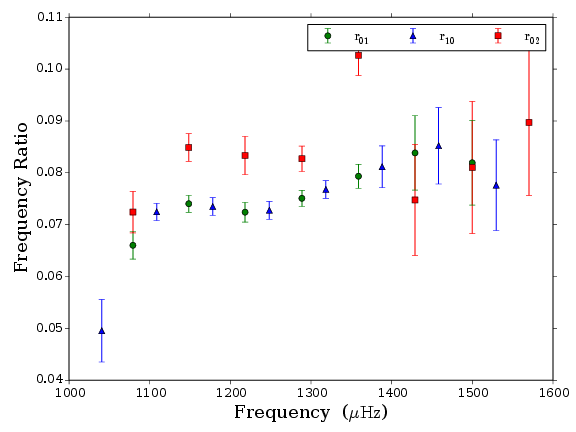}
   \caption{Ratios and $67 \%$ confidence intervals as a function of frequency for KIC 10514430.}   \label{fig::rat_10514430}\end{figure}
\begin{figure}
   \includegraphics[width=80mm]{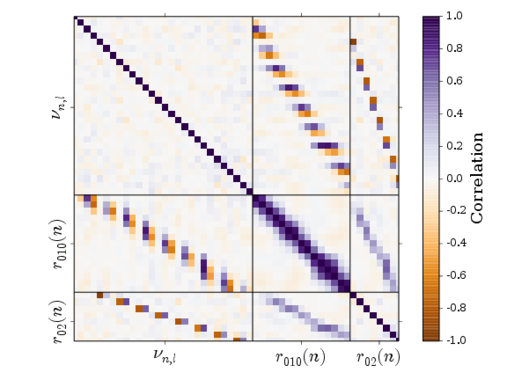}
   \caption{Correlation matrix of all frequencies and ratios for KIC 10514430.  The grid represents the matrix and hence the identity elements are all correlation 1.0.  The matrix is constructed so that frequencies and ratios are grouped separately.  If each matrix element is labelled $[i,j]$ then the first set of $i$ contains the mode frequencies, the second set contains the $r_{010}$, and the final set the $r_{02}$.}   \label{fig::cor_10514430}\end{figure}
\clearpage
\input{Results/10514430_freqs_table}
\input{Results/10514430_ratios_table}
\clearpage
\subsubsection{10586004}
\begin{figure}
   \includegraphics[width=80mm]{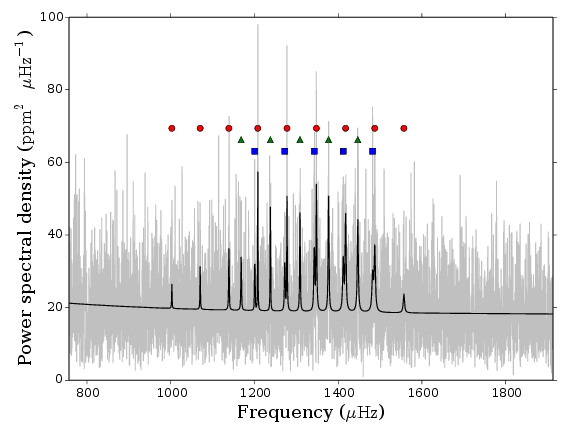}
   \includegraphics[width=80mm]{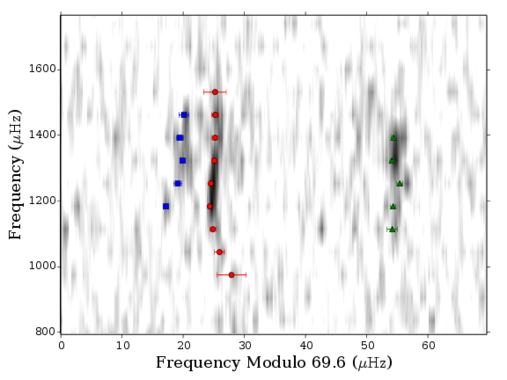}
   \caption{Power spectrum and echelle diagram for KIC 10586004.  Top: Power spectrum with data in grey smoothed over $3 \; \rm \mu Hz$ and best model in black.  Bottom: Echelle diagram with power in grey-scale.  Both: Mode frequencies are marked as: radial modes with red circles; dipole modes with green diamonds; quadrapole modes with blue squares; and octopole modes with yellow pentagons.}   \label{fig::10586004}\end{figure}
\begin{figure}
   \includegraphics[width=80mm]{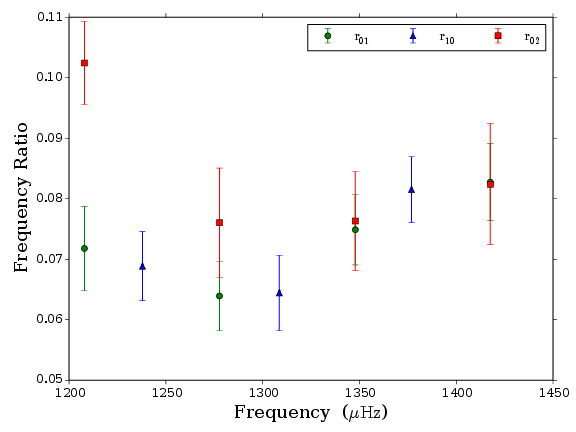}
   \caption{Ratios and $67 \%$ confidence intervals as a function of frequency for KIC 10586004.}   \label{fig::rat_10586004}\end{figure}
\begin{figure}
   \includegraphics[width=80mm]{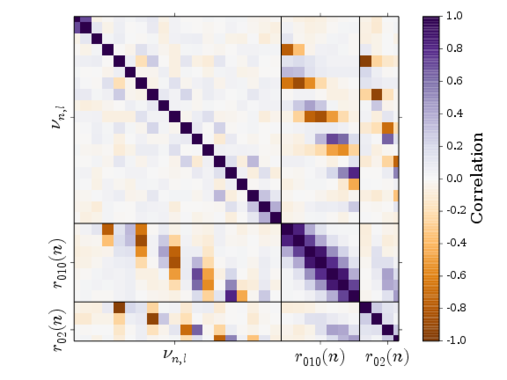}
   \caption{Correlation matrix of all frequencies and ratios for KIC 10586004.  The grid represents the matrix and hence the identity elements are all correlation 1.0.  The matrix is constructed so that frequencies and ratios are grouped separately.  If each matrix element is labelled $[i,j]$ then the first set of $i$ contains the mode frequencies, the second set contains the $r_{010}$, and the final set the $r_{02}$.}   \label{fig::cor_10586004}\end{figure}
\clearpage
\input{Results/10586004_freqs_table}
\input{Results/10586004_ratios_table}
\clearpage
\subsubsection{10666592}
\begin{figure}
   \includegraphics[width=80mm]{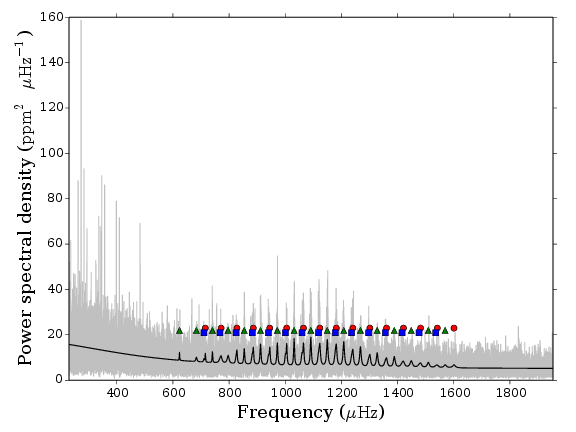}
   \includegraphics[width=80mm]{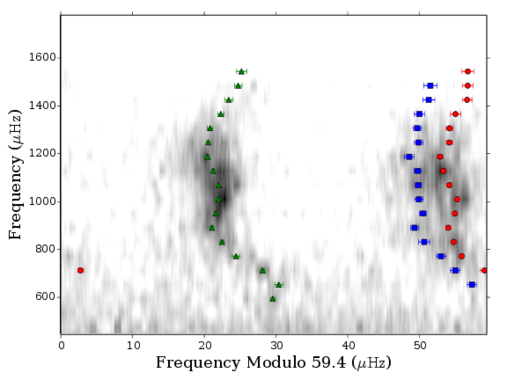}
   \caption{Power spectrum and echelle diagram for KIC 10666592.  Top: Power spectrum with data in grey smoothed over $3 \; \rm \mu Hz$ and best model in black.  Bottom: Echelle diagram with power in grey-scale.  Both: Mode frequencies are marked as: radial modes with red circles; dipole modes with green diamonds; quadrapole modes with blue squares; and octopole modes with yellow pentagons.}   \label{fig::10666592}\end{figure}
\begin{figure}
   \includegraphics[width=80mm]{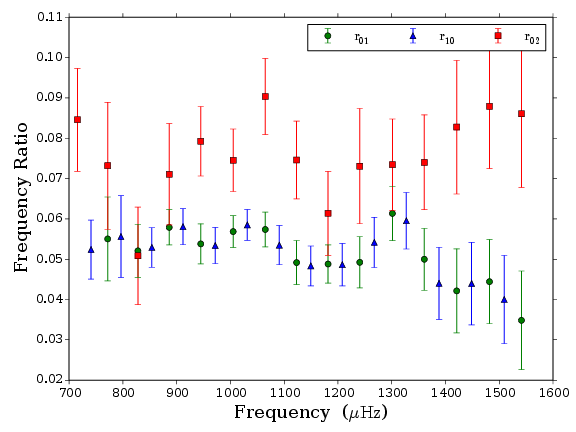}
   \caption{Ratios and $67 \%$ confidence intervals as a function of frequency for KIC 10666592.}   \label{fig::rat_10666592}\end{figure}
\begin{figure}
   \includegraphics[width=80mm]{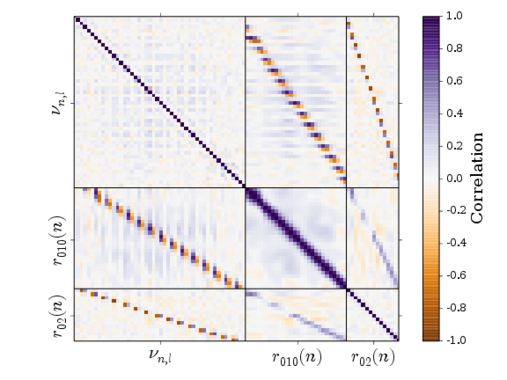}
   \caption{Correlation matrix of all frequencies and ratios for KIC 10666592.  The grid represents the matrix and hence the identity elements are all correlation 1.0.  The matrix is constructed so that frequencies and ratios are grouped separately.  If each matrix element is labelled $[i,j]$ then the first set of $i$ contains the mode frequencies, the second set contains the $r_{010}$, and the final set the $r_{02}$.}   \label{fig::cor_10666592}\end{figure}
\clearpage
\input{Results/10666592_freqs_table}
\input{Results/10666592_ratios_table}
\clearpage
\subsubsection{10963065}
\begin{figure}
   \includegraphics[width=80mm]{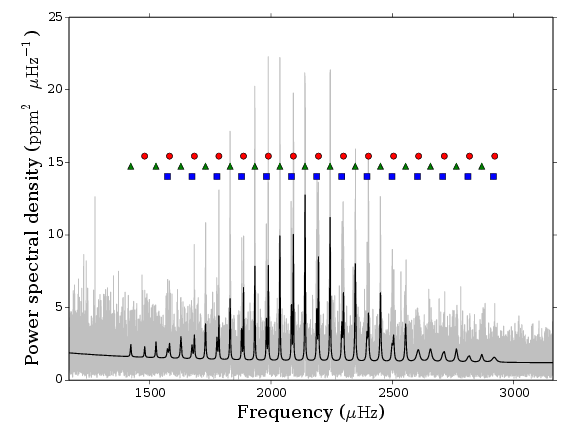}
   \includegraphics[width=80mm]{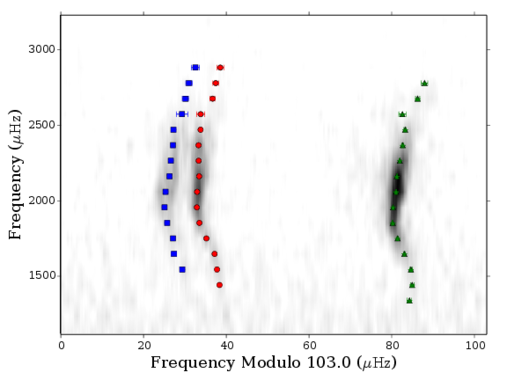}
   \caption{Power spectrum and echelle diagram for KIC 10963065.  Top: Power spectrum with data in grey smoothed over $3 \; \rm \mu Hz$ and best model in black.  Bottom: Echelle diagram with power in grey-scale.  Both: Mode frequencies are marked as: radial modes with red circles; dipole modes with green diamonds; quadrapole modes with blue squares; and octopole modes with yellow pentagons.}   \label{fig::10963065}\end{figure}
\begin{figure}
   \includegraphics[width=80mm]{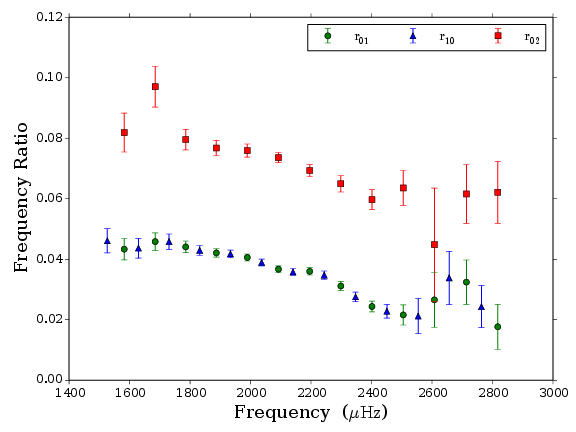}
   \caption{Ratios and $67 \%$ confidence intervals as a function of frequency for KIC 10963065.}   \label{fig::rat_10963065}\end{figure}
\begin{figure}
   \includegraphics[width=80mm]{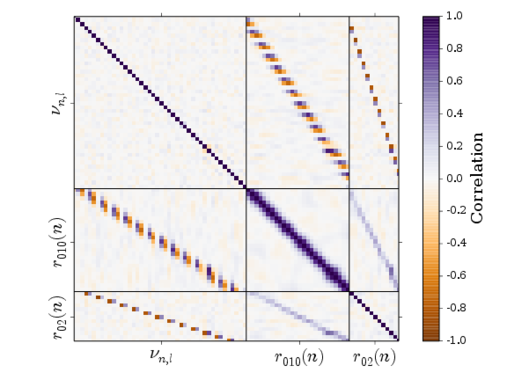}
   \caption{Correlation matrix of all frequencies and ratios for KIC 10963065.  The grid represents the matrix and hence the identity elements are all correlation 1.0.  The matrix is constructed so that frequencies and ratios are grouped separately.  If each matrix element is labelled $[i,j]$ then the first set of $i$ contains the mode frequencies, the second set contains the $r_{010}$, and the final set the $r_{02}$.}   \label{fig::cor_10963065}\end{figure}
\clearpage
\input{Results/10963065_freqs_table}
\input{Results/10963065_ratios_table}
\clearpage
\subsubsection{11133306}
\begin{figure}
   \includegraphics[width=80mm]{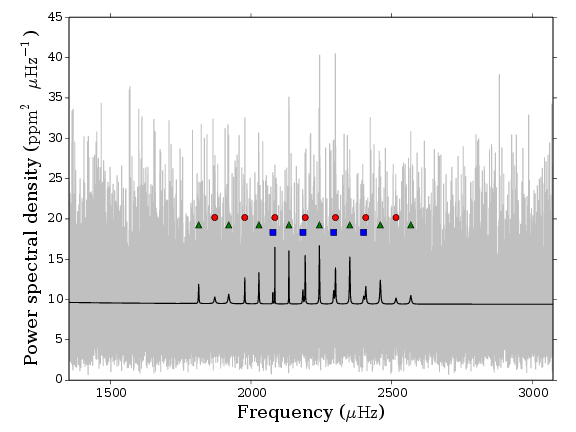}
   \includegraphics[width=80mm]{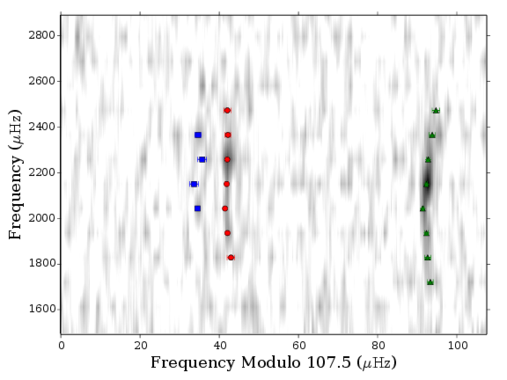}
   \caption{Power spectrum and echelle diagram for KIC 11133306.  Top: Power spectrum with data in grey smoothed over $3 \; \rm \mu Hz$ and best model in black.  Bottom: Echelle diagram with power in grey-scale.  Both: Mode frequencies are marked as: radial modes with red circles; dipole modes with green diamonds; quadrapole modes with blue squares; and octopole modes with yellow pentagons.}   \label{fig::11133306}\end{figure}
\begin{figure}
   \includegraphics[width=80mm]{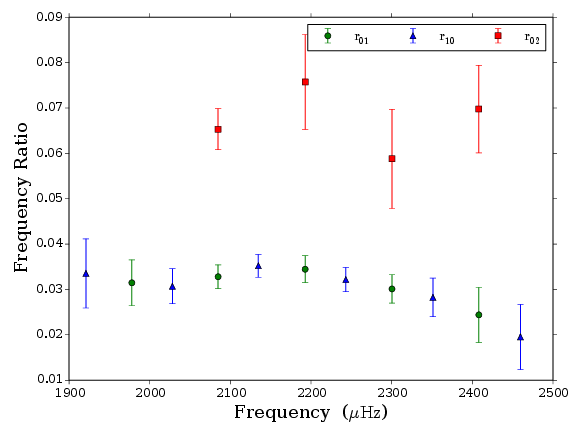}
   \caption{Ratios and $67 \%$ confidence intervals as a function of frequency for KIC 11133306.}   \label{fig::rat_11133306}\end{figure}
\begin{figure}
   \includegraphics[width=80mm]{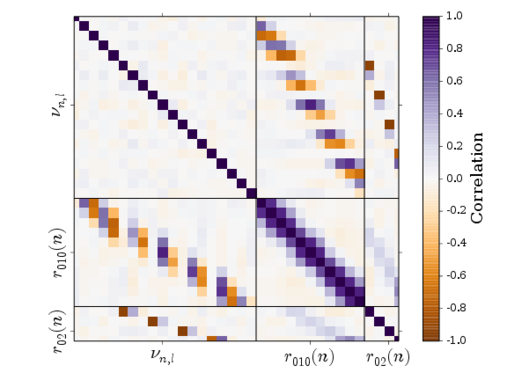}
   \caption{Correlation matrix of all frequencies and ratios for KIC 11133306.  The grid represents the matrix and hence the identity elements are all correlation 1.0.  The matrix is constructed so that frequencies and ratios are grouped separately.  If each matrix element is labelled $[i,j]$ then the first set of $i$ contains the mode frequencies, the second set contains the $r_{010}$, and the final set the $r_{02}$.}   \label{fig::cor_11133306}\end{figure}
\clearpage
\input{Results/11133306_freqs_table}
\input{Results/11133306_ratios_table}
\clearpage
\input{Results/11295426_freqs_table}
\input{Results/11295426_ratios_table}
\clearpage
\subsubsection{11401755}
\begin{figure}
   \includegraphics[width=80mm]{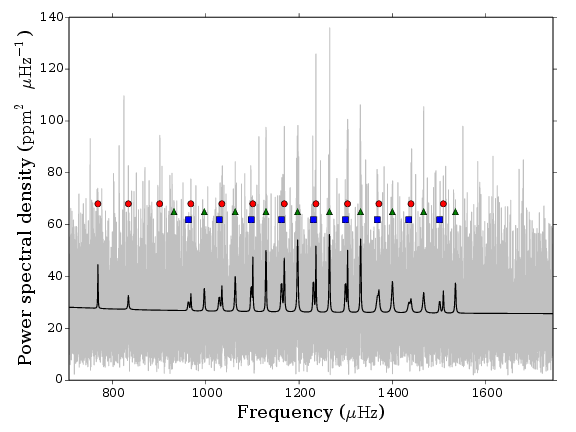}
   \includegraphics[width=80mm]{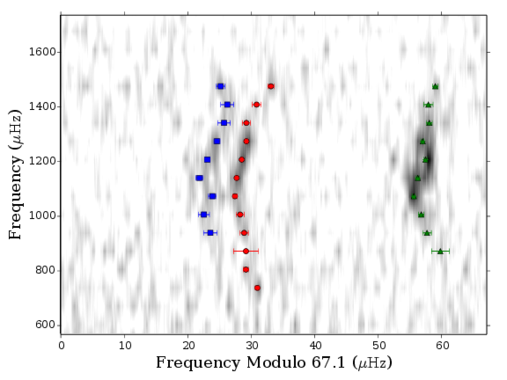}
   \caption{Power spectrum and echelle diagram for KIC 11401755.  Top: Power spectrum with data in grey smoothed over $3 \; \rm \mu Hz$ and best model in black.  Bottom: Echelle diagram with power in grey-scale.  Both: Mode frequencies are marked as: radial modes with red circles; dipole modes with green diamonds; quadrapole modes with blue squares; and octopole modes with yellow pentagons.}   \label{fig::11401755}\end{figure}
\begin{figure}
   \includegraphics[width=80mm]{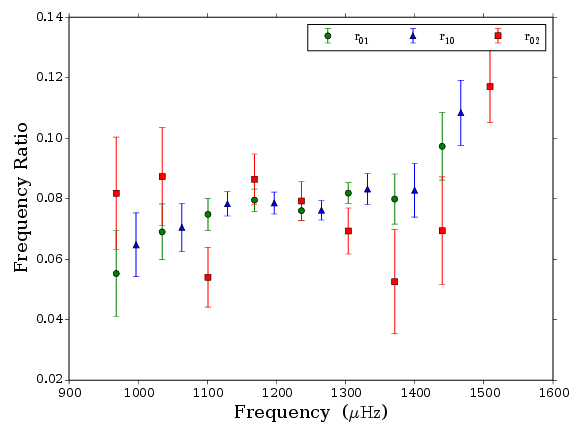}
   \caption{Ratios and $67 \%$ confidence intervals as a function of frequency for KIC 11401755.}   \label{fig::rat_11401755}\end{figure}
\begin{figure}
   \includegraphics[width=80mm]{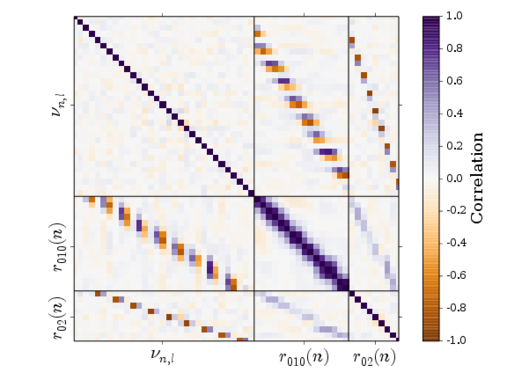}
   \caption{Correlation matrix of all frequencies and ratios for KIC 11401755.  The grid represents the matrix and hence the identity elements are all correlation 1.0.  The matrix is constructed so that frequencies and ratios are grouped separately.  If each matrix element is labelled $[i,j]$ then the first set of $i$ contains the mode frequencies, the second set contains the $r_{010}$, and the final set the $r_{02}$.}   \label{fig::cor_11401755}\end{figure}
\clearpage
\input{Results/11401755_freqs_table}
\input{Results/11401755_ratios_table}
\clearpage
\subsubsection{11807274}
\begin{figure}
   \includegraphics[width=80mm]{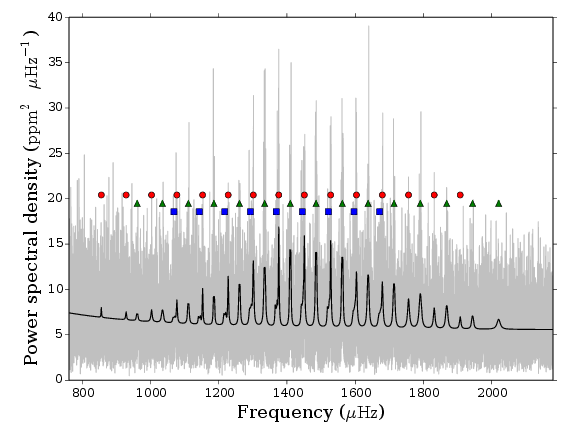}
   \includegraphics[width=80mm]{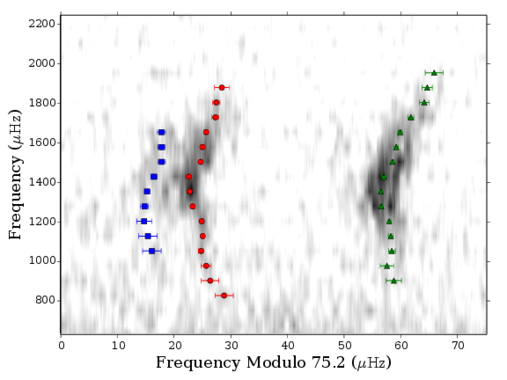}
   \caption{Power spectrum and echelle diagram for KIC 11807274.  Top: Power spectrum with data in grey smoothed over $3 \; \rm \mu Hz$ and best model in black.  Bottom: Echelle diagram with power in grey-scale.  Both: Mode frequencies are marked as: radial modes with red circles; dipole modes with green diamonds; quadrapole modes with blue squares; and octopole modes with yellow pentagons.}   \label{fig::11807274}\end{figure}
\begin{figure}
   \includegraphics[width=80mm]{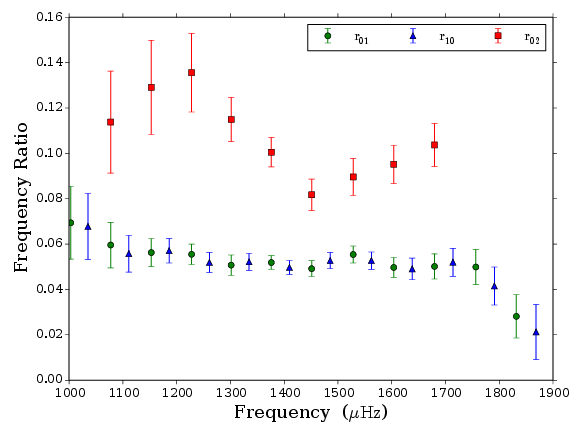}
   \caption{Ratios and $67 \%$ confidence intervals as a function of frequency for KIC 11807274.}   \label{fig::rat_11807274}\end{figure}
\begin{figure}
   \includegraphics[width=80mm]{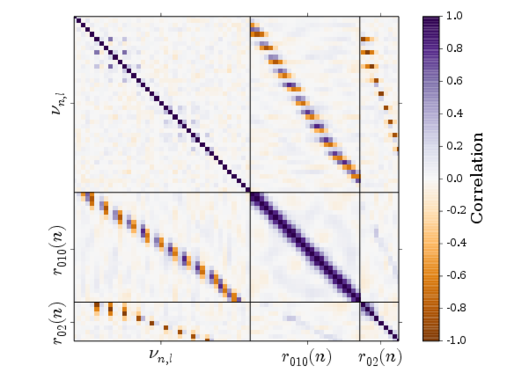}
   \caption{Correlation matrix of all frequencies and ratios for KIC 11807274.  The grid represents the matrix and hence the identity elements are all correlation 1.0.  The matrix is constructed so that frequencies and ratios are grouped separately.  If each matrix element is labelled $[i,j]$ then the first set of $i$ contains the mode frequencies, the second set contains the $r_{010}$, and the final set the $r_{02}$.}   \label{fig::cor_11807274}\end{figure}
\clearpage
\input{Results/11807274_freqs_table}
\input{Results/11807274_ratios_table}
\clearpage
\subsubsection{11853905}
\begin{figure}
   \includegraphics[width=80mm]{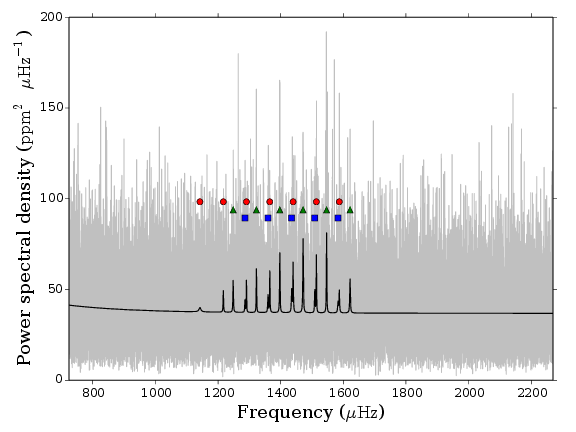}
   \includegraphics[width=80mm]{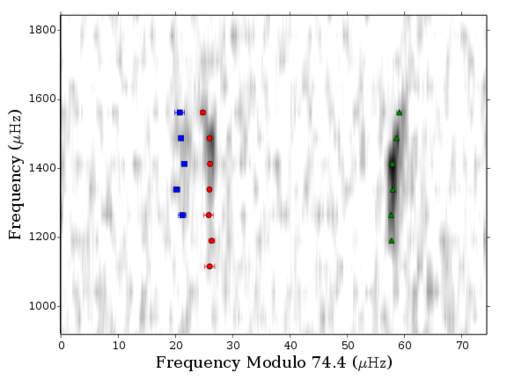}
   \caption{Power spectrum and echelle diagram for KIC 11853905.  Top: Power spectrum with data in grey smoothed over $3 \; \rm \mu Hz$ and best model in black.  Bottom: Echelle diagram with power in grey-scale.  Both: Mode frequencies are marked as: radial modes with red circles; dipole modes with green diamonds; quadrapole modes with blue squares; and octopole modes with yellow pentagons.}   \label{fig::11853905}\end{figure}
\begin{figure}
   \includegraphics[width=80mm]{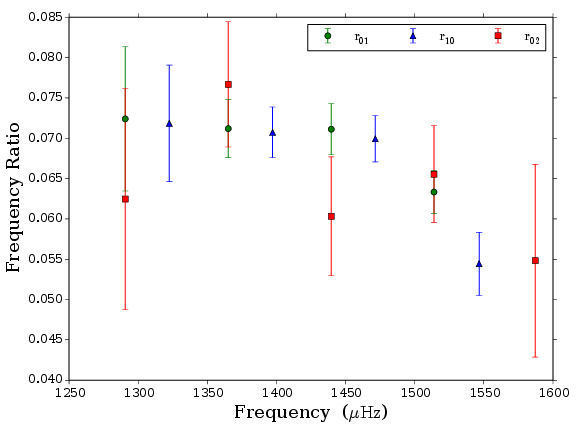}
   \caption{Ratios and $67 \%$ confidence intervals as a function of frequency for KIC 11853905.}   \label{fig::rat_11853905}\end{figure}
\begin{figure}
   \includegraphics[width=80mm]{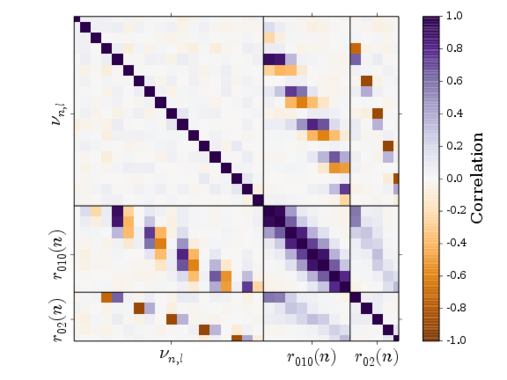}
   \caption{Correlation matrix of all frequencies and ratios for KIC 11853905.  The grid represents the matrix and hence the identity elements are all correlation 1.0.  The matrix is constructed so that frequencies and ratios are grouped separately.  If each matrix element is labelled $[i,j]$ then the first set of $i$ contains the mode frequencies, the second set contains the $r_{010}$, and the final set the $r_{02}$.}   \label{fig::cor_11853905}\end{figure}
\clearpage
\input{Results/11853905_freqs_table}
\input{Results/11853905_ratios_table}
\clearpage
\subsubsection{11904151}
\begin{figure}
   \includegraphics[width=80mm]{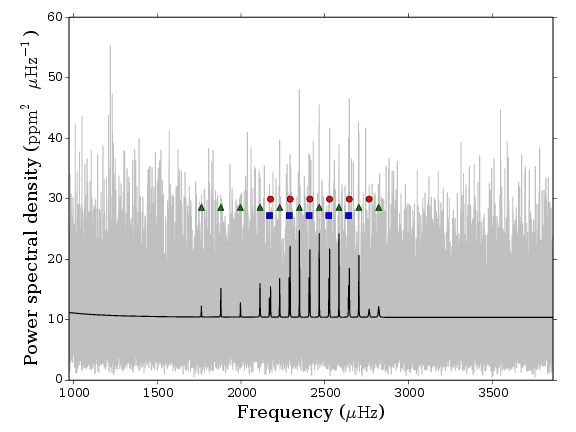}
   \includegraphics[width=80mm]{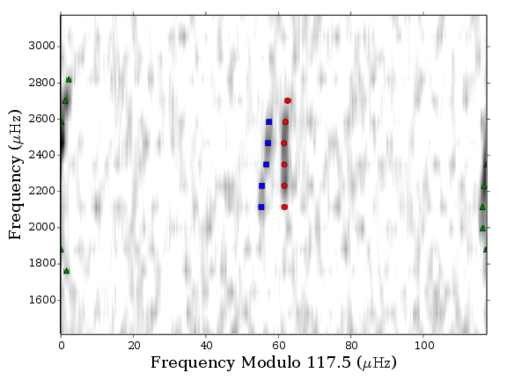}
   \caption{Power spectrum and echelle diagram for KIC 11904151.  Top: Power spectrum with data in grey smoothed over $3 \; \rm \mu Hz$ and best model in black.  Bottom: Echelle diagram with power in grey-scale.  Both: Mode frequencies are marked as: radial modes with red circles; dipole modes with green diamonds; quadrapole modes with blue squares; and octopole modes with yellow pentagons.}   \label{fig::11904151}\end{figure}
\begin{figure}
   \includegraphics[width=80mm]{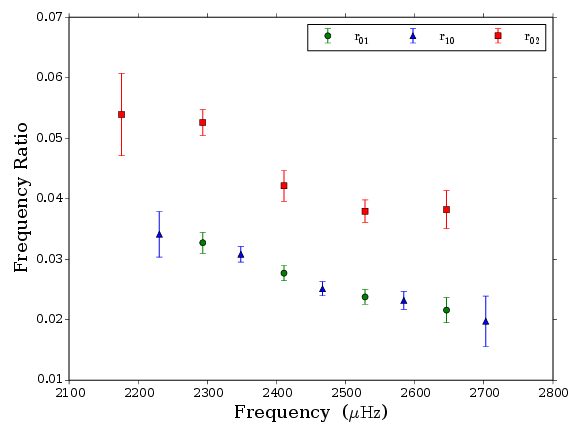}
   \caption{Ratios and $67 \%$ confidence intervals as a function of frequency for KIC 11904151.}   \label{fig::rat_11904151}\end{figure}
\begin{figure}
   \includegraphics[width=80mm]{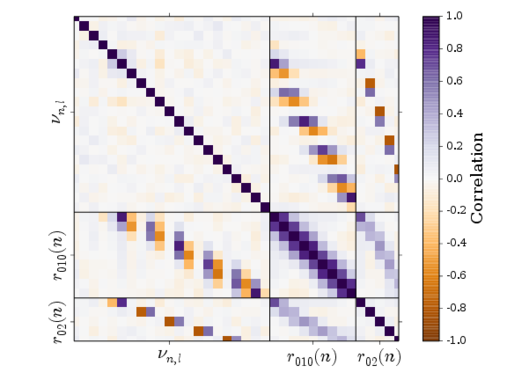}
   \caption{Correlation matrix of all frequencies and ratios for KIC 11904151.  The grid represents the matrix and hence the identity elements are all correlation 1.0.  The matrix is constructed so that frequencies and ratios are grouped separately.  If each matrix element is labelled $[i,j]$ then the first set of $i$ contains the mode frequencies, the second set contains the $r_{010}$, and the final set the $r_{02}$.}   \label{fig::cor_11904151}\end{figure}
\clearpage
\input{Results/11904151_freqs_table}
\input{Results/11904151_ratios_table}
\clearpage

\label{lastpage}
\end{document}